\documentclass[twocolumn,superscriptaddress,amsmath,amssymb,letter]{revtex4-2}
\usepackage{mathtools}    
\usepackage{amsthm}
\usepackage{graphicx}
\usepackage{dcolumn}
\usepackage{bm}
\usepackage{array} 
\usepackage[colorlinks=true, allcolors=blue]{hyperref}
\usepackage{physics}
\begin{document}
\title{A Unified Framework for the Non-Hermitian Localization: Boundary-Insensitive Modes and Electric-Magnetic Analogy}
\author{Zheng Wei}
\thanks{These authors contributed equally to the work}
\affiliation{Center for Advanced Quantum Studies, School of Physics and Astronomy, Beijing Normal University, Beijing 100875, China}
\affiliation{Key Laboratory of Multiscale Spin Physics(Ministry of Education), Beijing Normal University, Beijing 100875, China}

\author{Ji-Yao Fan}
\thanks{These authors contributed equally to the work}
\affiliation{Center for Advanced Quantum Studies, School of Physics and Astronomy, Beijing Normal University, Beijing 100875, China}
\affiliation{Key Laboratory of Multiscale Spin Physics(Ministry of Education), Beijing Normal University, Beijing 100875, China}

\author{Kui Cao}
\affiliation{Center for Advanced Quantum Studies, School of Physics and Astronomy, Beijing Normal University, Beijing 100875, China}
\affiliation{Key Laboratory of Multiscale Spin Physics(Ministry of Education), Beijing Normal University, Beijing 100875, China}

\author{Xin-Ran Ma}
\affiliation{Center for Advanced Quantum Studies, School of Physics and Astronomy, Beijing Normal University, Beijing 100875, China}
\affiliation{Key Laboratory of Multiscale Spin Physics(Ministry of Education), Beijing Normal University, Beijing 100875, China}

\author{Cui-Xian Guo}
\affiliation{Beijing Key Laboratory of Optical Detection Technology for Oil and Gas, China University of Petroleum-Beijing, Beijing 102249, China}
\affiliation{Basic Research Center for Energy Interdisciplinary, College of Science, China University of Petroleum-Beijing, Beijing 102249, China}

\author{Xue-Ping Ren}
\affiliation{Center for Advanced Quantum Studies, School of Physics and Astronomy, Beijing Normal University, Beijing 100875, China}
\affiliation{Key Laboratory of Multiscale Spin Physics(Ministry of Education), Beijing Normal University, Beijing 100875, China}

\author{Su-Peng Kou}
\email{spkou@bnu.edu.cn}
\affiliation{Center for Advanced Quantum Studies, School of Physics and Astronomy, Beijing Normal University, Beijing 100875, China}
\affiliation{Key Laboratory of Multiscale Spin Physics(Ministry of Education), Beijing Normal University, Beijing 100875, China}

\begin{abstract}
The non-Hermitian skin effect is fundamentally characterized by its sensitivity to boundary conditions, reflected in changes to the energy spectrum and boundary-localized eigenstates. Here, we demonstrate that a spatially inhomogeneous imaginary scalar potential field induces a skin effect that is insensitive to boundary conditions. Both the spectrum and eigenstate distribution remain invariant, a behavior not captured by existing theories. We attribute this anomaly to translational symmetry breaking induced by spatially varying imaginary potentials in finite systems. We further formulate a theory that universally predicts localization in single-particle non-Hermitian systems. This framework classifies skin effects into two fundamental types: electric, driven by imaginary scalar potentials, and magnetic, driven by imaginary vector potentials, and reveals a phase transition between them, where eigenstates become fully delocalized. Our work provides a unified theory for non-Hermitian localization, allowing full control over skin modes via potential engineering in various platforms like photonic crystals and cold-atom systems.
\end{abstract}

\maketitle

The field of non-Hermitian physics has recently experienced explosive growth, uncovering a series of phenomena that transcend the boundaries of traditional Hermitian systems \cite{el-ganainy_NonHermitianPhysicsPT_2018,gong_TopologicalPhasesNonHermitian_2018,kunst_BiorthogonalBulkBoundaryCorrespondence_2018,kawabata_SymmetryTopologyNonHermitian_2019,ashida_NonHermitianPhysics_2020}. Among these phenomena, one of the most notable examples is the Non-Hermitian Skin Effect (NHSE). This effect disrupts the extended Bloch wave behavior and is defined by a significant accumulation of eigenstates at the system’s boundaries. Owing to its extreme sensitivity to boundary conditions, the NHSE leads to substantial differences in spectra and eigenstates under varying boundary conditions, which challenges the conventional bulk-boundary correspondence principle \cite{yao_EdgeStatesTopological_2018,song_NonHermitianSkinEffect_2019,li_CriticalNonHermitianSkin_2020,yi_NonHermitianSkinModes_2020,yang_NonHermitianBulkBoundaryCorrespondence_2020,zhang_CorrespondenceWindingNumbers_2020,borgnia_NonHermitianBoundaryModes_2020,zhang_UniversalNonHermitianSkin_2022,peng_ManipulatingNonHermitianSkin_2022,li_ScalefreeLocalizationPT_2023,guo_ScaletailoredLocalizationIts_2024,wang_AmoebaFormulationNonBloch_2024}. By introducing nonreciprocal coupling, on-site dissipation \cite{lee_AnomalousEdgeState_2016,yi_NonHermitianSkinModes_2020,jiang_TunableNonHermitianSkin_2024,li_GainLossInducedHybridSkinTopological_2022}, or specific lattice geometries \cite{bhargava_NonHermitianSkinEffect_2021,lin_SteeringNonHermitianSkin_2021,jiang_ReciprocatingBipolarNonHermitian_2023,manna_InnerSkinEffects_2023}, one-dimensional systems can exhibit a variety of NHSE subtypes, including nonreciprocal skin effect (NSE) \cite{guo_ExactSolutionNonHermitian_2021,wang_GeneratingArbitraryTopological_2021}, critical non-Hermitian skin effect (critical NHSE) \cite{li_CriticalNonHermitianSkin_2020,liu_HelicalDampingDynamical_2020,yokomizo_ScalingRuleCritical_2021}, scale-free localization (SFL) \cite{li_ImpurityInducedScalefree_2021,guo_AccumulationScalefreeLocalized_2023,li_ScalefreeLocalizationPT_2023,xie_ObservationScalefreeLocalized_2024,zhang_ScalefreeLocalizationAnderson_2025}, scale-tailored localization (STL) \cite{guo_ScaletailoredLocalizationIts_2024,zhao_TentaclelikeSpectraBound_2025}, and bipolar non-Hermitian skin effect (bipolar NHSE) \cite{song_NonHermitianTopologicalInvariants_2019,jiang_ReciprocatingBipolarNonHermitian_2023}, among others \cite{schindler_DislocationNonHermitianSkin_2021,longhi_StochasticNonHermitianSkin_2020,yuce_NonlinearNonHermitianSkin_2021,xu_InteractioninducedDoublesidedSkin_2021,longhi_ErraticNonHermitianSkin_2025}. These subtypes are now rigorously described by established frameworks such as the generalized Brillouin zone (GBZ) \cite{yao_EdgeStatesTopological_2018,kunst_BiorthogonalBulkBoundaryCorrespondence_2018,song_NonHermitianTopologicalInvariants_2019,ashida_NonHermitianPhysics_2020,zhang_CorrespondenceWindingNumbers_2020,yokomizo_ScalingRuleCritical_2021,wang_GeneratingArbitraryTopological_2021,bhargava_NonHermitianSkinEffect_2021,guo_ExactSolutionNonHermitian_2021,zhang_DynamicalDegeneracySplitting_2023,wang_AmoebaFormulationNonBloch_2024} and have been experimentally verified \cite{helbig_GeneralizedBulkBoundary_2020,weidemann_TopologicalFunnelingLight_2020,xiao_NonHermitianBulkBoundary_2020,zou_ObservationHybridHigherorder_2021,li_ImpurityInducedScalefree_2021,zhang_ObservationHigherorderNonHermitian_2021,zhang_AcousticNonHermitianSkin_2021,lin_TopologicalPhaseTransitions_2022,xie_ObservationScalefreeLocalized_2024}.

Both NSE and bipolar NHSE are induced by nonreciprocal hoppings \cite{yao_EdgeStatesTopological_2018,song_NonHermitianTopologicalInvariants_2019}, which result in markedly distinct energy spectra and eigenstate profiles when periodic boundary conditions (PBC) and open boundary conditions (OBC) are compared. STL arises from long-range asymmetric couplings, producing skin modes whose localization lengths scale linearly with the coupling range (i.e., the distance from the boundary); such long-range couplings can be interpreted as nonlocal non-Hermitian impurities \cite{guo_ScaletailoredLocalizationIts_2024}. SFL manifests in two distinct scenarios. First, a local non-Hermitian perturbation (impurity) causes continuum eigenstates to accumulate near the impurity, with the localization length proportional to the system (or subsystem) size, while the distance between the impurity and the boundary strongly modulates this length \cite{guo_AccumulationScalefreeLocalized_2023,li_ScalefreeLocalizationPT_2023}. Second, boundary impurities tune the nonreciprocal coupling strengths at the end sites, thereby effectively engineering the boundary conditions and significantly altering the skin modes \cite{li_ImpurityInducedScalefree_2021}. Critical NHSE emerges from the coupling between subsystems with differing degrees of nonreciprocity, where the presence or absence of critical behavior is directly governed by the choice of boundary conditions \cite{li_CriticalNonHermitianSkin_2020}. 

Overall, the sensitivity of the NHSE to boundary conditions has gradually become one of the key research paradigms in the field of non-Hermitian physics. At the same time, existing explanations for the mechanisms underlying these skin effects remain highly fragmented.

\begin{figure*}[ptb]
 \centering
{\includegraphics[width=1\linewidth]{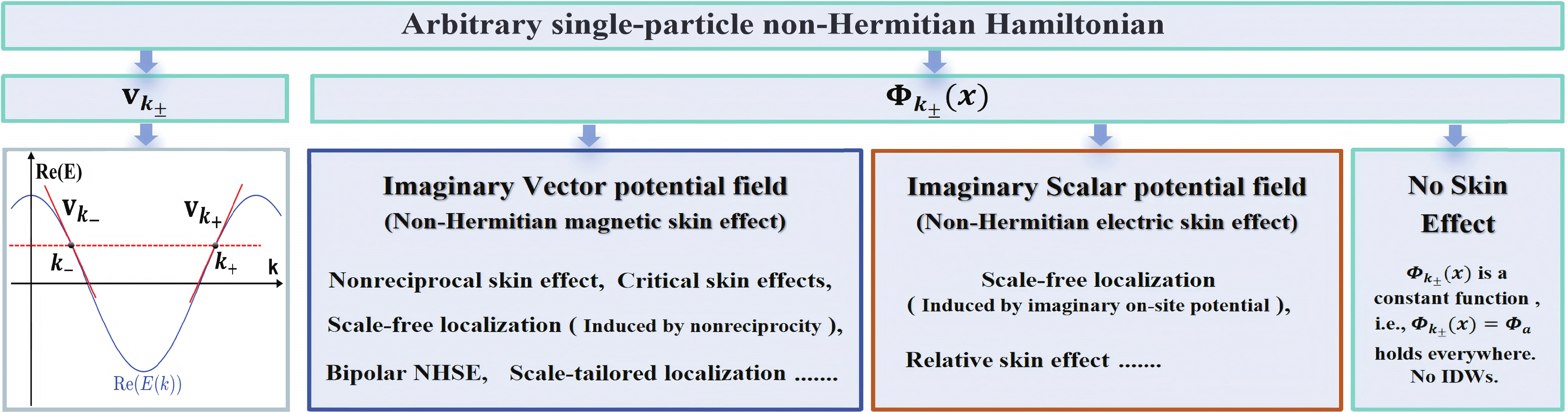}}  
\caption{Classification of known non-Hermitian skin effects (NHSEs) within the generalized non-Hermitian skin effect (GNHSE) framework. Any single-particle Hamiltonian can be analyzed within the GNHSE framework to determine its localization properties, requiring only the effective velocity $\mathbf{v}_{k_\pm}$ and potential field $\Phi_{k_\pm}(x)$. Existing NHSEs can be universally constructed from distinct types of imaginary vector or scalar potential fields. The effective velocity $\mathbf{v}_{k_\pm} = \frac{\partial \mathrm{Re}(E(k))}{\partial k}\big|_{k_\pm}$ represents the linearized dispersion near mode $k_\pm$, such that $\operatorname{sgn}(\mathbf{v}_{k_+}\cdot\mathbf{v}_{k_-})<0$. Notably, both imaginary scalar \cite{li_ScalefreeLocalizationPT_2023} and vector \cite{li_ImpurityInducedScalefree_2021} potentials can independently induce SFL. For instance, SFL induced by a Dirac delta scalar potential $\delta(x-d)$ \cite{li_ScalefreeLocalizationPT_2023}, representing a type of NHESE, differs fundamentally from the RSE in its boundary sensitivity: $d$ is the distance from the boundary, which significantly influences the localization length of both the energy spectrum and the eigenstates.}
\label{fig:unified model}
\end{figure*}

In this Letter, we unveil a novel NHSE, the relative skin effect (RSE), in which spectra and eigenstates remain nearly invariant under boundary changes, contradicting the conventional NHSE paradigm. We show this effect stems from spatially varying imaginary scalar potential field $\mathrm{i}\Phi(x)$, which locally break translational symmetry and anchor localization to the geometric deviation of $\Phi(x)$ from its global spatial average $\Phi_a$. Guided by this insight, we formulate the generalized non-Hermitian skin effect (GNHSE) theory, which predicts localization in any single-particle non-Hermitian Hamiltonian via two key quantities: effective velocities $\mathbf{v}_{k_\pm}$ and potential fields $\Phi_{k_\pm}(x)$. Based on $\Phi_{k_\pm}(x)$, we classify all one-dimensional NHSEs into two fundamental types: magnetic, induced by imaginary vector potential fields, and electric, driven by imaginary scalar potential fields, as illustrated in Fig.~\ref{fig:unified model}. We further demonstrate a transition between them, where at the critical point with inverse decay length $\eta=0$, eigenstates become fully delocalized and exhibit plane-wave character. Within the GNHSE framework, different types of NHSE act like pigments of distinct colors on the blank canvas of the system, allowing one to engineer virtually any desired localization pattern by judiciously combining them.\\

\noindent{\large{\textbf{Results}}}\\
\textbf{Relative skin effect}\\
We begin with a single-particle Hermitian lattice Hamiltonian $\hat{H}_h = \sum_{i,j} t_{ij} \hat{c}_{i}^{\dagger} \hat{c}_j$ (with $t_{ij} = t_{ji}^\ast$), where $\hat{c}_j$ ($\hat{c}_j^\dagger$) is the annihilation (creation) operator at site $j$. To this, we introduce a non-uniform imaginary potential $\mathrm{i}\Phi(j)$. Here, $\Phi(j)$ is the lattice discretization of a Riemann-integrable field $\Phi(x)$ defined on $x \in [0, L]$, having finitely many jump discontinuities. The system size is $L=Na$, where $N$ is the number of sites and we set $a=1$. The full Hamiltonian is
\begin{equation}\label{eq:RSE}
\hat{H} = \hat{H}_h + \mathrm{i}\lambda\sum_{j=0}^{N} \Phi(j)\hat{n}_j + \delta_e \left( \hat{c}^{\dagger}_{0} \hat{c}_N + \hat{c}^{\dagger}_{N} \hat{c}_0 \right),
\end{equation}
where $\hat{n}_j = \hat{c}_{j}^{\dagger} \hat{c}_j$ is the number operator. The boundary condition is controlled by $\delta_e$: OBC for $\delta_e=0$ and PBC for $\delta_e=1$. The parameter $\lambda \ll t_{ij}$ ensures weak non-Hermiticity, treated perturbatively. The spatial average of the potential is $\Phi_a \coloneqq \int_{0}^{L} \Phi(x) dx/L$.

\begin{figure}[ptb]
 \centering
{\includegraphics[width=1\linewidth]{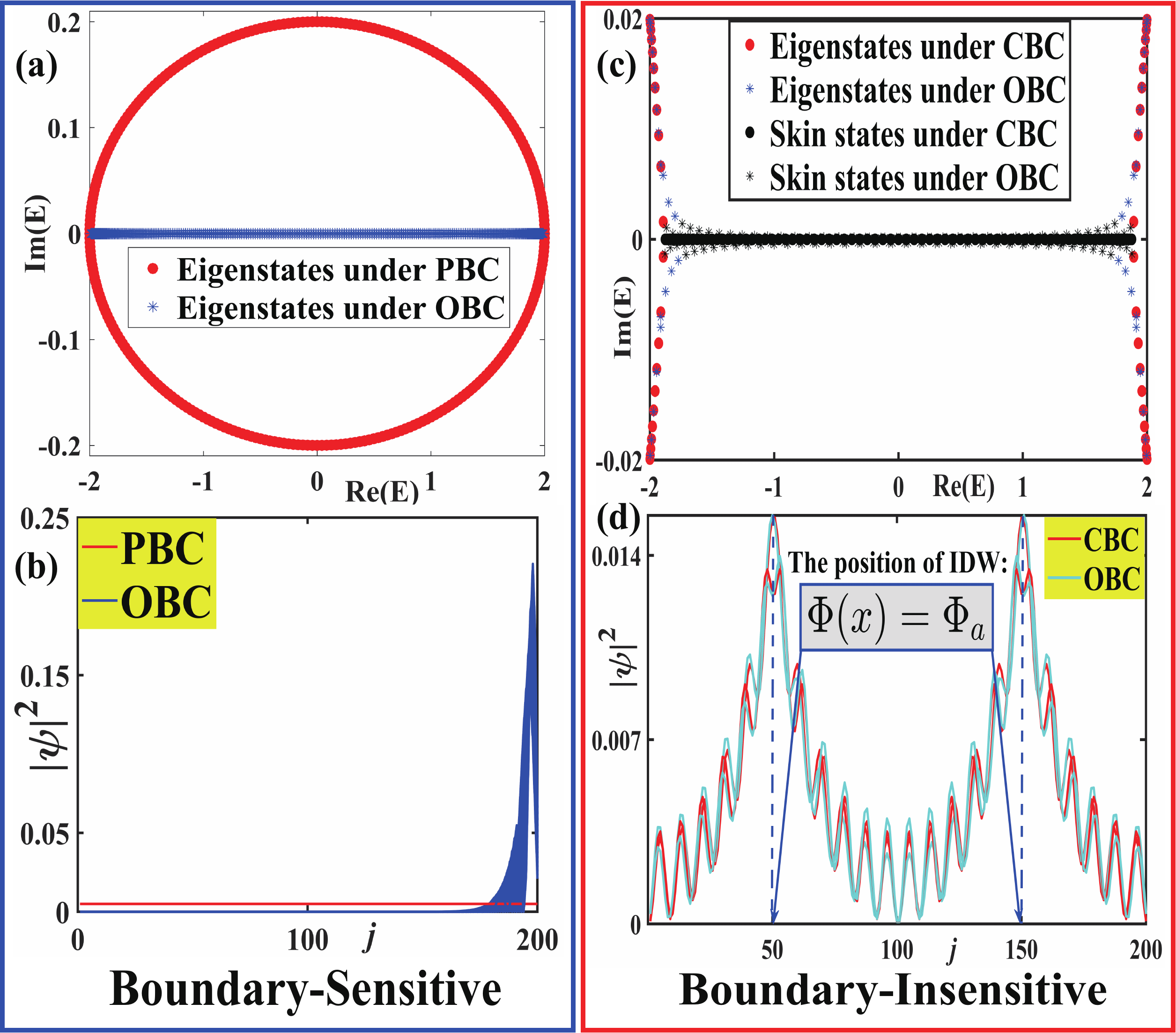}}  
\caption{Comparison of the nonreciprocal skin effect and the relative skin effect in terms of spectra and eigenstate distributions. OBC and PBC denote open and periodic boundary conditions, respectively. (a)-(b) Spectrum and eigenstate distributions for the nonreciprocal skin effect under different boundary conditions, with the Bloch Hamiltonian $H(k) = 2t\cos k - 2\mathrm{i}\gamma\sin k$, where the strength of asymmetric hopping $\gamma=0.1$ and the nearest-neighbor hopping amplitude $t=1$. (c)-(d) Spectrum and eigenstate distributions for the relative skin effect under different boundary conditions, with $\Phi(x) = \operatorname{sgn}\big((x - L/4)(x - 3L/4)\big)$ and $\lambda = 0.02$. Black: RSE skin modes; red and blue: excluded localized states \cite{du_NonHermitianTearingDissipation_2024}. For (a)-(d), $N=200$.}
\label{fig:RSE}
\end{figure}

Boundary insensitivity signifies that changes in boundary conditions induce no appreciable modifications to the spectrum or eigenstate distributions. To illustrate this clearly, we consider the simplest model and potential: the Bloch Hamiltonian ${H}_h (k)=2t\cos k$ and $\Phi(x)=\operatorname{sgn}\big((x - L/4)(x-3L/4)\big)$, where the nearest-neighbor hopping amplitude $t=1$. The spectrum is shown in Fig.~\ref{fig:RSE}(c), where black dots represent eigenstates exhibiting localization, with wavefunction distributions displayed in Fig.~\ref{fig:RSE}(d). These eigenstates localize precisely at the two points where $\Phi(x)$ intersects $\Phi_a$, namely at sites $j=50$ and $j=150$, which we term imaginary-domain walls (IDWs). Notably, varying $\delta_e$ has negligible effect on both the spectrum and eigenstate: even under OBC, the boundaries do not serve as additional localization centers, in stark contrast to the conventional NHSE, which is highly boundary sensitive (see Figs.~\ref{fig:RSE}(a) and (b)). 

\begin{figure}[ptb]
 \centering
{\includegraphics[width=1\linewidth]{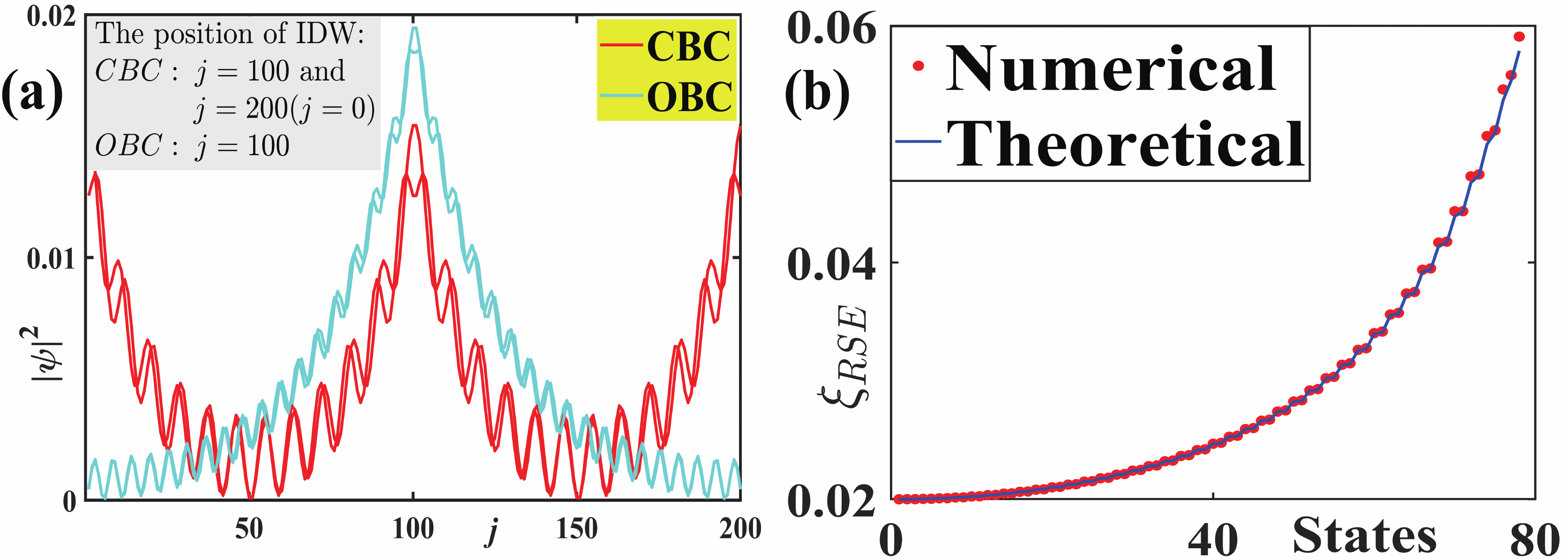}}  
\caption{Eigenstate distributions under different boundary conditions and numerical fitting of the localization length for $\Phi(x) = \operatorname{sgn}\big((x - L/2)(x - L)\big)$. (a) Eigenstate distribution when the imaginary domain wall coincides with the boundary. Under open boundary conditions, the boundary imaginary domain wall vanishes, and the associated localization disappears. Here $N=200,\lambda = 0.02$. (b) Numerical fitting of the localization length, using only modes with $\operatorname{Re}(E) > 0$, ordered by decreasing energy.}
\label{fig:RSE-IDW}
\end{figure}

When we align a boundary with an IDW by choosing $\Phi(x) = \operatorname{sgn}\big((x - L/2)(x-L)\big)$, the wavefunction distribution [Fig.~\ref{fig:RSE-IDW}(a)] shows boundary localization under PBC; however, upon switching to OBC, the IDW at the boundary vanishes and so does the boundary localization, leaving only the single IDW at $x=L/2$ as the localization center. Thus, the localization position of skin modes is dictated exclusively by IDWs and is independent of boundary conditions. Since skin mode localization depends on the \emph{relative} geometry of $\Phi(x)$ and $\Phi_a$, we term this boundary-insensitive phenomenon the relative skin effect. Moreover, due to the absence of translational symmetry and well-defined spectral winding numbers, established theories \cite{yao_EdgeStatesTopological_2018,zhang_CorrespondenceWindingNumbers_2020,yang_NonHermitianBulkBoundaryCorrespondence_2020,yokomizo_NonBlochBandTheory_2022} fail to describe the RSE, necessitating a new theoretical framework for non-Hermitian skin phenomena.

To explain the RSE, we propose an effective model derived from any system $\hat{H_R}$ exhibiting this phenomenon:
\begin{equation}\label{eq:RSE model}
\hat{H}_{k_\pm} = \mathbf{v}_{k_\pm}\hat{k} + \mathrm{i}\Phi(x),\;x \in [0, L],
\end{equation}
where $\hat{k}=-\mathrm{i}\partial_x$, and the effective velocities $\mathbf{v}_{k_\pm}= \frac{\partial E(k) }{\partial k}|_{k_\pm}$ are defined via linearization near a pair of reduced wave vectors $k_\pm$ sharing the same $E(k)$, such that $\operatorname{sgn}(\mathbf{v}_{k_+}\cdot\mathbf{v}_{k_-})<0$. $E(k)$ denotes the energy spectrum under Hermitian conditions. We consider only cases with $\mathbf{v}_{k_\pm}\neq0$. The field $\Phi(x)$ corresponds to the perturbative scalar potential in Eq.~\eqref{eq:RSE}, with the strength parameter $\lambda$ absorbed into its definition. Solving the eigenvalue equation under PBC, i.e., $\psi_{k_\pm}(0) = \psi_{k_\pm}(L)$ (details in the Methods), yields the wavefunction probability distribution:
\begin{equation}\label{eq:RSE model psi}
|\psi_{k_\pm}(x)|^2 = \frac{1}{\mathcal{N}}  \exp\left( \frac{2}{\mathbf{v}_{k_\pm}}\int_{0}^{x} [ \Phi(x') - \Phi_a ] \,dx' \right),
\end{equation}
where $\mathcal{N}$ is a normalization constant. The exponent $\Phi^s(x)/\mathbf{v}_{k_\pm}$ drives the skin localization properties, where $\Phi^s(x)=\int_{0}^{x} [ \Phi(x') - \Phi_a]\;dx'$. Skin modes vanish when $\Phi(x') \equiv \Phi_a$ (preserving translational symmetry), so localization arises only from deviations of $\Phi(x')$ from $\Phi_a$. For $\Phi^s(x)$ linear in $x$, we have $2\Phi^s(x)/\mathbf{v}_{k_\pm} = x/\xi_{k_\pm}$, where $\xi_{k_\pm}$ is the localization length. Notably, $\operatorname{sgn}(\xi_{k_\pm}) = \operatorname{sgn}(\mathbf{v}_{k_\pm})$, and since $\operatorname{sgn}(\xi_{k_+}\xi_{k_-}) < 0$, the $k_+$ and $k_-$ modes localize at opposite ends. This property holds even for $\Phi^s(x)$ nonlinear in $x$, providing the second rationale for the term \emph{relative}. To validate the theory, we set the Bloch Hamiltonian ${H}_h (k)=2t\cos k$ and $\Phi(x)=\lambda\operatorname{sgn}\big((x - L/2)(x-L)\big)$; as shown in Fig.~\ref{fig:RSE-IDW}(b), the numerically extracted localization lengths agree excellently with theoretical predictions (details in the Methods).

In summary, the essential distinction between PBC and OBC lies in whether translational symmetry is preserved. As evident from Eq.~\eqref{eq:RSE}, the nonuniform potential $\Phi(x)$ itself already breaks translational symmetry, rendering the choice of boundary conditions inconsequential. In finite systems, different forms of $\Phi(x)$ in real space break translational symmetry in distinct ways, thereby fixing the localization positions at the IDWs, in contrast to OBC, which breaks symmetry only at the physical boundaries. This highlights the crucial role of real-space non-Hermitian terms in understanding the RSE.\\

\noindent{\textbf{Generalized Non-Hermitian Skin Effect}}\\
\begin{figure}[ptb]
 \centering
{\includegraphics[width=1\linewidth]{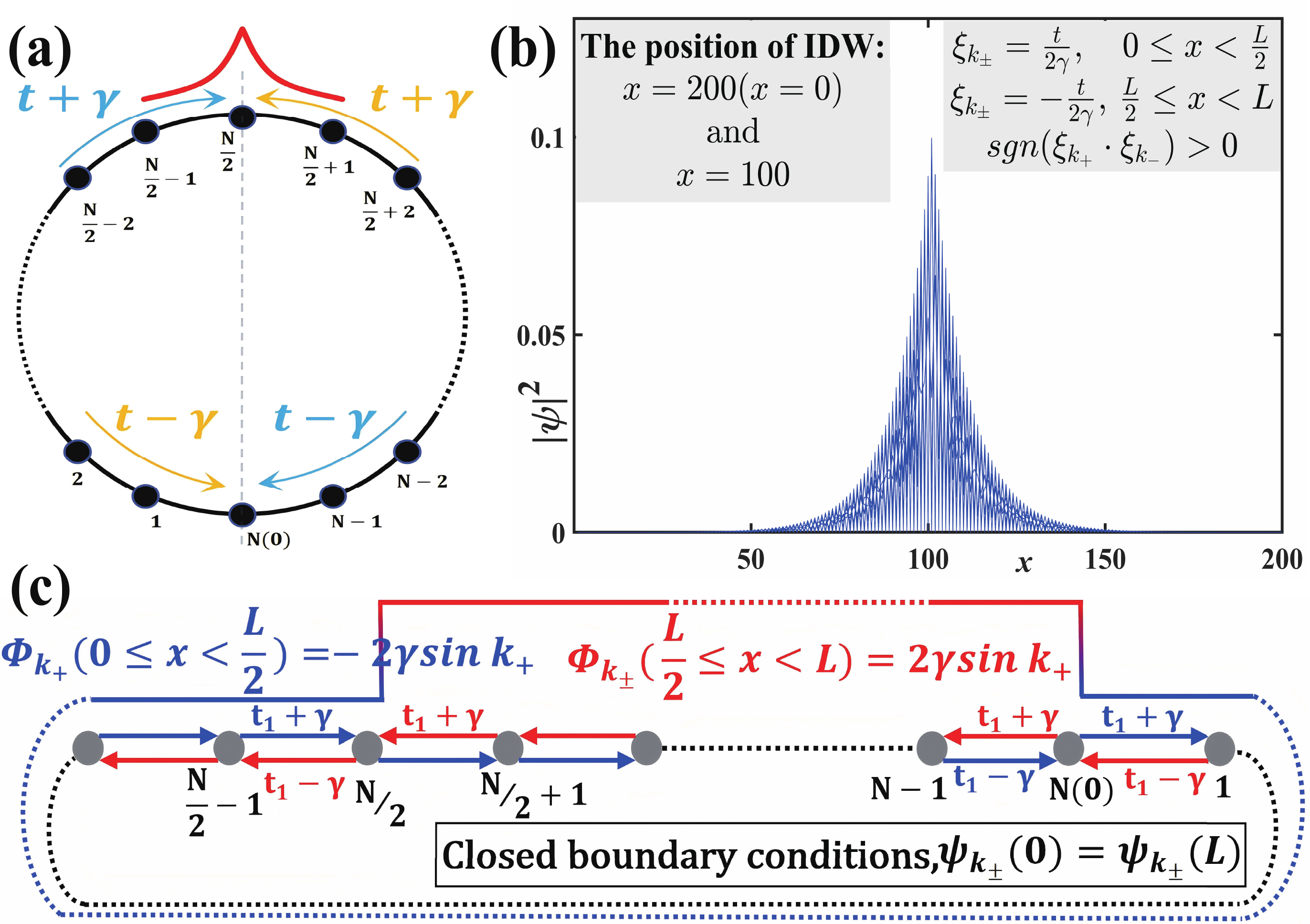}}  
\caption{The Hatano–Nelson model within the generalized non-Hermitian skin effect framework. (a) Real-space schematic of the lattice model. Arrows of different colors indicate distinct directions of nonreciprocal hopping, and red peaks denote the localization positions. (b) Eigenstate distribution of the Hatano–Nelson model, where the localization positions for the $k_+$ and $k_-$ modes coincide. $N=200,\gamma=0.05$. (c) Schematic of the field $\Phi_{k_+}(x)$.}
\label{fig:NSE}
\end{figure}

Building upon the above framework, we extend the theory to establish a more universal description of NHSEs. The Weyl equation describes massless Dirac fermions, with the Hamiltonian in an electromagnetic field given by \cite{jiang_QuantumSimulationTwoDimensional_2022}:
\begin{equation}\label{eq:weyl model}
H_{\text{EM}} = \pm c  \mathbf{\sigma} \cdot ( \mathbf{p} - q\mathbf{A} ) + q\Phi,
\end{equation}
where $c$ is the speed of light, $q$ the particle charge, $\mathbf{p}$ the momentum operator, $\mathbf{A}$ the magnetic vector potential, $\Phi$ the electric scalar potential, and $\mathbf{\sigma}=(\sigma_x,\sigma_y,\sigma_z)$ the Pauli matrices. The first term represents the kinetic energy in the electromagnetic field, and the $\pm$ signs correspond to two chiralities. 

It is widely recognized that existing NHSE theories can be effectively described by analytically continuing the wavevector $k$ into the complex plane \cite{yao_EdgeStatesTopological_2018,zhang_CorrespondenceWindingNumbers_2020,yang_NonHermitianBulkBoundaryCorrespondence_2020,yokomizo_NonBlochBandTheory_2022}, which provides key inspiration. For instance, in the one-dimensional nonreciprocal Su-Schrieffer-Heeger (SSH) model, the Bloch Hamiltonian reads $H(k)=(t_1+t_2\cos k)\sigma_x+(t_2\sin k+\mathrm{i}\gamma)\sigma_y$, where $t_1$ and $t_2$ denote intra- and inter-cell hopping, respectively, and $\gamma$ quantifies the nonreciprocity. Near the gap-closing point ($t_1+t_2=0$), the low-energy effective model is: $H_{low}(k) =(v_Fk+\gamma )\sigma_y=v_F(k+\mathrm{i}k_0)\sigma_y$, where the Fermi velocity $v_F=t_2$ characterizes the linear dispersion, and the non-Hermitian effect is captured by an imaginary vector potential $\mathrm{i}k_0=\gamma/v_F$. This differs from Eq.~\eqref{eq:RSE model} in two aspects: (i) the linear dispersion holds only near the band-touching point, with $v_F$ fixed; (ii) while the RSE is driven by an imaginary scalar potential $\mathrm{i}\Phi(x)$, here $\mathrm{i}k_0$ acts as an imaginary vector potential. 

Regarding the first point, since the NHSE is predominantly a bulk phenomenon, we perform a linear expansion around a single band of the two-band system without including boundary states to define $\mathbf{v}_{k_\pm}$ analogously to Eq.~\eqref{eq:RSE model}. The second point highlights a key distinction between the RSE and most known skin effects. By analogy with Eq.~\eqref{eq:weyl model}, non-Hermiticity can be introduced by making the magnetic potential $\mathbf{A}$ purely imaginary, which we term the non-Hermitian magnetic skin effect (NHMSE), or by making the electric potential $\Phi$ purely imaginary, which we term the non-Hermitian electric skin effect (NHESE). Together, they constitute the generalized non-Hermitian skin effect (GNHSE). We thus propose the following effective one-dimensional model for the GNHSE (details in the Methods):
\begin{equation}\label{eq:unified model}
 \begin{aligned}
\hat{H}^{\text{G}}_{k_\pm} &= \mathbf{v}_{k_\pm} \big(\,\hat{k}+ \mathrm{i}\mathbf{A}(x) \,\big) + \mathrm{i}\Phi(x)\\
                &=\mathbf{v}_{k_\pm}\hat{k}+\mathrm{i}\Phi_{k_\pm}(x),\;x \in [0, L].
  \end{aligned}
\end{equation}
Any single-particle non-Hermitian Hamiltonian $H_s$ exhibiting NHSE can be mapped to $\hat{H}^{\text{G}}_{k_\pm}$, where $\Phi_{k_\pm}(x)=\mathbf{v}_{k_\pm}\cdot\mathbf{A}(x)+\Phi(x)$. The first term captures the contribution from the imaginary vector potential (NHMSE), in which $\Phi_{k_\pm}(x)$ depends on $\mathrm{Im}(E(k))$, and $E(k)$ is the spectrum of $H_s$ under periodic boundary conditions. The second term corresponds to the imaginary scalar potential (NHESE), encompassing both $\Phi(x)$ in Eq.~\eqref{eq:RSE model} and potentials of $\delta$-function forms (details in Supplementary Section II.B). The effective velocity $\mathbf{v}_{k_\pm} = \frac{\partial \mathrm{Re}(E(k))}{\partial k}\big|_{k_\pm}$ represents the linearized dispersion near mode $k_\pm$, as schematically illustrated in Fig.~\ref{fig:unified model}; for the NHESE, it coincides with the expression given in Eq.~\eqref{eq:RSE model}.

We illustrate the construction of $\Phi_{k_\pm}(x)$ in systems exhibiting the NHMSE using the Hatano-Nelson model $H_{HN}$ \cite{yokomizo_NonBlochBandTheory_2022} (see Figs.~\ref{fig:RSE}(a) and (b)). Under OBC, eigenstates localize exponentially at the edges as $|\psi(x)|^2 \sim e^{x/\xi}$. The localization length from the GBZ is $\xi = 1 / \ln \frac{t+\gamma}{t-\gamma} \sim t / 2\gamma$ for small $\gamma$. 
\begin{figure*}[ptb]
 \centering
{\includegraphics[width=1\linewidth]{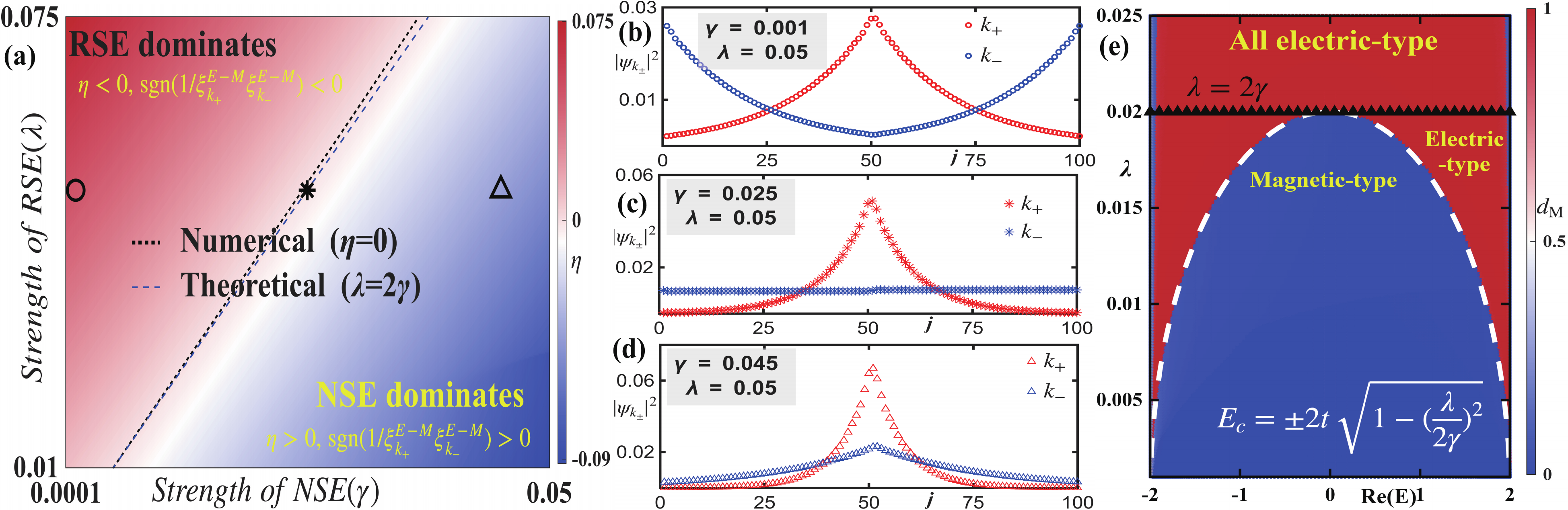}}  %
\caption{Localization phase transition between the relative skin effect (RSE) and the nonreciprocal skin effect (NSE). (a) Phase diagram of the order parameter $\eta$ for NSE and RSE, where the black dotted line and blue dashed line indicate numerical and theoretical phase boundaries for the $\sin k_- =1$ mode, respectively. The warm-colored region corresponds to RSE dominance; the cold-colored region, to NSE dominance. At $\eta=0$, $\xi^{\text{E-M}}_{k_-}\rightarrow\infty$, resulting in fully delocalized eigenstates. Parameters: $j\in[N+1,2N]$, $N=50$. (b)-(d) Evolution of the wavefunction probability distribution across the phase transition, where the three distinct markers correspond to the same markers in panel (a). (e) Visualization of the mobility edge. The real part of the energy $\mathrm{Re}(E)$ is plotted as a function of $\lambda$, with the color encoding the normalized distance $d_{\rm M} = \min(|x_{\max} - N|,\, 2N - |x_{\max} - N|)/N$ from the wavefunction peak of the $k_-$ mode to $j=N$. Since the magnetic-type skin modes localize at $j=N$, blue ($d_{\rm M}\approx 0$) represents the magnetic phase, in which $k_-$ and $k_+$ localize on the same side; red ($d_{\rm M}\approx N$) represents the electric phase, in which they localize at different IDWs. The white dashed line and the black triangle line ($\lambda=2\gamma$) denote the theoretically predicted mobility edge $E_c$ and the critical threshold, respectively. The mobility edge $E_c$ exists only for $\lambda<2\gamma$, and the white dashed line coincides perfectly with the numerically obtained red--blue boundary. Parameters: $t=1$, $\gamma=0.01$, $N=150$.}
\label{fig:phase}
\end{figure*}

Within the GNHSE framework, we replace OBC with PBC and define a piecewise imaginary vector potential: $\Phi_{k_\pm}(0\leq x< L/2)=-2\gamma \sin k_\pm$, $\Phi_{k_\pm}(L/2 \leq x< L)=2\gamma \sin k_\pm$. At fixed $k_\pm$, $\Phi_{k_\pm}(x)$ is piecewise constant with spatial average $\Phi_a^\pm = 0$. This setup corresponds to joining an $H_{HN}$ chain and its Hermitian conjugate $H^\dag_{HN}$, both originally defined under OBC, into a closed loop, leading to the emergence of IDWs at the junctions, i.e., at sites $j=1$ and $j=N/2$ [Fig.~\ref{fig:NSE}(a)]. The spatial eigenstate distribution and schematic of $\Phi_{k_+}(x)$ are shown in Figs.~\ref{fig:NSE}(b) and (c). From Eq.~\eqref{eq:unified model}, we obtain:
\begin{equation}\label{eq:MSE xi}
 \begin{aligned}
 \xi^{\text{M}}_{k_\pm}= \frac{\mathbf{v}_{k_\pm}}{-4\gamma \sin k_\pm }=\frac{t}{2\gamma}, \; 0\leq x <\frac{L}{2}.
         \end{aligned}
\end{equation}
This matches the GBZ prediction in the weak non-Hermiticity limit, and the localization positions for the $k_+$ and $k_-$ modes coincide, i.e., $\operatorname{sgn}(\xi^ {\text{M}}_{k_+}\xi^{\text{M}}_{k_-})>0$, which constitutes the most significant distinction between NHMSE and NHESE (details in the Methods).

By designing different forms of the imaginary potential field $\mathrm{i}\Phi_{k_\pm}(x)$, we break translational symmetry in real space, giving rise to various types of NHSE, as illustrated in Fig.~\ref{fig:unified model} and detailed in Supplementary Section II. The boundary sensitivity inherent to conventional NHSEs fundamentally arises from their dependence on where in real-space translational invariance is disrupted \cite{bhargava_NonHermitianSkinEffect_2021,schindler_DislocationNonHermitianSkin_2021,guo_ExactSolutionNonHermitian_2021}. Within the GNHSE framework, this sensitivity can be directly attributed to the spatial structure of the non-Hermitian terms, and can also be understood through the introduction of a mode-dependent, time-independent imaginary gauge transformation. The imaginary vector potential $A_k(x)$ can be completely eliminated from $\hat{H}^{\text{G}}_{k_\pm}$ via the transformation $\psi_k(x) = e^{\chi_k(x)}\tilde{\psi}_k(x)$ (where $\chi'_k = A_k$), but boundary conditions respond differently to this transformation (see Supplementary Section.~III for details). In contrast, the imaginary scalar potential $\Phi(x)$ remains strictly invariant under the transformation, a structure analogous to time-independent gauge transformations in electromagnetism. This real-space perspective provides a more comprehensive and physically intuitive understanding than traditional momentum-space approaches \cite{yao_EdgeStatesTopological_2018,borgnia_NonHermitianBoundaryModes_2020,yang_NonHermitianBulkBoundaryCorrespondence_2020,zhang_DynamicalDegeneracySplitting_2023,yokomizo_NonBlochBandTheory_2022}.\\

\noindent{\textbf{Phase transition between NHMSE and NHESE}}\\
We demonstrate the robust localization phase transition between NHMSE and NHESE through the competition between RSE and NSE:
\begin{equation}
\hat{H}_{\text{E-M}} = \sum_{j=1}^{2N} \left[ 
 t_1 \hat{c}_{j+1}^{\dagger} \hat{c}_j +t_2 \hat{c}_j^{\dagger} \hat{c}_{j+1}+\mathrm{i}\Phi(j) \hat{c}_{j}^{\dagger}\hat{c}_j
\right],
\end{equation}
where $\hat{c}_{2N+1}=\hat{c}_1$, $t_1=t+\gamma s_j,t_2=t - \gamma s_j,s_j = -\operatorname{sgn}\left( j - N - 0.5 \right)$. This can be viewed as the NSE model shown in Fig.~\ref{fig:NSE} supplemented by an imaginary scalar potential field $\mathrm{i}\Phi(j)$.

We first consider $\Phi(j) = \lambda\operatorname{sgn}\left(j - N\right)$, where $\lambda$ and $\gamma \ll t$ ensure that both the imaginary scalar and vector potentials are perturbative. According to the GNHSE theory, the IDWs of $A(x)$ and $\Phi(x)$ coincide spatially (at $x = N$ and $2N$), and the wavefunction probability distribution is given by Eq.~\eqref{eq:unN unified wavefunction}, with the exponential factor being a linear function of $x$ within each subinterval (see Supplementary Section.~IV.A for details). We define the logarithmic slope $\kappa_{k_\pm}(x) \equiv (d/dx)\ln|\psi(x)|$. Since $\Phi_a^\pm = 0$, we have $\kappa_{k_\pm}(x) = \Phi_{k_\pm}(x)/\mathbf{v}_{k_\pm}$. At the IDWs, $\kappa_{k_\pm}(x)$ undergoes a discontinuous jump $\Delta\kappa$, where a wavefunction peak (maximum) requires $\Delta\kappa < 0$ and a trough (minimum) requires $\Delta\kappa > 0$. Theoretical analysis reveals that the $k_+$ mode ($\mathbf{v}_{k_+} > 0$) always satisfies $\Delta\kappa(N)<0$ and $\Delta\kappa(2N)>0$ at the IDWs, meaning the $k_+$ mode is invariably localized at $x = N$ with the magnetic and electric skin effects reinforcing each other, exhibiting no phase transition. For the $k_-$ mode ($\mathbf{v}_{k_-} < 0$), the jump at the IDWs is:
\begin{equation}\label{eq:Deltakappa}
\Delta\kappa_{k_-}(N) = \frac{\lambda-2\gamma s}{ts},\,\Delta\kappa_{k_-}(2N) = \frac{2\gamma s - \lambda}{ts},
\end{equation}
where $s \equiv \sin k_- > 0$. When $2\gamma s > \lambda$, $\Delta\kappa_{k_-}(N) < 0$, so the $k_-$ mode localizes at $x = N$ (magnetic-type dominant, on the same side as $k_+$); when $2\gamma s < \lambda$, it localizes at $x = 2N$ (electric-type dominant, on the opposite side from $k_+$). Thus, the peak position jumps from one IDW to the other at $2\gamma s = \lambda$. This naturally defines the order parameter of the phase transition:
\begin{equation}\label{eq:order}
\eta \equiv \Delta\kappa_{k_-}(2N) =\frac{1}{\xi^{\text{E-M}}_{k_-}} = \frac{1}{\xi^{\rm M}} - \frac{1}{\xi_{k_-}^{\rm E}},
\end{equation}
where $1/\xi^{\rm M}$ and $1/\xi_{k_-}^{\rm E}$ are the magnetic and electric inverse localization lengths, respectively. At the transition point $\eta = 0$, the magnetic and electric skin effects cancel each other exactly, yielding an exponential factor $2\Phi_{k_-}^s(x)/\mathbf{v}_{k_-} = 0$, and the $k_-$ mode becomes fully delocalized. Moreover, since the transition condition $\lambda = 2\gamma \sin k_-$ is mode-dependent, it naturally defines a mobility edge in the energy spectrum:
\begin{equation}\label{eq:mobility_edge}
E_c = \pm 2t\sqrt{1 - \left(\frac{\lambda}{2\gamma}\right)^2},
\end{equation}
which separates the $k_-$ modes into a band-edge electric phase ($|{\rm Re}(E)| > |E_c|$, where $k_-$ and $k_+$ localize on opposite sides) and a band-center magnetic phase ($|{\rm Re}(E)| < |E_c|$, where they localize on the same side). Notably, the $k_+$ mode remains unaffected throughout the entire parameter space, rendering this mobility edge chirally selective: only one of the two degenerate modes participates in the phase transition. The two types of phase diagrams and the corresponding wavefunction distributions are shown in Fig.~\ref{fig:phase}.

The robustness of the above phase transition can be further verified by rigidly shifting $\Phi(j)$ by $\delta$ lattice sites ($0 < \delta < N$), which displaces the electric-type IDWs to $x = \delta$ and $N + \delta$, so that they no longer coincide with the magnetic-type IDWs. Theoretical analysis shows that the expressions for $\eta$ and $E_c$ remain identical to those at $\delta = 0$ (see Supplementary Section.~IV.B for details). The phase transition persists robustly as a codimension-1 phase boundary, although the critical state becomes a partially delocalized plateau state over the interval $[\delta, N]$. Complete delocalization requires that the spatial envelopes of the imaginary electric and magnetic potentials (after subtracting their respective spatial averages) satisfy the pointwise proportionality condition $[A(x) - A_a] = -[\Phi(x) - \Phi_a]/\mathbf{v}_{k_\pm}$ for all positions $x$. Furthermore, in Supplementary Section.~IV.C, we examine the case where $\Phi(j)$ is a smooth sinusoidal function and find that, due to the mismatch in envelope shapes, a coexistence region emerges between the magnetic and electric phases. The peak position switches via continuous drift rather than abrupt jumps between fixed locations, and the critical state takes the form of a modulated wave with three peaks of equal height. These results collectively demonstrate that the electric--magnetic localization phase transition (i.e., the reversal of the localization direction) is a universal and robust phenomenon, while the degree of delocalization and the transition pathway at the critical point depend on whether the spatial envelopes of the imaginary potential fields are matched.
\\

\noindent{\large{\textbf{Discussion}}}\\
In summary, we demonstrate that a Riemann-integrable, inhomogeneous imaginary scalar potential $\mathrm{i}\Phi(x)$ induces a boundary-insensitive NHSE, termed the relative skin effect. In this regime, the localization properties are determined solely by intrinsic features of $\Phi(x)$, affording precise and flexible control over the localization length and position. This phenomenon cannot be predicted by existing theoretical frameworks. Historically, most studies have adapted momentum-space concepts to interpret non-Hermitian localization, rather than building theories directly from real-space non-Hermitian structures. Following this latter approach, we develop a generalized non-Hermitian skin effect (GNHSE) theory that fully accounts for this anomaly. Crucially, when applied to conventional NHSEs, our framework extends the conventional picture of translational symmetry breaking only at the physical boundaries under OBC to a broader class of symmetry-breaking patterns under PBC induced by various forms of imaginary vector potential fields $\Phi_{k_\pm}(x)$. This enables flexible control over localization behavior in traditional NHSEs, rather than fixed localization positions and lengths.

Thus, the discovery of this new class of NHSE, together with the GNHSE theory, opens a significant new area of research in non-Hermitian physics. We propose experimental verification across multiple platforms, including electrical circuits \cite{helbig_GeneralizedBulkBoundary_2020,Halder_CircuitRealizationTwoorbital_2024,rafi-ul-islam_CriticalNonHermitianSkin_2025}, cold atom systems \cite{Zhang_ExceptionalPointHysteresis_2025,qin_KinkedLinearResponse_2024,li_TopologicalSwitchNonHermitian_2020}, photonic quantum walks \cite{xiao_NonHermitianBulkBoundary_2020,Longhi_IncoherentNonHermitianSkin_2024,zhu_ObservationNonHermitianEdge_2024,gao_QuantumWalksCorrelated_2024}, and metamaterials \cite{Wang_NonHermitianTopologyStatic_2023,lu_NonHermitianTopologicalPhononic_2023}, among others. Future work may explore couplings between imaginary scalar and vector potentials and extend the GNHSE framework to quasiperiodic and disordered systems, thus offering new routes to uncover exotic non-Hermitian physics. \\

\noindent{\large{\textbf{Methods}}}\\
\textbf{Solution of Eq.~\eqref{eq:RSE model}/\eqref{eq:unified model} under PBC}\\
In this section, we solve the eigenvalue equation for Eq.~\eqref{eq:RSE model}/\eqref{eq:unified model} and present several examples corresponding to different scalar potential fields $\Phi(x)$.

The distinction between Eq.~\eqref{eq:RSE model} and Eq.~\eqref{eq:unified model} lies in replacing the scalar potential $\Phi(x)$ with $\Phi_{k_\pm}(x)$, which incorporates contributions from both scalar and vector potentials. This modification does not affect the eigenstates under PBC. Thus, we directly solve the eigenvalue equation for Eq.~\eqref{eq:unified model}:
\begin{equation}\label{eq:unified model-M}
\hat{H}^{\text{G}}_{k_\pm} = \mathbf{v}_{k_\pm} \hat{k} + \mathrm{i}\Phi_{k_\pm}(x),
\end{equation}
for $x\in[0,L]$, yielding:
\[
[ \mathrm{-i} \mathbf{v}_{k_\pm}  \frac{d}{dx} + \mathrm{i}\Phi_{k_\pm}(x)]\psi_k(x)=E\psi_k(x),
\]
where $E$ is a complex eigenvalue. Assuming $\psi_k(x)\neq0$, we obtain:
\[
  \frac{d}{dx} \ln\psi_k(x)=\frac{1}{\mathbf{v}_{k_\pm}}[\mathrm{i}E+\Phi_{k_\pm}(x)].
\]
The unnormalized wave function is then:
\begin{equation}\label{eq:unN unified wavefunction}
  \psi_{k_\pm}(x)=\psi_{k_\pm}(0)\exp\left(\frac{\mathrm{i}Ex+\int^x_0\Phi_{k_\pm}(x')dx'}{\mathbf{v}_{k_\pm}}\right).
\end{equation}
Imposing the PBC $\psi_k(0) = \psi_k(L)$ gives:
\[
  \frac{1}{\mathbf{v}_{k_\pm}}(\mathrm{i}EL+\int^L_0\Phi_{k_\pm}(x)dx)=\mathrm{i}2\pi n,
\]
with $n\in\mathbb{Z}$. Simplifying yields the eigenvalues:
\begin{equation}\label{eq:unified eig}
 \begin{aligned} 
 E_n=&\mathbf{v}_{k_\pm} k-\mathrm{i}\Phi_a^\pm,\\
 k=\frac{2\pi n}{L},\;& \Phi_a^\pm=\frac{1}{L}\int^L_0\Phi_{k_\pm}(x)dx.
  \end{aligned}
\end{equation}
Substituting into Eq.~\eqref{eq:unN unified wavefunction} and normalizing, we obtain:
\begin{equation}\label{eq:unified wavefunction}
  \psi_{k_\pm}(x)=\frac{1}{\mathcal{\sqrt{N}}}\exp\left(\mathrm{i}kx+\frac{1}{\mathbf{v}_{k_\pm}}\int^x_0[\Phi_{k_\pm}(x')-\Phi_a^\pm]dx'\right),
\end{equation}
where $\mathcal{\sqrt{N}}$ is the normalization factor. The squared modulus of this wave function reproduces the eigenfunction of Eq.~\eqref{eq:unified model}. When $\Phi_{k_\pm}(x)$ reduces to $\Phi(x)$, it recovers Eq.~\eqref{eq:RSE model psi}.

\begin{figure}[ptb]
\centering
{\includegraphics[width=1\linewidth]{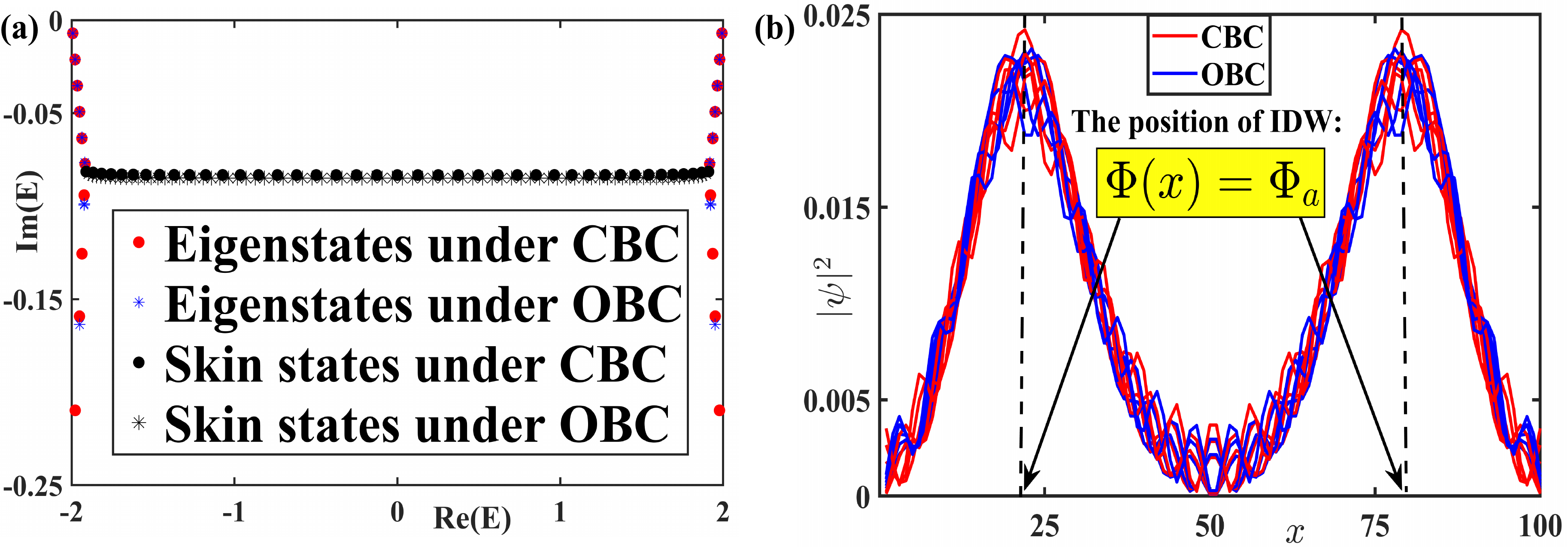}}
\caption{Numerically obtained spectrum and eigenstate distributions of skin modes for the same Bloch Hamiltonian ${H}_h (k)=2t\cos k$ as in Fig.~\ref{fig:RSE-IDW}, with the scalar potential $\Phi(x) = -\lambda(x-L/2)^2$ added, $L=100,\lambda=10^{-4}$.}  
\label{fig:L2}
\end{figure}

As discussed in Eq.~\eqref{eq:RSE model psi}, the skin modes associated with $v_{k_+}$ and $v_{k_-}$ propagate in opposite directions, with their localization positions determined by the slope of the dispersion at mode $k$. Furthermore, as illustrated by the schematic of $\mathbf{v}_{k_\pm}$ in Fig.~\ref{fig:unified model}, the modes $k_\pm$ share the same real part of the energy, leading to degeneracy. While each non-degenerate mode exhibits a single skin peak at distinct positions, the degenerate eigenstates display two simultaneous skin peaks, as illustrated in Fig.~\ref{fig:RSE}(d) and Fig.~\ref{fig:L2}(b). \\

\noindent{Example 1: $\mathbf{ \Phi(x) = -\lambda(x-L/2)^2}$}\\
Consider the scalar potential $\Phi(x) = -\lambda(x-L/2)^2$. The resulting spectrum and skin mode wave function distributions are shown in Fig.~\ref{fig:L2}. The spatial average is:
\[
  \Phi_a=\frac{1}{L} \int_0^L -\lambda(x-\frac{L}{2})^2  dx=-\frac{\lambda L^2}{12},
\]
and the accumulated potential deviation is:
\[
\Phi^s(x) = \int_{0}^{x} [\Phi(x) - \Phi_a]\,dx =  \lambda \left( -\frac{x^3}{3} + \frac{L x^2}{2} - \frac{L^2}{6} x \right).
\]
Since $\Phi^s(x)$ is nonlinear in $x$, a simple localization length $\xi$ cannot be defined. However, analytical insight is still possible. From Eq.~\eqref{eq:RSE model psi}, the exponential factor in the wave function modulus is:
\[
 \frac{2\Phi^s(x)}{\mathbf{v}_{k_\pm}} =\frac{\lambda }{\mathbf{v}_{k_\pm}} \left( -\frac{x^3}{3} + \frac{L x^2}{2} - \frac{L^2}{6} x \right).
\]
Define $g(x) = -x^3/3 + L x^2/2 - L^2 x/6$, so that:
\[
f(x) = \frac{2\Phi^s(x)}{\mathbf{v}_{k_\pm}} = \frac{\lambda }{\mathbf{v}_{k_\pm}} g(x).
\]
Let $c = \lambda/\mathbf{v}_{k_\pm}$, so $f(x) = c g(x)$. Since $e^{f(x)}$ is monotonic, its localization peak coincides with the maximum of $f(x)$, which depends on the sign of $c$ and the shape of $g(x)$.

The first derivative of $g(x)$ is:
\[
g'(x) = -x^2 + L x - \frac{L^2}{6}.
\]
Setting $g'(x)=0$ yields two critical points:
\[
 \begin{aligned} 
x_1 = \frac{L}{2} \left(1 - \frac{1}{\sqrt{3}}\right) \approx 0.211L,\\ x_2 = \frac{L}{2} \left(1 + \frac{1}{\sqrt{3}}\right) \approx 0.789L,
\end{aligned}
\]
which correspond to the positions of the imaginary domain walls (IDWs), as shown in Fig.~\ref{fig:L2}(b). The second derivative is:
\[
g''(x) = -2x + L.
\]
At $x_1$, $g''(x_1) = L/\sqrt{3} > 0$, indicating a local minimum. At $x_2$, $g''(x_2) = -L/\sqrt{3} < 0$, indicating a local maximum.

Since $f(x) = c g(x)$, the position of its maximum depends on the sign of $c = \lambda/\mathbf{v}_{k_\pm}$. If $c > 0$, $f(x)$ peaks at $x_2$; if $c < 0$, it peaks at $x_1$. Given that $\operatorname{sgn}(v_{k_+}v_{k_-})<0$, the localization positions for $k_+$ and $k_-$ are opposite, consistent with the conclusion for linear $\Phi^s(x)$.\\

\noindent{Example 2: $\mathbf{\Phi(x) =\lambda \operatorname{sgn}\big((x - L/2)(x-L)\big)}$}\\
Consider $\Phi(x) =\lambda\operatorname{sgn}\big((x - L/2)(x-L)\big)$, as in Fig.~\ref{fig:RSE-IDW}(a), with $\mathbf{v}_{k_\pm}=-2t \sin k_\pm$, $\Phi(x)=\lambda$ for $0 < x < L/2$, and $\Phi(x)=-\lambda$ for $L/2 < x < L$. The spatial average is:
\[
\Phi_a=0, \quad \Phi^s(x)=\int_{0}^{x} \operatorname{sgn}\big((x' - L/2)(x'-L)\big)dx'.
\]
Thus:
\begin{equation}
\frac{2\Phi^s(x)}{\mathbf{v}_{k_\pm}} =
       \begin{cases}
 \frac{2\lambda x}{\mathbf{v}_{k_\pm}}=-\frac{\lambda}{t\sin k_\pm}x,  & 0 < x < \frac{L}{2} .\\[1.5mm]
  -\frac{2\lambda(x-L)}{\mathbf{v}_{k_\pm}}=\frac{\lambda}{t\sin k_\pm}(x-L),  & \frac{L}{2}  < x< L.
       \end{cases}  
\end{equation}
Since $\Phi^s(x)$ is piecewise linear in $x$, we identify $\frac{x}{\xi_{k_\pm}} = \frac{2\Phi^s(x)}{\mathbf{v}_{k_\pm}}$, yielding:
\begin{equation}
\xi_{k_\pm}=
       \begin{cases}
 -\frac{t\sin k_\pm}{\lambda}, \; & 0< x <L/2.\\[1.5mm]
  \frac{t\sin k_\pm}{\lambda}, \; & L/2 < x < L.
       \end{cases}  
\end{equation}
This recovers the non-Hermitian electric skin effect localization length $\xi_{k_\pm}^{\text{E}}$ used in Eq.~\eqref{eq:order}. \\

\noindent{\textbf{Key distinction between NHMSE and NHESE}}\\
We analyze the fundamental difference between the non-Hermitian magnetic skin effect (NHMSE) and the non-Hermitian electric skin effect (NHESE). We begin with the simplest case in which $\Phi^s(x)$ is linear in $x$, yielding $2\Phi^s(x)/\mathbf{v}_{k_\pm} = x/\xi_{k_\pm}$. This allows direct extraction of the localization lengths $\xi_{k_\pm}$. For NHMSE, the localization lengths satisfy:
\[
\operatorname{sgn}(\xi^{\text{M}}_{k_+} \xi^{\text{M}}_{k_-}) > 0.
\]
For NHESE, they satisfy:
\[
\operatorname{sgn}(\xi^{\text{E}}_{k_+} \xi^{\text{E}}_{k_-}) < 0.
\]
It can be easily seen that the core difference between NHMSE and NHESE lies in whether the localization positions of the $k_+$ and $k_-$ modes are opposite (NHESE) or the same (NHMSE). We now analyze the origin of this difference from the perspective of the generalized non-Hermitian skin effect (GNHSE).

From Eq.~\eqref{eq:unified wavefunction}, the exponential part governing skin localization is:
\begin{equation}\label{eq:GNHSE exp}
\frac{x}{\xi_{k_\pm}} = \frac{1}{\mathbf{v}_{k_\pm}} \int_0^x \left[ \Phi_{k_\pm}(x') - \Phi_a^\pm \right] dx',
\end{equation}
where $\Phi_{k_\pm}(x) = \mathbf{v}_{k_\pm} \cdot \mathbf{A}(x) + \Phi(x)$. The first term, $\mathbf{v}_{k_\pm} \cdot \mathbf{A}(x)$, captures the NHMSE contribution; the second term, $\Phi(x)$, corresponds to NHESE. Their key difference is that the first term depends explicitly on the mode velocity $\mathbf{v}_{k_\pm}$.

Substituting the first term into Eq.~\eqref{eq:GNHSE exp} yields:
\[
\frac{x}{\xi^{\text{M}}_{k_\pm}} = \frac{\int_0^x \left[ \mathbf{v}_{k_\pm} \cdot \mathbf{A}(x') - \frac{1}{L} \int_0^L \mathbf{v}_{k_\pm} \cdot \mathbf{A}(x'')\, dx'' \right] dx'}{\mathbf{v}_{k_\pm}}.
\]
Since $\mathbf{v}_{k_\pm}$ is independent of $x$, it factors out of the integral. In one dimension, vector direction is encoded in sign, so we obtain:
\[
\frac{x}{\xi^{\text{M}}_{k_\pm}} = \int_0^x \left[ A(x') - \frac{1}{L} \int_0^L A(x'')\, dx'' \right] dx'.
\]
It is evident that the sign of $\xi^{\text{M}}_{k_\pm}$ no longer depends on $\operatorname{sgn}(\mathbf{v}_{k_\pm})$. Thus, the $k_+$ and $k_-$ modes localize at the same boundary.

In contrast, the skin localization induced by $\Phi(x)$ retains explicit dependence on $\mathbf{v}_{k_\pm}$. Given that $\operatorname{sgn}(\mathbf{v}_{k_+} \mathbf{v}_{k_-}) < 0$, it follows naturally that:
\[
\operatorname{sgn}(\xi^{\text{E}}_{k_+} \xi^{\text{E}}_{k_-}) < 0, \quad \operatorname{sgn}(\xi^{\text{M}}_{k_+} \xi^{\text{M}}_{k_-}) > 0.
\]

\noindent {\large{\textbf{Data availability}}}\\ 
The data used in this study are available in the Supplementary Section VI \cite{Wei_SupplementalMaterial_2024}.  \\

\noindent {\large{\textbf{Code availability}}}\\ 
The code used in this study is available in the Supplementary Section VI \cite{Wei_SupplementalMaterial_2024}. \\

\noindent {\large{%

\begin{acknowledgments}
\noindent {\large{\textbf{Acknowledgments}}}\\
This work was supported by the National Key R$\&$D Program of China (Grant No.~2023YFA1406704) and Natural Science Foundation of China (Grant No.~12174030).\\
\end{acknowledgments}

\noindent {\large{\textbf{Author contributions}}}\\
S.-P.K conceived the work. Z.W and J.-Y.F did the major part of the theoretical derivation and numerical calculation. K.C, X.-R.M, C.-X.G and X.-P.R helped to improve the design of lattice models. All authors discussed the results and participated in the writing of the manuscript.

\noindent {\large{\textbf{Competing interests}}}\\
The authors declare no competing interests.

\end{document}


{\title{Supplementary Material for "A Unified Framework for the Non-Hermitian Localization: Boundary-Insensitive Modes and Electric-Magnetic Analogy"} 
\author{Zheng Wei}
\thanks{These authors contributed equally to the work}
\affiliation{Center for Advanced Quantum Studies, School of Physics and Astronomy, Beijing Normal University, Beijing 100875, China}
\affiliation{Key Laboratory of Multiscale Spin Physics(Ministry of Education), Beijing Normal University, Beijing 100875, China}

\author{Ji-Yao Fan}
\thanks{These authors contributed equally to the work}
\affiliation{Center for Advanced Quantum Studies, School of Physics and Astronomy, Beijing Normal University, Beijing 100875, China}
\affiliation{Key Laboratory of Multiscale Spin Physics(Ministry of Education), Beijing Normal University, Beijing 100875, China}

\author{Kui Cao}
\affiliation{Center for Advanced Quantum Studies, School of Physics and Astronomy, Beijing Normal University, Beijing 100875, China}
\affiliation{Key Laboratory of Multiscale Spin Physics(Ministry of Education), Beijing Normal University, Beijing 100875, China}

\author{Xin-Ran Ma}
\affiliation{Center for Advanced Quantum Studies, School of Physics and Astronomy, Beijing Normal University, Beijing 100875, China}
\affiliation{Key Laboratory of Multiscale Spin Physics(Ministry of Education), Beijing Normal University, Beijing 100875, China}

\author{Cui-Xian Guo}
\affiliation{Beijing Key Laboratory of Optical Detection Technology for Oil and Gas, China University of Petroleum-Beijing, Beijing 102249, China}
\affiliation{Basic Research Center for Energy Interdisciplinary, College of Science, China University of Petroleum-Beijing, Beijing 102249, China}

\author{Xue-Ping Ren}
\affiliation{Center for Advanced Quantum Studies, School of Physics and Astronomy, Beijing Normal University, Beijing 100875, China}
\affiliation{Key Laboratory of Multiscale Spin Physics(Ministry of Education), Beijing Normal University, Beijing 100875, China}

\author{Su-Peng Kou}
\email[Email: ]{spkou@bnu.edu.cn}
\affiliation{Center for Advanced Quantum Studies, School of Physics and Astronomy, Beijing Normal University, Beijing 100875, China}
\affiliation{Key Laboratory of Multiscale Spin Physics(Ministry of Education), Beijing Normal University, Beijing 100875, China}
\thanks{Corresponding author: spkou@bnu.edu.cn}
\maketitle
\tableofcontents

\section{Solution of Eq.~(2) under open boundary conditions}\label{eq2-obc}
Before proceeding, we first list several key formulas from the main text that will be used below.

The unnormalized wavefunction of the effective model for generalized non-Hermitian skin effect, before imposing any boundary conditions, is given by:
\begin{equation}\label{eq:unN unified wavefunction}
  \psi_{k_\pm}(x)=\psi_{k_\pm}(0)\exp\left(\frac{\mathrm{i}Ex+\int^x_0\Phi_{k_\pm}(x')dx'}{\mathbf{v}_{k_\pm}}\right).
\end{equation}
Under periodic boundary conditions (PBC), the normalized wavefunction and corresponding eigenvalues are:
\begin{equation}\label{eq:unified wavefunction}
  \psi_{k_\pm}(x)=\frac{1}{\sqrt{\mathcal{N}}}\exp\left(\mathrm{i}kx+\frac{1}{\mathbf{v}_{k_\pm}}\int^x_0[\Phi_{k_\pm}(x')-\Phi_a^\pm]dx'\right),
\end{equation}
\begin{equation}\label{eq:unified eig}
 \begin{aligned} 
 E_n=&\mathbf{v}_{k_\pm}k-\mathrm{i}\Phi_a^\pm,\\
 k=\frac{2\pi n}{L},\quad & \Phi_a^\pm=\frac{1}{L}\int^L_0\Phi_{k_\pm}(x)dx,
  \end{aligned}
\end{equation}
where $\mathcal{N}$ is the normalization constant and $\Phi_a^\pm$ denotes the spatial average of $\Phi_{k_\pm}(x)$ over the system.

In the continuum limit, open boundary conditions (OBC) typically correspond to Dirichlet boundary conditions:
  \[
  \lim_{x \to +\infty} \psi_k(x) = 0, \quad \lim_{x \to -\infty} \psi_k(x) = 0.
  \]
However, since the general solution of the first-order differential operator $\mathrm{i}\dv{}{x}$ is a single-parameter exponential, it cannot simultaneously satisfy both boundary conditions. Such solutions typically describe unidirectional waves (e.g., chiral modes), decaying only at one infinity and generally failing to form standing waves within a finite region. There are two special cases that require separate treatment ( (1) scale-free localization in the non-Hermitian electric skin effect, and  
(2) scalar potentials of the form $\Phi(x) = \lambda\operatorname{sgn}\big((x - L/2)(x-L)\big)$.).

\subsection{General case}
To construct solutions under OBC, we split the interval $[0,L]$ into two half-infinite domains: $[0,+\infty)$ and $(-\infty,L]$. We solve the equation separately on each domain and then form a symmetric superposition ($\alpha = \beta$) over the overlap region $[0,L]$, where $C_R$ and $C_L$ are normalization constants:

\noindent{1. Right half-line ($x \geq 0$, $v_R > 0$):}
    \[
    \left[-\mathrm{i} v_R \frac{d}{dx} + \mathrm{i} \Phi(x)\right]\psi_R(x) = E \psi_R(x)
    \]
Boundary conditions: $\psi_R(0) = C_R,\lim_{x \to +\infty} \psi_R(x) = 0$.
    
\noindent{2. Left half-line ($x \leq L$, $v_L < 0$):}
    \[
    \left[-\mathrm{i} v_L \frac{d}{dx} + \mathrm{i} \Phi(x)\right]\psi_L(x) = E \psi_L(x)
    \]
Boundary conditions: $\lim_{x \to -\infty} \psi_L(x) = 0,\psi_L(L) = C_L$.
    
\noindent{3. Global wave function:}
    \begin{equation}
    \psi(x) = 
    \begin{cases} 
    \psi_L(x) & x < 0 \\
    \alpha\psi_L(x) + \beta\psi_R(x) & 0 \leq x \leq L \\
    \psi_R(x) & x > L 
    \end{cases}
    \label{eq:global_wavefunction_general}
    \end{equation}
Here, $\operatorname{sgn}(v_{R}v_{L})<0$. Since $\Phi(x)$ is Riemann integrable over $[0,L]$ with finitely many discontinuities of the first kind, it can be extended to $[0,+\infty)$ and $(-\infty,L]$ such that the decay conditions are satisfied:
\begin{itemize}
\item As $x \to +\infty$, $\int_0^x \Phi(x')  dx' \to +\infty$, ensuring $\lim_{x \to +\infty} \psi_R(x) = 0$. 
\item As $x \to -\infty$, $\int_x^L \Phi(x')  dx' \to -\infty$, ensuring $\lim_{x \to -\infty} \psi_L(x) = 0$. 
\end{itemize}
Since we are only concerned with the behavior of the wave function within $[0,L]$, the asymptotic behavior outside this interval is mathematically imposed and carries no direct physical significance.

The general solution for the right half-line ($x \geq 0$) is:
\begin{equation}
\psi_R(x) = C_R \exp\left( \frac{\mathrm{i} E}{v_R} x - \frac{1}{v_R} \int_0^x \Phi(t)  dt \right)
\label{eq:sol_right_general}
\end{equation}

The general solution for the left half-line ($x \leq L$) is:
\begin{equation}
\psi_L(x) = C_L \exp\left( \frac{\mathrm{i} E}{v_L} (x - L) - \frac{1}{v_L} \int_L^x \Phi(t)  dt \right).
\label{eq:sol_left_general}
\end{equation}

Substituting Eqs.~\eqref{eq:sol_right_general} and \eqref{eq:sol_left_general} into Eq.~\eqref{eq:global_wavefunction_general} gives:
\[
\begin{aligned}
\psi(x < 0) &= C_L \exp\left( \dfrac{\mathrm{i} E}{v_L} (x - L) - \dfrac{1}{v_L} \displaystyle\int_L^x \Phi(t)  dt \right),  \\[10pt]
\psi(0 \leq x \leq L) &= \alpha C_L \exp\left( \dfrac{\mathrm{i} E}{v_L} (x - L) - \dfrac{1}{v_L} \displaystyle\int_L^x \Phi(t)  dt \right)  \\
&+ \beta C_R \exp\left( \dfrac{\mathrm{i} E}{v_R} x - \dfrac{1}{v_R} \displaystyle\int_0^x \Phi(t)  dt \right), \\
\psi(x > L)&=C_R \exp\left( \dfrac{\mathrm{i} E}{v_R} x - \dfrac{1}{v_R} \displaystyle\int_0^x \Phi(t)  dt \right).
\end{aligned}
\]

Continuity at $x = 0$ requires:
\[
\begin{aligned}
\psi(0^-) &= \psi_L(0) = C_L \exp\left( -\frac{\mathrm{i} E}{v_L} L + \frac{1}{v_L} \int_0^L \Phi(t)  dt \right),\\
\psi(0^+) &= \alpha\psi_L(0) + \beta \psi_R(0) \\
&= \alpha C_L \exp\left( -\frac{\mathrm{i} E}{v_L} L + \frac{1}{v_L} \int_0^L \Phi(t)  dt \right) + \beta C_R.
\end{aligned}
\]
Equating them yields:
\begin{align}
C_L \exp\left( g_L \right) &= \alpha C_L \exp\left( g_L \right) + \beta C_R \nonumber \\
(1-\alpha) C_L \exp\left( g_L \right) &= \beta C_R \nonumber \\
C_R = \frac{1-\alpha}{\beta} C_L \exp\left( g_L \right) \label{eq:continuity0_general}
\end{align}
where $g_L = -\dfrac{\mathrm{i} E}{v_L} L + \dfrac{1}{v_L} \displaystyle\int_0^L \Phi(t)  dt$.

Continuity at $x = L$ requires:
\[
\begin{aligned}
\psi(L^-) &= \alpha \psi_L(L) + \beta \psi_R(L) \\
&= \alpha C_L + \beta C_R \exp\left( \frac{\mathrm{i} E}{v_R} L - \frac{1}{v_R} \int_0^L \Phi(t)  dt \right),\\
\psi(L^+) &= \psi_R(L) = C_R \exp\left( \frac{\mathrm{i} E}{v_R} L - \frac{1}{v_R} \int_0^L \Phi(t)  dt \right).
\end{aligned}
\]
Equating them yields:
\begin{align}
\alpha C_L + \beta C_R \exp\left( g_R \right) &= C_R \exp\left( g_R \right) \nonumber \\
\alpha C_L &= (1-\beta) C_R \exp\left( g_R \right) \nonumber \\
C_L = \frac{1-\beta}{\alpha}C_R \exp\left( g_R \right) \label{eq:continuityL_general}
\end{align}
where $g_R = \dfrac{\mathrm{i} E}{v_R} L - \dfrac{1}{v_R} \displaystyle\int_0^L \Phi(t)  dt$.

Combining Eqs.~\eqref{eq:continuity0_general} and \eqref{eq:continuityL_general} with $\alpha = \beta = 1/2$ yields:
\begin{align}
C_L &= \left[ C_L \exp\left( g_L \right) \right] \exp\left( g_R \right) \nonumber \\
1 &= \exp\left( g_L + g_R \right). \label{eq:exp_condition_general}
\end{align}

Expanding the exponent:
\[
\begin{aligned}
g_L + g_R &= \left( -\frac{\mathrm{i} E}{v_L} L + \frac{1}{v_L} \int_0^L \Phi(t)  dt \right) \\
&+\left( \frac{\mathrm{i} E}{v_R} L - \frac{1}{v_R} \int_0^L \Phi(t)  dt \right) \\
&= \mathrm{i} E L \left( \frac{1}{v_R} - \frac{1}{v_L} \right) + \int_0^L \Phi(t)  dt \left( \frac{1}{v_L} - \frac{1}{v_R} \right).
\end{aligned}
\]
Defining $\Delta = \frac{1}{v_R} - \frac{1}{v_L} > 0$ (since $v_R > 0$, $v_L < 0$) and $\Phi_a=\frac{1}{L} \int_0^L \Phi(t)  dt$, we obtain:
\[
g_L + g_R = \mathrm{i} E L \Delta - \Phi_a L \Delta = L\Delta (\mathrm{i} E  - \Phi_a).
\]

From Eq.~\eqref{eq:exp_condition_general}:
\begin{align}
\exp\left[ L\Delta (\mathrm{i} E - \Phi_a ) \right] &= 1 \nonumber \\
 L\Delta (\mathrm{i} E - \Phi_a ) &= 2\pi \mathrm{i} n, \quad n \in \mathbb{Z} \nonumber \\
E  &= \frac{2\pi n}{L \Delta} - \mathrm{i} \Phi_a. \nonumber
\end{align}

The eigenvalues are thus:
\begin{equation}
E_n =v_a k- \mathrm{i}\frac{1}{L} \int_0^L \Phi(t)  dt, \quad n \in \mathbb{Z}
\label{eq:eigenvalues_general}
\end{equation}
where $v_a= \dfrac{v_Rv_L}{v_L-v_R }$, $k=\dfrac{2\pi n}{L }$. The imaginary part matches that under PBC (Eq.~\eqref{eq:unified eig}), while the real part replaces $\mathbf{v}_{k_\pm}$ with $v_a$. Identifying $v_L$ and $v_R$ with $v_{k-}$ and $v_{k+}$, respectively, we note that when $v_R = v_L$, the real part of the spectrum under OBC is smaller than under PBC for equal $n$, consistent with numerical results.

Substituting $E_n$ into the wave function for $x\in[0,L]$ yields:
\begin{equation}\label{eq:full_solution_general}
\begin{aligned}
\psi_n(x) &= \frac{1}{2} \exp\left( \mathrm{i}\dfrac{ v_a}{v_L}k (x - L) - \dfrac{1}{v_L} \displaystyle\int_L^x [\Phi(x')-\Phi_a]   dx' \right) +  \frac{1}{2} \exp\left(\mathrm{i}\dfrac{ v_a}{v_R}k x - \dfrac{1}{v_R} \displaystyle\int_0^x [\Phi(x')-\Phi_a]   dx' \right).
\end{aligned}
\end{equation}
Thus, under open boundaries, the localization positions coincide with those under PBC, determined by the intersections of $\Phi(x)$ and $\Phi_a$, i.e., the positions of the IDWs.

\subsection{Special Case}

\subsubsection{$\Phi(x) = \lambda \operatorname{sgn}\big((x - L/2)(x-L)\big)$}

Unlike the semi-infinite systems discussed earlier, since the piecewise constant function $\Phi(x) = \lambda \operatorname{sgn}\big((x - L/2)(x-L)\big)$ approaches distinct constants at infinity, we can directly impose full Dirichlet boundary conditions. 

In the continuum limit, assume $\Phi(x)$ asymptotically satisfies:
\[
\lim_{x \to +\infty} \Phi(x) = \lambda_+, \quad \lim_{x \to -\infty} \Phi(x) = \lambda_- \quad (\lambda_\pm \in \mathbb{R}).
\]
The integral then behaves asymptotically as:
\begin{align*}
  \int_{0}^{x} \Phi(x') dx' &\sim \lambda_+ x + o(1), \quad x \to +\infty, \\
  \int_{0}^{x} \Phi(x') dx' &\sim \lambda_- x + o(1), \quad x \to -\infty.
\end{align*}
Substituting into Eq.~\eqref{eq:unN unified wavefunction}, the exponential factor becomes:
\begin{align*}
\frac{\mathrm{i}E + \lambda_+}{v_{k}} x, \quad x \to +\infty,\\
\frac{\mathrm{i}E + \lambda_-}{v_{k}} x, \quad x \to -\infty.
\end{align*}
Taking the modulus of the wave function yields the exponential decay rates:
\begin{align*}
  \frac{\lambda_+ - E_i}{v_k} x , \quad x \to +\infty, \\
\frac{\lambda_- - E_i}{v_k} x , \quad x \to -\infty,
\end{align*}
where $E_i$ denotes the imaginary part of the eigenvalue. The boundary condition $\lim_{x \to +\infty} |\psi(x)| = 0$ implies:
\[\dfrac{\lambda_+ - E_i}{v_k} < 0.\]
Similarly, $\lim_{x \to -\infty} |\psi(x)| = 0$ (noting $x < 0$) implies:
\[\dfrac{\lambda_- - E_i}{v_k} > 0.\]
We analyze the cases separately:

\begin{enumerate}
  \item \textbf{For $v_k > 0$:}
  \[
  \lambda_+ - E_i < 0 \quad \text{and} \quad \lambda_- - E_i > 0 \implies \boxed{\lambda_+ < E_i < \lambda_-},
  \]
  which requires $\lambda_+ < \lambda_-$.
  
  \item \textbf{For $v_k < 0$:}
  \[
  \lambda_+ - E_i > 0 \quad \text{and} \quad \lambda_- - E_i < 0 \implies \boxed{\lambda_- < E_i < \lambda_+},
  \]
  which requires $\lambda_- < \lambda_+$.
\end{enumerate}

Thus, when $\Phi(x)$ approaches distinct constants at infinity, solutions satisfying full Dirichlet boundary conditions exist. Since $\lambda_+ < \lambda_-$ and $\lambda_- < \lambda_+$ cannot hold simultaneously, only one of the two modes corresponding to $\mathbf{v}_{k_\pm}$ survives. Consequently, the degeneracy in Eq.~\eqref{eq:unified wavefunction} and Eq.~\eqref{eq:full_solution_general} is lifted, and the wave function localizes at a single imaginary domain wall. This can also be directly illustrated by the following example:

\textbf{Example:} A piecewise constant profile under open boundary conditions can be approximated by $\Phi(x) = \lambda_0 \tanh(\alpha x)$ with $\alpha \gg 1$. For $\lambda_0 > 0$ and $\alpha > 0$, the boundary imaginary domain walls vanish, leaving only one at $x=0$:
\begin{equation}
\begin{aligned}
\lim_{x \to +\infty} \Phi(x) = \lambda_0, \quad \lim_{x \to -\infty} \Phi(x) = -\lambda_0.\\
\end{aligned}
\end{equation}
\begin{itemize}
  \item For $v_k > 0$: $\lambda_0 < E_i < -\lambda_0$ (no solution, since $\lambda_0 > 0$).
  \item For $v_k < 0$: $-\lambda_0 < E_i < \lambda_0$.
\end{itemize}

Due to the boundary insensitivity of RSE, the imaginary part of the eigenvalue matches that under periodic boundary conditions. Setting $\Phi_a = E_i = 0$, the solution reads:
\begin{align*}
  \psi_k(x) &= \frac{1}{\mathcal{\sqrt{N}}} \exp\left( \frac{\lambda_0}{v_k \alpha} \ln(\cosh(\alpha x)) + \mathrm{i} \frac{E_r}{v_k} x \right) \\
  &= \frac{1}{\mathcal{\sqrt{N}}} [\cosh(\alpha x)]^{\frac{\lambda_0}{\alpha v_k}} e^{\mathrm{i} \frac{E_r}{v_k} x},
\end{align*}
where $\mathcal{\sqrt{N}}$ is the normalization factor. When $v_k < 0$, $\frac{\lambda_0}{\alpha v_k} < 0$, ensuring the boundary condition:
\begin{equation}
\lim_{|x| \to \infty}|\psi_k(x)| \propto |\cosh(\alpha x)|^{\frac{\lambda_0}{\alpha v_k}} \to 0.
\end{equation}

Under open boundaries, since the two velocities $\mathbf{v}_{k_\pm}$ cannot simultaneously satisfy Dirichlet conditions (i.e., $\lambda_0 < E_i < -\lambda_0$ and $-\lambda_0 < E_i < \lambda_0$ are mutually exclusive), and the surviving mode always localizes at the imaginary domain wall, the skin localization position in non-Hermitian electric skin effect depends solely on the location of the imaginary domain wall, independent of boundary condition changes unless they shift the domain wall position.

\subsubsection{Scale-Free Localization in Non-Hermitian Electric Skin Effect}

The on-site effective dissipation is equivalent to the imaginary potential field being a Dirac delta function, formed by two imaginary domain walls infinitesimally close to each other on either side of the lattice site. Mathematically, this corresponds to $\Phi(x) \sim \delta(x - x_0)$, which is not Riemann integrable and thus incompatible with the general framework developed earlier for smooth or piecewise-continuous potentials. The presence of a Dirac delta introduces a discontinuity in the wave function, requiring additional jump conditions at the singularity --- an unnatural complication in the continuum model.

However, if we consider coupling between the two chiral branches ($v_{k_+}$ and $v_{k_-}$), a natural mechanism emerges: the hybridization of the two modes can produce a bound state localized precisely at the delta-function IDW. This hybridized state automatically satisfies Dirichlet boundary conditions at infinity without ad hoc matching conditions, as the interference between the two decaying exponentials (one from each chirality) forms a normalizable, exponentially localized profile centered at the IDW. Thus, while the single-mode picture breaks down for singular potentials, the two-mode coupled picture restores consistency and naturally explains why localization remains pinned to the IDW --- even under open boundary conditions. For the detailed derivation, see the solution under open boundary conditions for the scale-free localization in the non-Hermitian electric skin effect.

\section{Universality of Generalized Non-Hermitian Skin Effect}

In the main text, we demonstrated that for the Hatano-Nelson model exhibiting non-Hermitian magnetic skin effect, the skin localization length derived from the GNHSE theory coincides with that obtained from the generalized Brillouin zone formalism in the perturbative non-Hermitian regime. We now examine several other prototypical non-Hermitian skin effects, applying the GNHSE framework to classify and compute their localization properties, thereby illustrating its universality.

\subsection{Scale-tailored localization}

Following Ref.~\cite{guo_ScaletailoredLocalizationIts_2024}, we consider scale-tailored localization (STL), a novel wave localization phenomenon induced by long-range asymmetric couplings. The model Hamiltonian reads:
\begin{equation}\label{eq:STL model}
\hat{H}^{\text{STL}} = \sum_{j=1}^{N-1} t \hat{c}_{j}^{\dagger} \hat{c}_{j+1} + t \delta_t \hat{c}_{N}^{\dagger} \hat{c}_p,
\end{equation}
illustrated in Fig.~\ref{fig:stl}(a), where $N$ denotes the lattice length. The operators $\hat{c}_{j}^{\dagger}$ and $\hat{c}_{j}$ create and annihilate a particle at site $j$, respectively. The parameter $t$ is the hopping amplitude between nearest neighbors. The model features an additional long-range asymmetric hopping connecting the last site to an internal site $p$, with strength $\delta_t$ and range $l = N - p$. The special cases $p = 1$ with $\delta_t = 1$ or $\delta_t = 0$ correspond to periodic or open boundary conditions, respectively. For generic $\delta_t \neq 0$ and $p \neq 1$, the solutions of Eq.~\eqref{eq:STL model} fall into two distinct classes:

The first class comprises $P$-fold degenerate solutions, representing a $P$th-order exceptional point. Their eigenstates localize strictly at the first site, interpreted as remnants of the $N$th-order exceptional point under open boundaries, unaffected by the long-range coupling.

The second class consists of $l$ non-degenerate solutions with eigenvalues $E_m = t \sqrt[l]{\delta_t} e^{\mathrm{i}\theta_m}$, where $\theta_m = 2\pi m / l$ and $m = 1, 2, \ldots, l$. For $|\delta_t| < 1$, the spatial profile $|\psi^{m}(x)|^2 \sim e^{-(x-j_0)/\xi}$ yields a localization length $\xi = -l / (2 \ln |\delta_t|)$, with all states accumulating at site $j_0 = 1$. For $|\delta_t| > 1$, they localize at the last site $j_0 = N$ with $\xi = l / \ln |\delta_t|$. Stronger deviations of $|\delta_t|$ from unity enhance localization, as numerically verified in Fig.~\ref{fig:stl}(b).

Notably, the localization length of the second class scales linearly with the coupling range $l$ but is independent of the total system size $N$, as shown in Figs.~\ref{fig:stl}(c) and (d). This contrasts with scale-free localization (SFL), and we term this phenomenon scale-tailored localization (STL). In the thermodynamic limit, SFL modes delocalize, whereas STL modes retain a finite, fixed localization length.

\begin{figure}[ptb]
\centering
{\includegraphics[width=\linewidth]{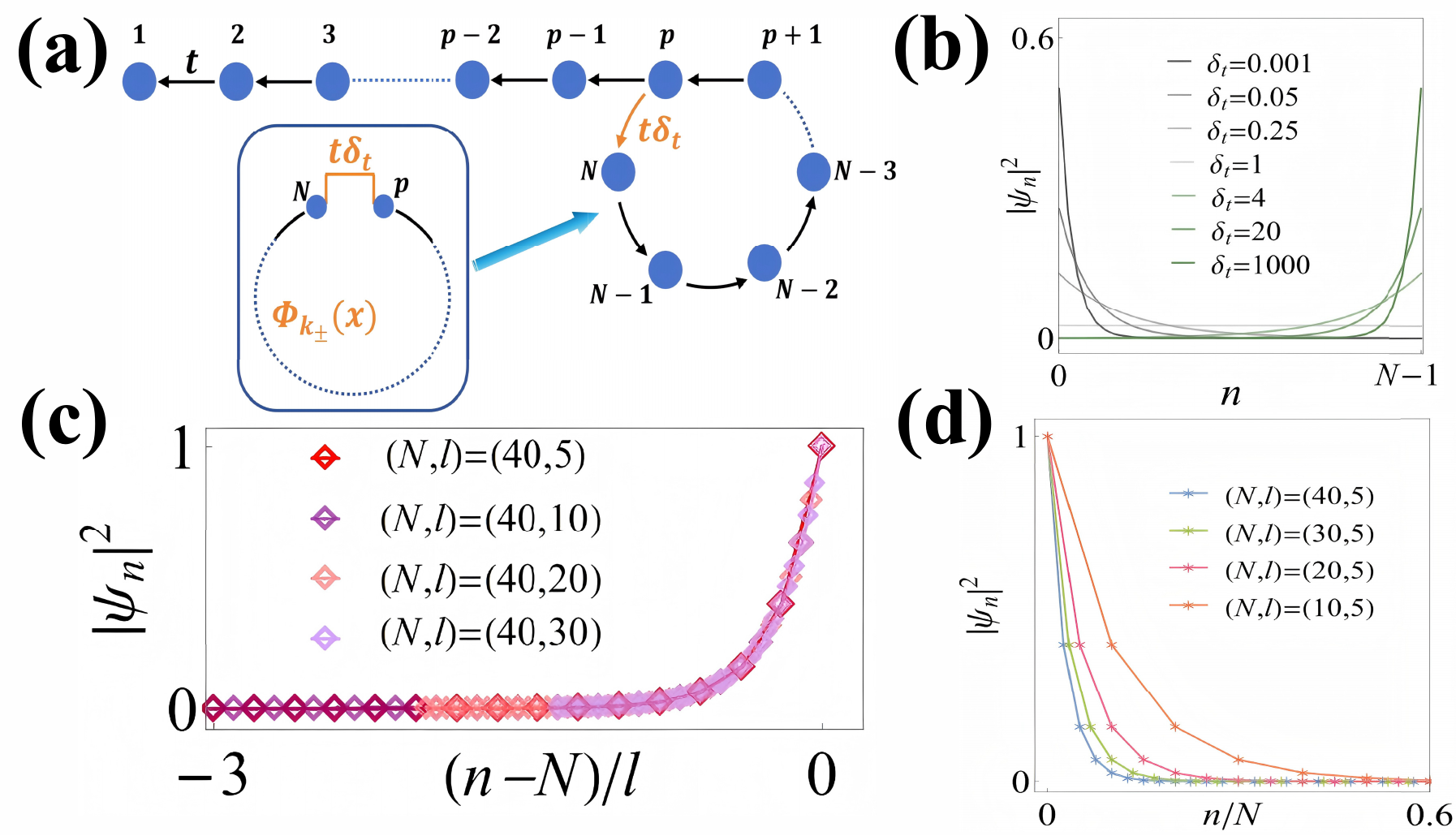}}
\caption{(a) Lattice model of STL \cite{guo_ScaletailoredLocalizationIts_2024}. The inset depicts the effective dissipation profile of STL. 
(b) Spatial distribution of the squared modulus of the second type of eigenstates for representative $\delta_t$ values, with fixed $(N, l) = (40, 20)$ \cite{guo_ScaletailoredLocalizationIts_2024}.  
(c) Rescaled spatial distributions of the squared modulus for varying $(N, l)$, demonstrating perfect data collapse \cite{guo_ScaletailoredLocalizationIts_2024}.  
(d) Rescaled spatial distribution as a function of total system size $N$, with $\delta_t = 0.1$ and $l = 5$ \cite{guo_ScaletailoredLocalizationIts_2024}.}
\label{fig:stl}
\end{figure}

We now reanalyze STL from the GNHSE perspective. As evident from Fig.~\ref{fig:stl}(a), the STL model inherently forms a local PBC, analogous to the scenario in Fig.~\ref{fig:sfl}(a), differing only in the nature of the potential $\Phi_{k_\pm}(x)$ --- scalar versus vector. For Eq.~\eqref{eq:STL model}, we focus solely on the segment forming this local closed loop. Given the existing nonreciprocal hopping $\hat{c}_{j}^{\dagger} \hat{c}_{j+1}$ along the chain, if $|\delta_t| < 1$ prevents localization at site $N$, nonreciprocity inevitably drives accumulation at site $j = 1$. Thus, for STL, analysis reduces to the small loop defined by the local boundary condition $|N+1\rangle = |p\rangle$.

Since STL fundamentally arises from a non-Hermitian vector potential, and for this unidirectional hopping model the dispersion is $E(k) = t \cos k + \mathrm{i} t \sin k$, we obtain $\mathbf{v}_{k_\pm} = -t \sin k_\pm$ and $\mathrm{Im}(E(k)) = t \sin k_\pm$. We construct $\Phi_{k_\pm}^s(x)$ as depicted in the inset of Fig.~\ref{fig:stl}(a):
\begin{equation}\label{eq:stl lambda}
\begin{aligned}
\Phi_a^\pm &= \frac{1}{l} \int_{0}^{l} \Phi_{k_\pm}(x)\;dx \approx \frac{1}{l} \sum_{j=1}^{l} \Phi_{k_\pm}(j) = \frac{t \sin k (l - 1 + \delta_t)}{l}, \\[1mm]
\Phi_{k_\pm}^s(x) &= \int_{0}^{x} [\Phi_{k_\pm}(x) - \Phi_a^\pm]\,dx \approx \sum_{j=1}^{x} [\Phi_{k_\pm}(j) - \Phi_a^\pm] = \frac{t \sin k (1 - \delta_t)}{l} x, \quad 1 \leq x \leq l.
\end{aligned}
\end{equation}
The exponential factor in the wave function modulus is then:
\begin{equation}\label{eq:stl xi}
\begin{aligned}
\frac{x}{\xi_{\text{stl},k_\pm}} &= \frac{2\Phi_k^s(x)}{\mathbf{v}_{k_\pm}} = 2\frac{\delta_t - 1}{l} x, \quad 1 \leq x \leq l, \\
\xi_{\text{stl}} &= \frac{l}{2(\delta_t - 1)}.
\end{aligned}
\end{equation}
In the perturbative regime $|\delta_t| \approx 1$, where $\ln \delta_t \sim \delta_t - 1$, the localization length $\xi = l / (2 \ln |\delta_t|)$ from Ref.~\cite{guo_ScaletailoredLocalizationIts_2024} matches $\xi_{\text{stl}} = l / (2(\delta_t - 1))$, in full agreement with the GNHSE prediction. The sign of $\delta_t - 1$ determines whether localization occurs at $j = 1$ or $j = L$. Furthermore, for $|\delta_t| \approx 1$, $\sqrt[l]{\delta_t} \sim 1 + (\delta_t - 1)/l$, yielding $\mathrm{Im} E_m \sim t \sin \theta_m (l + \delta_t - 1)/l$. With $k = 2\pi n / l$ ($n = 1, 2, \ldots, l$), we identify $\theta_m = k$ and $\mathrm{Im} E_m = \Phi_a^\pm$, consistent with Ref.~\cite{guo_ScaletailoredLocalizationIts_2024}.

\subsection{Scale-free localization}

From the perspective of GNHSE, both suitable imaginary vector potentials and imaginary scalar potentials can induce scale-free localization (SFL), corresponding respectively to non-Hermitian magnetic skin effect and non-Hermitian electric skin effect. We first examine SFL arising from the electric skin effect.

\subsubsection{SFL from non-Hermitian electric skin effect}
\begin{figure}[ptb]
\centering
{\includegraphics[width=\linewidth]{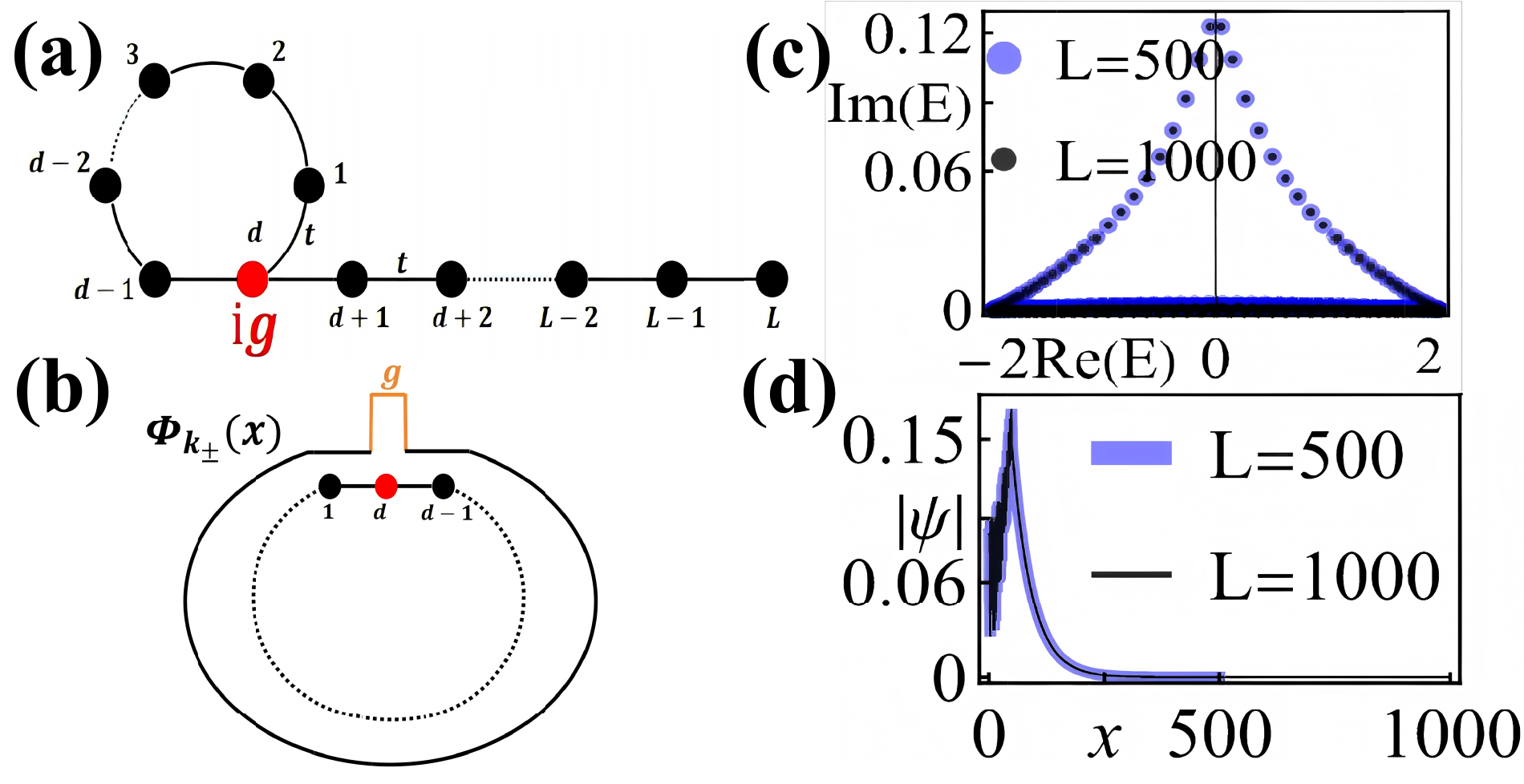}}
\caption{Scale-free localization induced by an imaginary scalar potential. (a) Lattice model of SFL under the GNHSE framework.  
(b) Schematic of the effective dissipation profile $\Phi(x)$ at the $j=d$ interface.  
(c)-(d) Eigenvalues and eigenstate profiles for varying system sizes, with $t = 1, d = 50$ \cite{li_ScalefreeLocalizationPT_2023}.}
\label{fig:sfl}
\end{figure}

Following Ref.~\cite{li_ScalefreeLocalizationPT_2023}, SFL manifests in two distinct scenarios: one with a localized non-Hermitian perturbation (e.g., gain/loss) applied at the boundary, and another with a non-Hermitian impurity located at a finite distance from the boundary. We first consider the simplest boundary case:
\begin{equation}\label{eq:SFL model 1}
H_{\text{SFL},\text{edge}}= \sum_{j=1}^{L-1}t\left(|j\rangle\langle j+1|+|j+1\rangle\langle j|\right)+\text{i}g|L\rangle\langle L|, 
\end{equation}
where eigenstates exhibit scale-free localization $|\psi_L(x)|\simeq e^{cx/L}$, with $c$ independent of system size $L$ (though typically mode-dependent). The localization length scales linearly with $L$. Additionally, the imaginary part of eigenvalues scales inversely with system size: $\mathrm{Im}E\sim 2c/L$.

In the second scenario, with the impurity located at distance $d$ from the boundary:
\begin{equation}\label{eq:SFL model 2}
H_{\text{SFL},\text{bulk}}= \sum_{j=1}^{L-1}t\left(|j\rangle\langle j+1|+|j+1\rangle\langle j|\right)+\text{i}g|d\rangle\langle d|, 
\end{equation}
a set of eigenmodes accumulates near the impurity, exhibiting imaginary eigenvalues and localization lengths independent of total system size $L$, as shown in Figs.~\ref{fig:sfl}(c) and (d). The localization length scales with the subsystem size $d$: $|\psi_L(x)|\sim e^{cx/d}$. For the subsystem between boundary and impurity, these modes are scale-free. When $d \ll L$, they also appear exponentially localized in the full system.

We now reanalyze SFL using the GNHSE framework. Two equivalent approaches exist: periodic-boundary and open-boundary analyses.

\textbf{Periodic boundary condition}---In Eq.~\eqref{eq:SFL model 2}, both the open boundary and the term $\text{i}g|d\rangle\langle d|$ break translational symmetry, partitioning the system into two subsystems: $j\in[1,d]$ and $j\in[d+1,L]$. Imposing a global periodic boundary condition $|L+1\rangle=|1\rangle$ renders Eqs.~\eqref{eq:SFL model 1} and \eqref{eq:SFL model 2} indistinguishable due to rotational symmetry, yet numerical results reveal distinct localization behaviors: $|\psi_L(x)|\simeq e^{cx/L}$ versus $|\psi_L(x)|\sim e^{cx/d}$. To resolve this, we adopt a local periodic boundary condition $|d\rangle=|0\rangle$, as illustrated in Fig.~\ref{fig:sfl}(a). When $d=L$, this reduces to the global periodic condition $|L+1\rangle=|1\rangle$.

Under this local boundary condition, the potential $\Phi(x)$, shown in Fig.~\ref{fig:sfl}(b), corresponds to a discretized Dirac delta function localized at a single site. We obtain:
\begin{equation}\label{eq:sfl lambda}
\begin{aligned}
\Phi_a &= \frac{1}{d} \int_{0}^{d} \Phi_{k_\pm}(x)\;dx \approx \frac{1}{d} \sum_{j=1}^{d} \Phi_{k_\pm}(j) = \frac{g}{d}, \\[1mm]
\Phi_{k_\pm}^s &= \int_{0}^{x} [\Phi_{k_\pm}(x) - \Phi_a]\,dx \approx \sum_{j=1}^{x} [\Phi_{k_\pm}(j) - \Phi_a] = -\frac{gx}{d}, \quad 1 \leq x \leq d.
\end{aligned}
\end{equation}
With $v_{k\pm}=-2t \sin k_\pm$, the exponential factor in the wave function modulus becomes:
\begin{equation}\label{eq:sfl xi}
\begin{aligned}
x\kappa_{\text{E-sfl},k_\pm} &= \frac{2\Phi_{k_\pm}^s}{\mathbf{v}_{k_\pm}} = \frac{c'x}{d}, \quad 1 \leq x \leq d, \\
c'&= \frac{g}{t \sin k_\pm}.
\end{aligned}
\end{equation}
When $d=L$, $c'x/d \sim c'x/L$, and Eq.~\eqref{eq:unified eig} yields $\mathrm{Im}E \sim \Phi_a = g/L$, consistent with numerical results in Ref.~\cite{li_ScalefreeLocalizationPT_2023}. Thus, the boundary impurity case is a special instance of the bulk impurity scenario. The parameter $c'$ is independent of system size but depends on the eigenmode $k$; modes with opposite $k$ exhibit localization on opposite sides of the impurity, characteristic of the electric skin effect. However, since the discretized imaginary domain walls are separated by only one lattice site, the localization appears symmetric around a single site, resembling the magnetic skin effect shown in Fig.~3 of the main text, though fundamentally distinct.

\textbf{Open boundary condition}---We can equivalently analyze SFL under open boundaries in the continuum limit, with $\Phi(x)$ as a Dirac delta function and Dirichlet boundary conditions $\psi_k(x) \to 0$ as $x \to \pm \infty$. Since the single-particle solutions of the first-order operator $\pm\mathrm{i}\dv{}{x}$ decay exponentially in only one direction, they cannot satisfy Dirichlet conditions at both infinities simultaneously. However, coupling two such modes enables full Dirichlet localization. Consider the coupled system:
\begin{equation}\label{eq:E SFL OBC}
\begin{cases}
\mathrm{i} v \dv{\psi_1(x)}{x} + \mathrm{i} g \delta(x - d) \psi_2(x) = (E_r + \mathrm{i} E_i) \psi_1(x) \\
\mathrm{i} (-v) \dv{\psi_2(x)}{x} + \mathrm{i} g \delta(x - d) \psi_1(x) = (-E_r + \mathrm{i} E_i) \psi_2(x)
\end{cases},
\end{equation}
where:
\begin{itemize}
    \item $v = v_k > 0$ is real,
    \item $g$ is real (coupling strength),
    \item $\delta(x - d)$ is the Dirac delta function (localized at $x = d$),
    \item $E = E_r + \mathrm{i} E_i$ is the complex eigenvalue,
    \item Dirichlet boundary conditions: $\psi_1(x), \psi_2(x) \to 0$ as $x \to \pm \infty$.
\end{itemize}
The two coupled modes possess opposite effective velocities $\pm v$. Eigenvalue constraints are inferred from Fig.~\ref{fig:sfl}(c).

For $x \neq d$, $\delta(x - d) = 0$, and the equations decouple:
\[
\begin{cases}
\mathrm{i} v \dv{\psi_1(x)}{x} = (E_r + \mathrm{i} E_i) \psi_1 \\
\mathrm{i} (-v) \dv{\psi_2(x)}{x} = (-E_r + \mathrm{i} E_i) \psi_2
\end{cases},
\]
with general solutions:
\[
\psi_1(x) = A \exp\left[ \left( \frac{E_i}{v} - \mathrm{i} \frac{E_r}{v} \right) x \right], \quad
\psi_2(x) = B \exp\left[ \left( -\frac{E_i}{v} - \mathrm{i} \frac{E_r}{v} \right) x \right].
\]
Boundary condition analysis yields:
\begin{itemize}
\item For $\psi_1$, the real exponent $\frac{E_i}{v}$ requires:
  \begin{itemize}
  \item Decay as $x \to -\infty$ implies $\frac{E_i}{v} > 0$ ($E_i > 0$).
  \item Decay as $x \to +\infty$ implies $\frac{E_i}{v} < 0$ ($E_i < 0$).
  \end{itemize}
\item For $\psi_2$, the real exponent $-\frac{E_i}{v}$ requires:
  \begin{itemize}
  \item Decay as $x \to -\infty$ implies $-\frac{E_i}{v} > 0$ ($E_i < 0$).
  \item Decay as $x \to +\infty$ implies $-\frac{E_i}{v} < 0$ ($E_i > 0$).
  \end{itemize}
\end{itemize}
At $x = d$, the Dirac delta introduces discontinuities. Defining $\psi_i(d) = \frac{\psi_i(d^+) + \psi_i(d^-)}{2}$, integration yields the jump conditions:
\begin{align*}
v [\psi_1(d^+) - \psi_1(d^-)] &= -g \cdot \frac{\psi_2(d^+) + \psi_2(d^-)}{2} \\
v [\psi_2(d^+) - \psi_2(d^-)] &= g \cdot \frac{\psi_1(d^+) + \psi_1(d^-)}{2}
\end{align*}

Two cases emerge:

\paragraph*{Case 1: $E_i > 0$}
\begin{itemize}
\item For $x < d$: $\psi_1$ decays, $\psi_2$ grows, so $\psi_2^- = 0$:
  \[
  \psi_1^-(x) = A \exp\left[ \left( \frac{E_i}{v} - \mathrm{i} \frac{E_r}{v} \right) (x - d) \right], \quad \psi_2^-(x) = 0
  \]
\item For $x > d$: $\psi_1$ grows, $\psi_2$ decays, so $\psi_1^+ = 0$:
  \[
  \psi_1^+(x) = 0, \quad \psi_2^+(x) = B \exp\left[ \left( -\frac{E_i}{v} - \mathrm{i} \frac{E_r}{v} \right) (x - d) \right]
  \]
\end{itemize}
Jump conditions:
\begin{align*}
v [0 - A] &= -g \cdot \frac{B + 0}{2} \implies v A = \frac{g B}{2} \\
v [B - 0] &= g \cdot \frac{0 + A}{2} \implies v B = \frac{g A}{2}
\end{align*}
This yields $g^2 = 4v^2$ (i.e., $g = \pm 2v$), with:
\[
B = 
\begin{cases} 
A & \text{if } g = 2v \\ 
-A & \text{if } g = -2v 
\end{cases}
\]

\paragraph*{Case 2: $E_i < 0$}
\begin{itemize}
\item For $x < d$: $\psi_1$ grows, $\psi_2$ decays, so $\psi_1^- = 0$:
  \[
  \psi_1^-(x) = 0, \quad \psi_2^-(x) = C \exp\left[ \left( -\frac{E_i}{v} - \mathrm{i} \frac{E_r}{v} \right) (x - d) \right]
  \]
\item For $x > d$: $\psi_1$ decays, $\psi_2$ grows, so $\psi_2^+ = 0$:
  \[
  \psi_1^+(x) = D \exp\left[ \left( \frac{E_i}{v} - \mathrm{i} \frac{E_r}{v} \right) (x - d) \right], \quad \psi_2^+(x) = 0
  \]
\end{itemize}
Jump conditions:
\begin{align*}
v [D - 0] &= -g \cdot \frac{0 + C}{2} \implies v D = -\frac{g C}{2} \\
v [0 - C] &= g \cdot \frac{D + 0}{2} \implies -v C = \frac{g D}{2}
\end{align*}
This again yields $g^2 = 4v^2$ (i.e., $g = \pm 2v$), with:
\[
D = 
\begin{cases} 
-C & \text{if } g = 2v \\ 
C & \text{if } g = -2v 
\end{cases}
\]

In summary, nontrivial solutions exist only if:
\[
g^{2} = 4 v^{2} \quad \text{or equivalently} \quad g = 2v \quad \text{or} \quad g = -2v.
\]
The eigenvalue is $E = E_{r} + \mathrm{i} E_{i}$, with $E_{r} \in \mathbb{R}$ and $E_{i} > 0$ or $E_{i} < 0$. For $E_{i} > 0$:
\begin{equation}
\psi_{1}(x) = 
\begin{cases} 
A \exp\left( \dfrac{E_{i} - \mathrm{i} E_{r}}{v} (x - d) \right) & x < d \\ 
0 & x > d 
\end{cases} ,\quad
\psi_{2}(x) = 
\begin{cases} 
0 & x < d \\ 
B \exp\left( \dfrac{ - E_{i} - \mathrm{i} E_{r}}{v} (x - d) \right) & x > d 
\end{cases} ,
\end{equation}
where:
\[
B = 
\begin{cases} 
A & \text{if } g = 2v \\ 
-A & \text{if } g = -2v 
\end{cases}.
\]
For $E_{i} < 0$:
\begin{equation}
\psi_{1}(x) = 
\begin{cases} 
0 & x < d \\ 
D \exp\left( \dfrac{E_{i} - \mathrm{i} E_{r}}{v} (x - d) \right) & x > d 
\end{cases} \\
\psi_{2}(x) = 
\begin{cases} 
C \exp\left( \dfrac{ - E_{i} - \mathrm{i} E_{r}}{v} (x - d) \right) & x < d \\ 
0 & x > d 
\end{cases},
\end{equation}
where:
\[
D = 
\begin{cases} 
-C & \text{if } g = 2v \\ 
C & \text{if } g = -2v 
\end{cases}.
\]
No nontrivial solutions exist if $g^{2} \neq 4v^{2}$.

The coupled eigenstates localize at $x=d$ and decay exponentially at infinity, with localization length $v/|E_i|$. Since $x$ extends to $\pm\infty$, the spectrum remains continuous. Following Eq.~\eqref{eq:sfl lambda}, we set $E_i = \Phi_a$, recovering the identical localization length derived under the periodic boundary condition in Eq.~\eqref{eq:sfl xi}.

\subsubsection{SFL from non-Hermitian magnetic skin effect}

\begin{figure*}[ptb]
\centering
{\includegraphics[width=\textwidth,height=7cm]{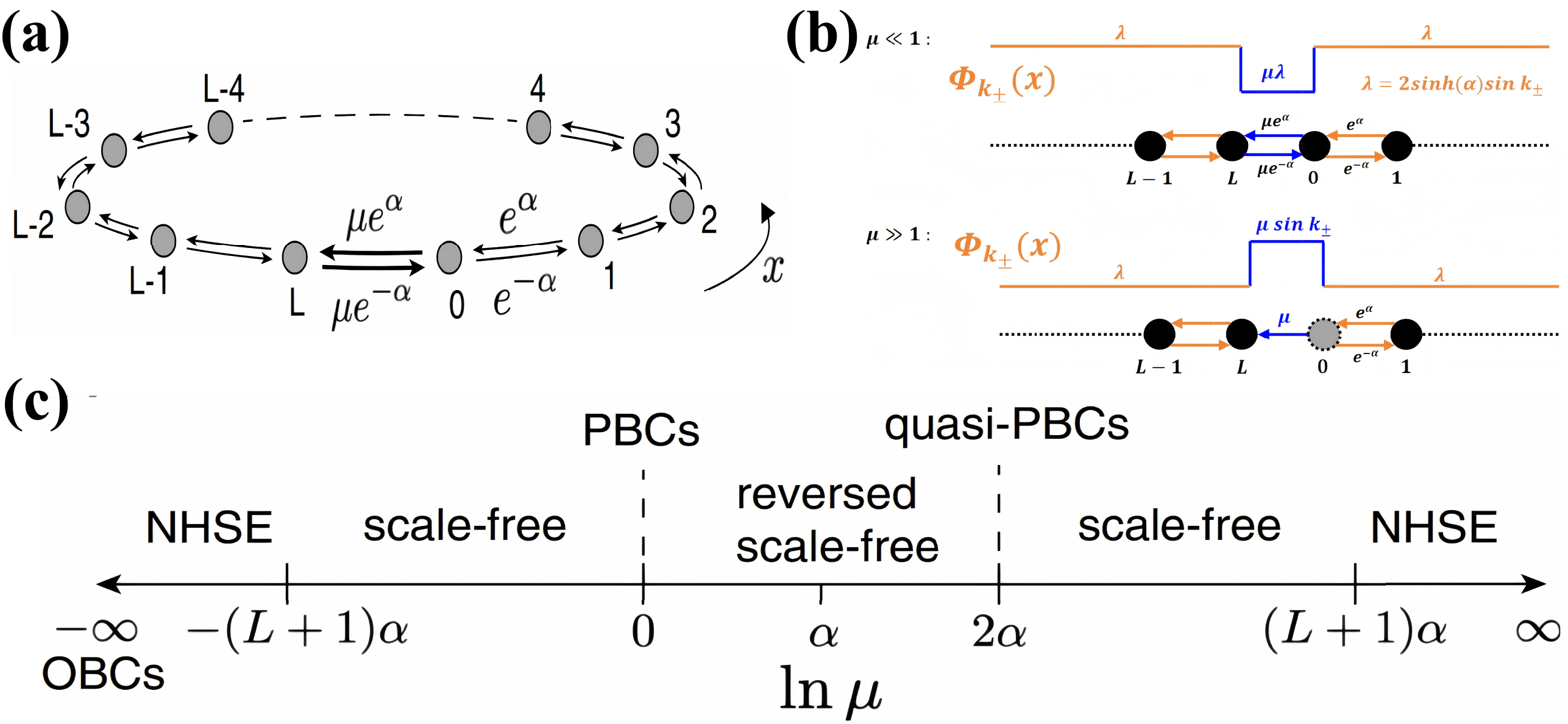}}
\caption{Scale-free localization induced by an imaginary vector potential. (a) Lattice model of SFL \cite{li_ImpurityInducedScalefree_2021}.  
(b) Schematic of the effective dissipation profile $\Phi_{k_\pm}(x)$ at the interface.  
(c) Distinct regimes across the full range of $\mu$, marked by different accumulation phenomena, with dualities relating strong and weak $\mu$ \cite{li_ImpurityInducedScalefree_2021}.}
\label{fig:M-sfl}
\end{figure*}

An imaginary vector potential can also induce scale-free localization (SFL). We begin with the model proposed in Ref.~\cite{li_ImpurityInducedScalefree_2021}, illustrated in Fig.~\ref{fig:M-sfl}(a):
\begin{equation}\label{eq:M-sfl model}
\mathcal{H}= \sum_{j=0}^{L-1}[e^\alpha\hat{c}_j^\dag\hat{c}_{j+1}+e^{-\alpha}\hat{c}_j^\dag\hat{c}_{j-1}]+\mu_+\hat{c}_L^\dag\hat{c}_{0}+\mu_-\hat{c}_0^\dag\hat{c}_{L},
\end{equation}
where $\mu_\pm=\mu e^{\pm\alpha}$, $\mu$ controls the local impurity strength, and the nonreciprocal hopping strength $\alpha>0$. When $\mu=1$, the model reduces to periodic boundary conditions; when $\mu=0$, it corresponds to open boundary conditions. The regime $0<\mu<1$ interpolates between periodic and open boundaries, as summarized in Fig.~\ref{fig:M-sfl}(c). As $\ln \mu$ varies, the skin behavior changes significantly, broadly dividing into two regimes: (i) $e^{-2\alpha} \ll \mu \ll 1$ (weak impurity), and (ii) $\mu \gg e^{\pm\alpha}$ (strong impurity). Notably, in the strong impurity regime, quasi-periodic boundary condition (quasi-PBC) behavior emerges; specifically, at $\mu = 2\alpha$, eigenstates become spatially uniform despite broken translational symmetry --- the effective boundary condition is not strictly periodic.

The physical origin of the quasi-PBC phenomenon is that strong impurity coupling $\mu$ compensates for the non-Hermitian skin effect (NHSE) induced by nonreciprocity ($e^{\pm\alpha}$), effectively restoring translational invariance. The impurity introduces a coupling that precisely cancels the “imaginary magnetic flux” accumulated at the boundaries due to bulk nonreciprocity \cite{li_ImpurityInducedScalefree_2021}, rendering the system equivalent to a homogeneous ring with zero net gauge field.

We now reanalyze this imaginary vector potential–induced SFL using the GNHSE framework. The model in Eq.~\eqref{eq:M-sfl model} inherently forms a closed loop, as shown in Fig.~\ref{fig:M-sfl}(a). Under periodic boundary conditions ($\mu=1$), the dispersion is:
\[
E(k)=2\cosh(\alpha)\cos k+\mathrm{i}2\sinh(\alpha)\sin k,
\]
yielding group velocity $\mathbf{v}_{k_\pm}=-2\cosh(\alpha)\sin k_\pm$ and imaginary part $\mathrm{Im}(E(k))=2\sinh(\alpha)\sin k_\pm$. Since $\Phi_{k_\pm}(x)$ is related to $\mathrm{Im}(E(k))$, we proceed to construct $\Phi_{k_\pm}(x)$. Due to the distinct skin behaviors in weak ($\mu \ll 1$) and strong ($\mu \gg 1$) impurity regimes, including the emergence of quasi-PBC in the latter, we must define two different $\Phi_{k_\pm}(x)$ profiles.

\textbf{Case 1: Weak impurity ($\mu \ll 1$)} --- Here, the impurity acts as a perturbation to the intrinsic nonreciprocal hopping $e^{\pm\alpha}$, as depicted in the upper panel of Fig.~\ref{fig:M-sfl}(b). We obtain:
\begin{equation}\label{eq:M-sfl lambda}
\begin{aligned}
\Phi_a^{\pm} &= \frac{1}{L} \int_{0}^{L} \Phi_{k_\pm}(x)\;dx \approx \frac{1}{L} \sum_{j=1}^{L} \Phi_{k_\pm}(j) = \frac{\lambda(L-1)+\lambda\mu}{L}, \\[1mm]
\Phi_{k_\pm}^s (x)&= \int_{0}^{x} [\Phi_{k_\pm}(x) - \Phi_a^{\pm}]\,dx \approx \sum_{j=1}^{x} [\Phi_{k_\pm}(j) - \Phi_a^{\pm}] = \frac{\lambda(1-\mu)}{L}x, \quad 0 \leq x \leq L,
\end{aligned}
\end{equation}
where $\lambda=2\sinh(\alpha)\sin k_\pm$. Given $v_{k\pm}=-2\cosh(\alpha)\sin k_\pm$, the exponential localization factor becomes:
\begin{equation}\label{eq:M-sfl xi}
\begin{aligned}
x\kappa_{\text{M-sfl},k_\pm} = \frac{2\Phi_k^s(x)}{\mathbf{v}_{k_\pm}} = -\frac{2x(1-\mu)}{L}\tanh(\alpha)\approx\frac{2\alpha(\mu-1)}{L}x, \quad 0\leq x \leq L.
\end{aligned}
\end{equation}
The final approximation holds for $\alpha \sim 0$, consistent with GNHSE as a perturbative effective theory. From Eq.~\eqref{eq:M-sfl xi}, we observe:
\begin{itemize}
    \item When $\mu = 1$, $x\kappa = 0$: no skin effect, corresponding to periodic boundaries.
    \item For $\mu < 1$ ($\ln \mu < 0$): SFL --- localization toward the right boundary.
    \item For $\mu > 1$ ($\ln \mu > 0$): reversed SFL ---localization toward the left boundary.
    \item When $\mu - 1 \propto -L$, $x\kappa_{\text{M-sfl},k_\pm} \sim -2\alpha x$, recovering the localization length of the open-boundary case.
\end{itemize}
All these behaviors are illustrated in Fig.~\ref{fig:M-sfl}(c).

\textbf{Case 2: Strong impurity ($\mu \gg 1$)} --- When the coupling between sites $j=0$ and $j=L$ becomes sufficiently strong ($\ln \mu > 2\alpha$, as shown in Fig.~\ref{fig:M-sfl}(c)), the wavefunction amplitude at $j=0$ is suppressed, mimicking an open boundary. Eigenstates localize near $j=1$, and a transition from reversed SFL to SFL occurs. In this regime, the strong impurity can be viewed as a unidirectional hopping of strength $\mu$ between $j=0$ and $j=L$, analogous to Eq.~\eqref{eq:stl lambda}. As shown in the lower panel of Fig.~\ref{fig:M-sfl}(b), we construct $\Phi_{k_\pm}(x)$ as:
\begin{equation}\label{eq:M-sfl lambda gg}
\begin{aligned}
\Phi_a^{\pm} &= \frac{1}{L} \int_{0}^{L} \Phi_{k_\pm}(x)\;dx \approx \frac{1}{L} \sum_{j=1}^{L} \Phi_{k_\pm}(j) = \frac{\lambda(L-1)+\mu\sin k_\pm}{L}, \\[1mm]
\Phi_{k_\pm}^s &= \int_{L}^{x} [\Phi_{k_\pm}(x) - \Phi_a^{\pm}]\,dx \approx \sum_{j=L}^{x} [\Phi_{k_\pm}(j) - \Phi_a^{\pm}] = \frac{\lambda-\mu\sin k_\pm}{L}(x-L), \quad 0 \leq x \leq L.
\end{aligned}
\end{equation}
Note that the lower integration limit in $\Phi_{k_\pm}^s$ is now $x=L$ instead of $x=0$, reflecting the suppression of the wavefunction at $j=0$ due to strong coupling. The exponential factor is:
\begin{equation}\label{eq:M-sfl xi gg}
\begin{aligned}
x\kappa_{\text{M-sfl},k_\pm} = \frac{2\Phi_k^s}{\mathbf{v}_{k_\pm}} = -\frac{2\sinh(\alpha)-\mu}{L\cosh(\alpha)}(x-L)\approx\frac{2\alpha-\mu}{L}(L-x), \quad 0\leq x \leq L.
\end{aligned}
\end{equation}
Again, the final approximation assumes $\alpha \sim 0$. From Eq.~\eqref{eq:M-sfl xi gg}, we find:
\begin{itemize}
    \item When $\mu = 2\alpha$ ($\ln \mu \sim 2\alpha - 1$), $x\kappa = 0$: no skin effect, corresponding to the quasi-PBC regime.
    \item For $\mu < 2\alpha$ ($\ln \mu < 2\alpha - 1$): reversed SFL --- localization near $j=L$.
    \item For $\mu > 2\alpha$ ($\ln \mu > 2\alpha - 1$): SFL --- localization near $j=1$.
\end{itemize}
While GNHSE is an effective theory and thus yields quantitative discrepancies with the exact numerical results in Ref.~\cite{li_ImpurityInducedScalefree_2021} (Fig.~\ref{fig:M-sfl}(c)), the qualitative behavior is fully consistent.

In summary, the GNHSE framework successfully captures the SFL phenomenon induced by an imaginary vector potential, revealing a rich phase diagram governed by impurity strength $\mu$, and unifying the understanding of weak-impurity SFL, strong-impurity quasi-PBC, and the transitions between them.

\subsection{Critical non-Hermitian skin effect}
\begin{figure*}[ptb]
 \centering
{\includegraphics[width=1\linewidth]{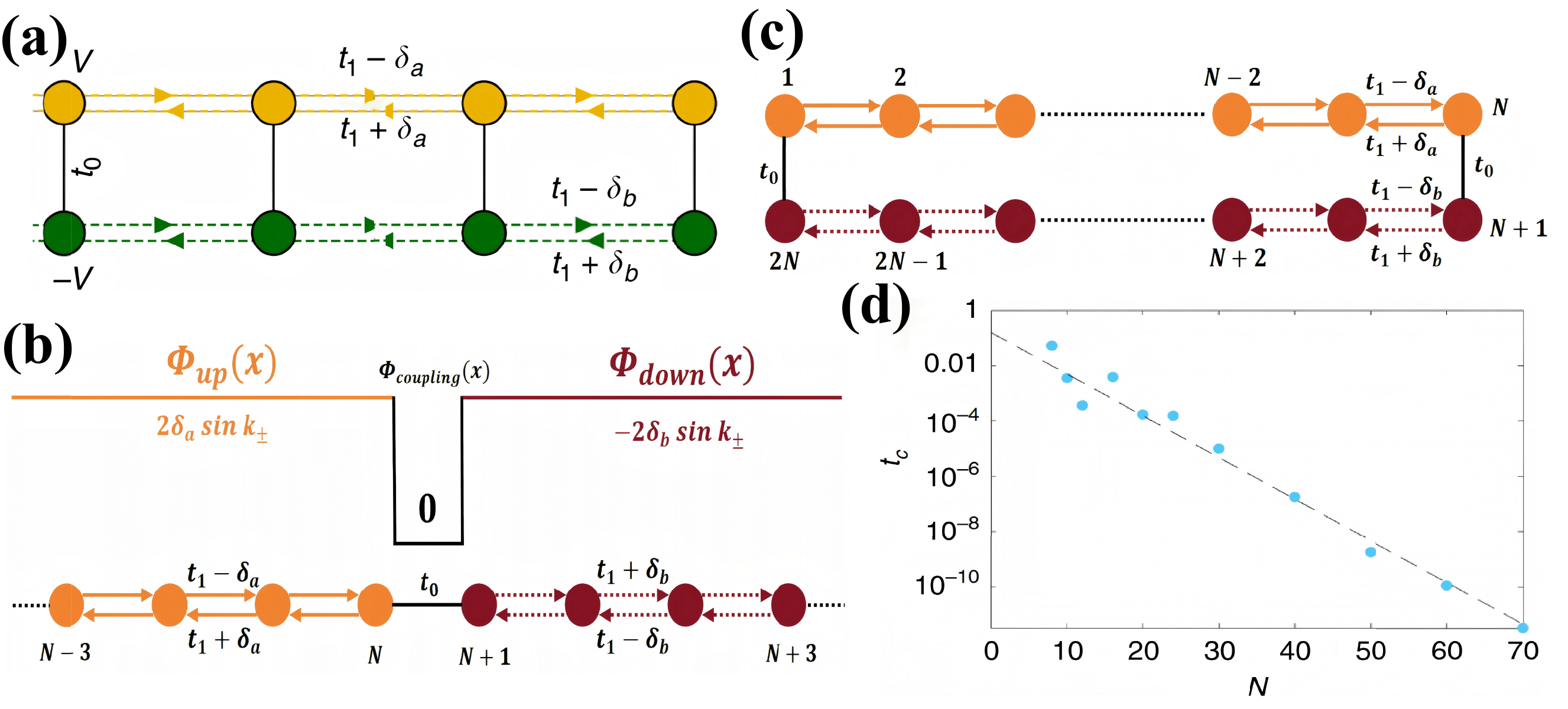}}  
\caption{(a) Lattice model of the critical non-Hermitian skin effect (critical NHSE) \cite{li_CriticalNonHermitianSkin_2020}.  
(b) Schematic of the dissipation term $\Phi_{k_\pm}(x)$ at the $j=N$ interface.  
(c) Reduced lattice model of the critical NHSE under the GNHSE theory.  
(d) Variation of the critical coupling strength $t_c$ with system size $N$ \cite{li_CriticalNonHermitianSkin_2020}.}
\label{fig:cse}
\end{figure*}

The original real-space Hamiltonian for the critical non-Hermitian skin effect (critical NHSE) studied in Ref.~\cite{li_CriticalNonHermitianSkin_2020} is schematically depicted in Fig.~\ref{fig:cse}(a). It consists of two coupled one-dimensional non-Hermitian Hatano-Nelson chains. The upper chain features nearest-neighbor nonreciprocal hoppings $t_1 \pm \delta_a$, while the lower chain has hoppings $t_1 \pm \delta_b$. Each site is coupled between chains by an interchain hopping $t_0$, and $\pm V$ denotes the on-site energy offset.

For this bilayer model under open boundary conditions on both chains, the inverse decay lengths are given by
\begin{equation*}
\kappa_{\text{up}} = \ln\sqrt{\frac{t_1 + \delta_a}{t_1 - \delta_a}}, \quad \kappa_{\text{down}} = \ln\sqrt{\frac{t_1 + \delta_b}{t_1 - \delta_b}}.
\end{equation*}
When $\delta_a \neq \delta_b$, the inverse decay lengths differ. Applying a similarity transformation with factor $e^{j\kappa_{\text{down}}}$ at each site $j$ renders the lower chain reciprocal ($\kappa'_{\text{down}} = 0$), while the upper chain acquires a modified inverse decay length $\kappa'_{\text{up}} = \kappa_{\text{up}} - \kappa_{\text{down}}$. If $\kappa'_{\text{up}} \neq 0$, the upper chain exhibits skin modes exponentially localized at one end with amplitude scaling as $e^{\kappa'_{\text{up}} N}$. Consequently, even an infinitesimal coupling $t_0$ significantly alters the spectrum and eigenstates as system size $N$ increases. Unlike conventional skin modes with fixed spatial decay lengths, the skin radius in critical NHSE at finite size scales with $N$, yielding scale-free exponential wavefunctions:
$$ 
|\psi(x)| \sim e^{-x\kappa'_{\text{up}}}, \quad \kappa'_{\text{up}} \sim N^{-1}.
$$ 
The dependence of the critical coupling strength $t_c$ on system size $N$ is shown in Fig.~\ref{fig:cse}(d). A detailed quantitative solution via the generalized Brillouin zone formalism is provided in Ref.~\cite{yokomizo_ScalingRuleCritical_2021}.

We now revisit the critical NHSE model from the perspective of GNHSE theory. The model in Ref.~\cite{li_CriticalNonHermitianSkin_2020} comprises two Hatano-Nelson chains, each independently exhibiting the non-Hermitian skin effect (NHSE) under open boundaries, unlike impurity-induced local dissipation in models such as SFL or STL. For such composite systems, we propose the following framework to construct the potential $\Phi_{k_\pm}(x)$:

\paragraph{NHSE operator space} Let $H_s$ denote the Hilbert space of a non-Hermitian system, and let $\mathfrak{A} \subset \mathcal{B}(H_s)$ be the set of bounded linear operators acting on it. Define the \textbf{NHSE realization functional} $P_{SE}: \mathfrak{A} \to \{0,1\}$ such that $P_{SE}(\mathcal{H}_{sym}) = 1$ if and only if $\mathcal{H}_{sym}$ exhibits the non-Hermitian skin effect.

\text{Axiomatic definition of elementary operators}---An operator $\mathcal{H}_0 \in \mathfrak{A}$ is elementary if it satisfies:
\begin{equation}
\begin{aligned}
&[1] \, P(\mathcal{H}_0) = 1 \quad \text{(self-consistency)}. \\
&[2] \, \nexists \, \{\mathcal{H}_1, \mathcal{H}_2\} \subset \mathfrak{A} \setminus \{0\} \text{ such that } 
\mathcal{H}_0 = \mathcal{C}\left( \mathcal{H}_1, \mathcal{H}_2 \right) \quad \text{(irreducibility)}.
\end{aligned}
\end{equation}
We denote such operators as $H_0 \in \mathfrak{A_f}$, with $\mathfrak{A_f}\subset \mathfrak{A}$.

\emph{Note: $\mathcal{C}: \mathfrak{A}^n \to \mathfrak{A}$ is a composition operator mapping $\mathcal{H}_1, \mathcal{H}_2, \dots$ to a new operator $\mathcal{H}_j$, where $\{\mathcal{H}_1, \mathcal{H}_2, \mathcal{H}_j \} \subset \mathfrak{A}$.}

For any $\mathcal{H}_{sym}\in \mathfrak{A}$ satisfying $P_{SE}(\mathcal{H}_{sym}) = 1$, $\mathcal{H}_{sym}$ can always be decomposed as $\mathcal{H}_{sym}=\mathcal{C}\left( \mathcal{H}_1, \mathcal{H}_2 \ldots \right)$. We further define a translation-symmetry-restoring operator $T: \mathcal{H}_{sym} \to \mathcal{H}_{sym}^p $ such that $P_{SE}(\mathcal{H}_{sym}^p) = 0$, i.e., $[\mathcal{H}_{sym}^p,\hat{\mathcal{T}}(\mathbf{x})]=0$, where $\hat{\mathcal{T}}(\mathbf{x})=\exp(-\frac{i}{\hbar}\mathbf{x}\cdot \hat{p})$ and $\hat{p}=-i\hbar\nabla$ in the position representation.

\paragraph{GNHSE superoperator $\mathcal{L}$} For any $\mathcal{H}_{sym}\in \mathfrak{A}$ decomposed as $\mathcal{H}_{sym}=\mathcal{C}\left( \mathcal{H}_1, \mathcal{H}_2 \ldots \mathcal{H}_j \right)$ with $ \mathcal{H}_n\in \mathfrak{A_f}$, applying $T$ yields $\mathcal{H}_n^p $, whose eigenvalue decomposition reads:
\begin{equation}
\mathcal{H}_n^p(\mathbf{k}) \ket{u(\mathbf{k})} = \varepsilon_n(\mathbf{k}) \ket{u(\mathbf{k})}, \quad 
\varepsilon_n(\mathbf{k}) = \varepsilon^R_n(\mathbf{k}) + i\varepsilon^I_n(\mathbf{k}).
\end{equation}
Here, $\varepsilon^R_n$ and $\varepsilon^I_n$ are real-valued functions representing the real and imaginary parts of the spectrum. The GNHSE superoperator $\mathcal{L}$ acts as:
\begin{equation}
\begin{aligned}
\mathcal{L}[\mathcal{H}_{sym}] &= \big\{  v_{sym}(\mathbf{k}), \Phi_{sym}(\mathbf{x}) \big\}, \\
v_{sym}(\mathbf{k}) &= \mathcal{C}_v\left( \nabla_{\mathbf{k}} \varepsilon^R_1(\mathbf{k}) ,\nabla_{\mathbf{k}} \varepsilon^R_2(\mathbf{k})\ldots \nabla_{\mathbf{k}} \varepsilon^R_j(\mathbf{k})\right), \\
\Phi_{sym}(\mathbf{x}) &= \mathcal{C}_\Phi\left(  \varepsilon^I_1(\mathbf{k}) ,  \varepsilon^I_2(\mathbf{k}) \ldots     \varepsilon^I_j(\mathbf{k})   \right).
\end{aligned}
\end{equation}
The operators $\mathcal{C}_v$ and $\mathcal{C}_\Phi$ combine the effective velocity $v_{sym}$ and potential $\Phi_{sym}$ analogously to $\mathcal{C}$. Most non-Hermitian systems encountered are elementary, as they constitute minimal units for NHSE. Exceptions such as the critical NHSE system in Ref.~\cite{li_CriticalNonHermitianSkin_2020} are reducible, being composed of two Hatano-Nelson chains.

From $\{  v_{sym}(\mathbf{k}), \Phi_{sym}(\mathbf{x})\}$, we derive the effective Hamiltonian for the non-Hermitian system $H_s$:
\begin{equation}\label{eq:unified H}
\begin{aligned}
\hat{H}_{eff}&=v_{sym}(\mathbf{k})\hat{k}  + \mathrm{i}\Phi_{sym}(\mathbf{x}),
\end{aligned}
\end{equation}

Applying the \emph{NHSE operator space} and \emph{GNHSE superoperator} to the system in Fig.~\ref{fig:cse}(a), we recognize its reducibility under open boundaries, as it decomposes into two elementary Hatano-Nelson operators, each independently capable of NHSE. Neglecting interchain coupling, the system reduces to that in Fig.~\ref{fig:cse}(c), subject to periodic boundary conditions $\psi(0) = \psi(2N)$. The effective velocity and potential are then:
\begin{equation*}
v_{sym}(\mathbf{k})= 
\begin{cases}
-2t \sin k_\pm, & 1 \leq x \leq N, \\
-2t \sin k_\pm, & N+1 \leq x \leq 2N,
\end{cases}, \;
\Phi_{sym}(\mathbf{x})=
\begin{cases}
\Phi_{\text{up}}(x)=2\delta_a \sin k_\pm, & 1 \leq x \leq N, \\
\Phi_{\text{coupling}}(x)=0, &  x\in(N,N+1)\cup(2N,1),\\
\Phi_{\text{down}}(x) = -2\delta_b \sin k_\pm, & N+1 \leq x \leq 2N.
\end{cases}
\end{equation*}
Denoting $v_{sym}(\mathbf{k})=\mathbf{v}_{k_\pm}$ and $\Phi_{sym}(\mathbf{x})=\Phi_{k_\pm}(x)$, we note that $\Phi_{k_\pm}(x)$ is spatially inhomogeneous, with its profile at the $j=N$ interface shown in Fig.~\ref{fig:cse}(b). Including the $j=2N$ interface, there are four interfacial dissipation walls (IDWs), forming two closely spaced pairs discretized to adjacent lattice sites at the boundaries. Setting $\delta_a=-\delta_b$ as in Ref.~\cite{li_CriticalNonHermitianSkin_2020}, we compute:
\begin{equation}\label{eq:CSE lambda}
\begin{aligned}
\Phi_a^\pm&= \frac{1}{2N} \left( \int_{0}^{N} \Phi_{\text{up}}(x)\;dx + \int_{N}^{2N} \Phi_{\text{down}}(x)\;dx \right)\approx \frac{1}{2N} \left( \sum_{j=1}^{N} \Phi_{\text{up}}(j) + \sum_{j=N+1}^{2N} \Phi_{\text{down}}(j) \right) = 2\delta_a \sin k_\pm \frac{N-1}{N}, \\[1mm]
\Phi_{k_\pm}^s(x) &= \int_{0}^{x} [\Phi_{k_\pm}(x) - \Phi_a]\,dx \approx 
\begin{cases}
\displaystyle 
\sum_{j=1}^{x} [\Phi_{\text{up}}(j) - \Phi_a^\pm] = 2\delta_a \sin k_\pm \frac{x}{N}, & 1 \leq x \leq N, \\[5mm]
\displaystyle 
2\delta_a \sin k + \sum_{j=N+1}^{x-N} [\Phi_{\text{down}}(j) - \Phi_a] = \mathcal{N}_1 + 2\delta_a \sin k_\pm \frac{x-N}{N}, & N+1 \leq x \leq 2N.
\end{cases}
\end{aligned}
\end{equation}
Since $\mathcal{N}_1$ is $x$-independent, it contributes only a normalization factor $e^{\mathcal{N}_1}$ to the wavefunction and does not affect skin behavior, hence is omitted. With $\mathbf{v}_{k_\pm}=-2t \sin k_\pm$, the exponential factor in the wavefunction modulus becomes:
\begin{equation}\label{eq:CSE xi}
\begin{aligned}
x\kappa_{\text{c},k_\pm} = \frac{2\Phi_{k_\pm}^s(x)}{\mathbf{v}_{k_\pm}} =
\begin{cases}
\displaystyle 
-\frac{2\delta_a}{tN}x, & 1 \leq x \leq N, \\[3mm]
\displaystyle 
-\frac{2\delta_a}{tN}(x-N), & N+1 \leq x \leq 2N.
\end{cases}
\end{aligned}
\end{equation}
Thus, the inverse decay length scales as $\kappa_{\text{c}}\sim N^{-1}$. From Eq.~\eqref{eq:unified eig}, the eigenenergy is $E=v_{k}k+\mathrm{i}\Phi_a^\pm$. In the thermodynamic limit $N\rightarrow\infty$, $\frac{N-1}{N}\sim 1$ and $\Phi_a^\pm\sim\mathrm{Im}E_{pbc}$, so the critical NHSE spectrum converges to that under periodic boundary conditions, consistent with numerical results in Ref.~\cite{li_CriticalNonHermitianSkin_2020}. Within the GNHSE framework, any $t_0\neq0$ induces inhomogeneous imaginary potentials and critical NHSE, equivalent to $t_c\sim0$ in the $N\rightarrow\infty$ limit (Fig.~\ref{fig:cse}(d)). This arises because GNHSE operates in the continuum limit, where discretization naturally implies $N\rightarrow\infty$; for instance, the error in approximating integrals by sums in Eq.~\eqref{eq:CSE lambda} vanishes as $N$ grows.

In summary, intrachain couplings, apart from the end-to-end coupling at $j=1$ and $j=N$, do not generate spatially inhomogeneous imaginary potentials. In essence, interchain coupling is not necessary for scale-free skin modes with $\kappa_{\text{c}}\sim N^{-1}$; end-to-end coupling alone suffices to establish periodic boundaries and induce spatially inhomogeneous imaginary potentials.

\subsection{Bipolar Non-Hermitian Skin Effect}

As an illustrative example of the GNHSE theory applied to multi-band systems, we consider the Bipolar Non-Hermitian Skin Effect (Bipolar NHSE) Hamiltonian introduced in Ref.~\cite{song_NonHermitianTopologicalInvariants_2019}:
\begin{equation}\label{eq:bi_NHSE model}
 \begin{aligned}
H_{\text{bi-NHSE}}(k)&= d_x(k)\sigma_x + d_y(k)\sigma_y, \\
d_x(k)&=t_1+(t_2+t_3)\cos k +\mathrm{i}\frac{\gamma}{2}\sin k,\\
d_y(k)&=(t_2-t_3)\sin k +\mathrm{i}\frac{\gamma}{2}\cos k,
 \end{aligned}
\end{equation}
where $\sigma_{x,y}$ are Pauli matrices acting on sublattices $A$ and $B$. This model describes a Su-Schrieffer-Heeger (SSH) chain with nonreciprocal next-nearest-neighbor hoppings $t_3 \pm \gamma/2$. The energy dispersion is $E_\pm(k)=\pm\sqrt{d_x^2 + d_y^2}$. The real-space lattice obtained via Fourier transform is shown in Fig.~\ref{fig:biNHSE}(a). The novelty of this skin effect lies in its bipolar nature: eigenstates localize at both boundaries simultaneously, as illustrated in Figs.~\ref{fig:biNHSE}(b) and (c). Specifically, eigenstates with $|E| \lesssim 0.8$ localize near the right edge, while those with $|E| \gtrsim 0.8$ localize near the left edge. This behavior is also reflected in the generalized Brillouin zone (GBZ), where part of the GBZ lies outside the unit circle (BZ), and the rest lies inside. The intersections between GBZ and BZ are called Bloch points, occurring at $k = \pm \arccos(-t_1/2t_2) \equiv k_B$, where eigenstates revert to plane waves.

\begin{figure*}[ptb]
 \centering
{\includegraphics[width=0.75\textwidth]{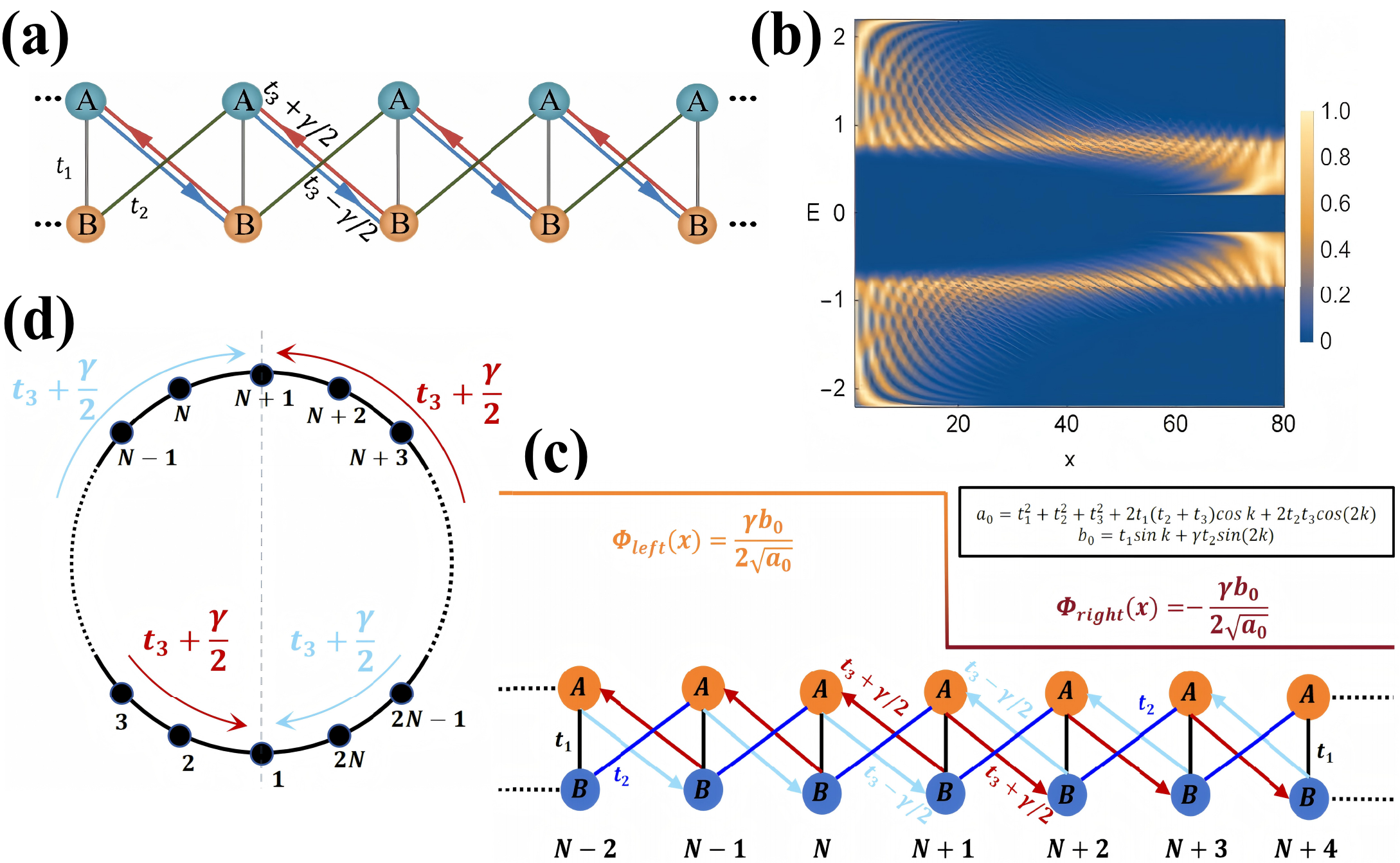}}   
\caption{(a) Real-space lattice model obtained by Fourier transforming Eq.~\eqref{eq:bi_NHSE model} \cite{song_NonHermitianTopologicalInvariants_2019}.
(b) Modulus squared of all bulk eigenstates plotted against eigenenergy $E$; $x$ denotes the site index in an open chain. Eigenstate profiles are normalized to peak value 1. Extended states exist only at discrete Bloch points ($E \approx 0.8$); all others are skin-localized. Parameters: $t_1 = t_2 = 1$, $t_3 = 0.2$, $\gamma = 0.1$ \cite{song_NonHermitianTopologicalInvariants_2019}.
(c) Schematic of the dissipation profile $\Phi_{k_\pm}(x)$ across the $j=N+1$ interface.
(d) Bipolar NHSE lattice model reinterpreted within the GNHSE framework.}
\label{fig:biNHSE}
\end{figure*}

We now reanalyze the Bipolar NHSE from the perspective of GNHSE theory. It can be recognized as another manifestation of NHSE induced by an imaginary vector potential---specifically, a non-Hermitian magnetic skin effect. However, due to the multi-band nature of the model, the effective dissipation profile $\Phi_{k_\pm}(x)$ is not as straightforward as in single-band cases. Since GNHSE is applied band-by-band, we focus on the upper band $E_+(k)$ for concreteness:
\begin{equation}\label{eq:bi NHSE Ek}
 \begin{aligned}
E_+(k)&=\sqrt{d_x^2+d_y^2}=\sqrt{a+\mathrm{i}b},\\
a &= t_1^2 + t_2^2 + t_3^2 - \frac{\gamma^2}{4} + 2t_1(t_2 + t_3)\cos k + 2t_2 t_3 \cos(2k), \\
b &= \gamma t_1 \sin k + \gamma t_2 \sin(2k), \\
\mathrm{Re}(E_+(k)) &= \sqrt{\frac{\sqrt{a^2 + b^2} + a}{2}}, \quad
\mathrm{Im}(E_+(k)) = \mathrm{sgn}(b)\sqrt{\frac{\sqrt{a^2 + b^2} - a}{2}}.
 \end{aligned}
\end{equation}
From this, we identify:
\begin{equation}
 \begin{aligned}
\Phi_{Bi}(x)&=\mathrm{sgn}(b)\cdot\mathrm{Im}(E_+(k)),\\
\mathbf{v}_{k_\pm} &= \frac{\partial \, \mathrm{Re}(E_+)}{\partial k}  \\
&=\frac{ -2t_1(t_2+t_3)\sin k \left(a + \sqrt{a^2 + b^2}\right) -4t_2t_3\sin 2k \left(a + \sqrt{a^2 + b^2}\right) + \gamma t_1 \cos k \cdot b + 2\gamma t_2\cos 2k \cdot b }{4 \sqrt{a^2 + b^2} \cdot \sqrt{\frac{\sqrt{a^2 + b^2} + a}{2}}}.
 \end{aligned}
\end{equation}

For Bipolar NHSE, we adopt a symmetric configuration analogous to the Hatano-Nelson model, two mirror-symmetric subsystems connected at sites $j=1$ and $j=N$, forming two interface domain walls (IDWs), as shown in Fig.~\ref{fig:biNHSE}(d). In this case, $\Phi_a^\pm = 0$, so $E_i = \Phi_a^\pm$. Since GNHSE is a perturbative theory valid for $\gamma \to 0$, we expand $\sqrt{a^2 + b^2}$ to leading order in $\gamma$, neglecting $\mathcal{O}(\gamma^2)$ and higher:
\begin{equation}
 \begin{aligned}
a &=a_0- \frac{\gamma^2}{4},\quad a_0=t_1^2 + t_2^2 + t_3^2 + 2t_1(t_2 + t_3)\cos k + 2t_2 t_3 \cos(2k), \\
b &= \gamma b_0,\quad b_0=t_1 \sin k + t_2 \sin(2k), \\
\sqrt{a^2 + b^2}&\approx a_0+\frac{\gamma^2}{4}\left(\frac{2b_0^2}{a_0}-1\right),
 \end{aligned}
\end{equation}
using the expansion $\sqrt{x+\epsilon}\approx \sqrt{x}+\frac{\epsilon}{2\sqrt{x}}$. Substituting and simplifying (assuming $\gamma>0$ and keeping only $\mathcal{O}(\gamma)$ terms):
\begin{equation}
 \begin{aligned}
\sqrt{a^2 + b^2}-a&\approx \frac{\gamma^2 b_0^2}{2a_0},\quad
\sqrt{a^2 + b^2}+a\approx2a_0,\\
\Phi_{Bi}(x)&\approx \frac{\gamma b_0}{2\sqrt{a_0}},\\
\mathbf{v}_{k_\pm}&\approx -\frac{t_1(t_2+t_3)\sin k+2t_2t_3\sin 2k}{\sqrt{a_0}}.
 \end{aligned}
\end{equation}
Substituting into the GNHSE localization length expression:
\begin{equation}\label{eq:bi NHSE xi}
 \begin{aligned}
\Phi_{k_\pm}^s(x) &= \int_{0}^{x} [ \Phi_{Bi}(x') - \Phi_a^\pm ] \,dx' \approx \frac{\gamma b_0}{2\sqrt{a_0}}x, \quad 1\leq x \leq N, \\
\frac{x}{\xi_{Bi}} &= \frac{2\Phi_{k}^s}{\mathbf{v}_{k_\pm}} \quad \Rightarrow \quad
\xi_{Bi} = \frac{ -t_1(t_2 + t_3)\sin k - 2t_2 t_3 \sin(2k) }{ \gamma \,(t_1 \sin k + t_2 \sin(2k)) }.
 \end{aligned}
\end{equation}
Crucially, $\xi_{Bi}$ depends on momentum $k$, and its sign can flip for different $k$ values --- suggesting localization at different boundaries. However, since GNHSE is perturbative ($\gamma \to 0$), $\xi_{Bi}$ diverges rather than vanishes. Divergence occurs when the denominator vanishes --- corresponding to Bloch points where eigenstates become extended (plane-wave-like), as seen in Figs.~\ref{fig:biNHSE}(b) and (c). Setting the denominator to zero:
\begin{equation}
 \begin{aligned}
\gamma \,(t_1 \sin k + t_2 \sin(2k)) &= 0 \\
\Rightarrow \gamma \sin k (t_1 + 2 t_2 \cos k) &= 0 \\
\Rightarrow k &= \arccos(-t_1/2t_2),\; 2\pi - \arccos(-t_1/2t_2),\; \text{or}\; k = 0, \pi, 2\pi.
 \end{aligned}
\end{equation}
Solutions $k = 0, \pi, 2\pi$ are discarded because the numerator also vanishes there, yielding finite $\xi_{Bi}$. The remaining solutions are precisely the Bloch points $k_B = \pm \arccos(-t_1/2t_2)$. Substituting $k_B$ into Eq.~\eqref{eq:bi NHSE Ek} yields:
\[
E(k_B) = \pm \sqrt{(t_2 - t_3)^2 - \gamma^2/4}.
\]
With parameters $t_1 = t_2 = 1$, $t_3 = 0.2$, $\gamma = 0.1$, we obtain $|E| \approx 0.8$, in perfect agreement with Ref.~\cite{song_NonHermitianTopologicalInvariants_2019}.

Importantly, although $\xi_{Bi}$ changes sign with $k$ --- suggesting localization at different IDWs, reminiscent of Reciprocal Skin Effect (RSE) --- Bipolar NHSE remains fundamentally a non-Hermitian magnetic skin effect. The sign change in $\xi_{Bi}$ arises from $k$-dependence in the dissipation profile, not from the $\mathbf{v}_{k_\pm}$. For RSE, a pair of momenta $k_\pm$ sharing the same $\mathrm{Re}(E(k))$ must have opposite velocities ($\mathrm{sgn}(v_{k_+} v_{k_-}) < 0$) and opposite localization positions ($\mathrm{sgn}(\xi_{k_+} \xi_{k_-}) < 0$). Here, the bipolar localization stems from the global structure of $\Phi_{k}(x)$ and the band dispersion, not from reciprocal velocity pairs.

\subsection{Summary}

In summary, all the non-Hermitian skin effect variants discussed --- including SFL, quasi-PBC, and Bipolar NHSE --- can be universally understood as arising from spatially inhomogeneous imaginary potentials $\Phi_{k_\pm}(x)$ in real space. The distinct phenomenology of each type stems solely from the specific form of $\Phi_{k_\pm}(x)$. This demonstrates the remarkable universality of the GNHSE framework in unifying and explaining diverse one-dimensional non-Hermitian skin effects under a single theoretical umbrella.

\section{Mode-dependent imaginary gauge transformation and boundary sensitivity}
In this section, we introduce a mode-dependent imaginary gauge transformation to provide a more intuitive understanding of boundary sensitivity within the GNHSE framework. This transformation is mathematically equivalent to a mode-dependent similarity transformation, whose essential role is to decompose the total potential field $\Phi_{k_\pm}(x) = v_{k_\pm}A(x) + \Phi(x)$ in the effective Hamiltonian into a removable velocity-dependent component $v_{k_\pm}A_{k_\pm}(x)$ and an irremovable velocity-independent component $\Phi(x)$. By analyzing the distinct physical consequences of eliminating the velocity-dependent component under different boundary conditions, we establish a mechanistic explanation for the fundamentally different responses of the electric and magnetic skin effects to boundary conditions within the GNHSE theory.

Since the GNHSE effective Hamiltonian $\hat{H}^G_k$ itself depends on the mode $k$, we can define an imaginary gauge transformation separately for each mode: $\psi_k(x) \to \tilde{\psi}_k(x) = e^{-\chi_k(x)}\psi_k(x)$, where $\chi_k(x)$ is a real-valued function. The transformed Hamiltonian is $\tilde{H} = e^{-\chi_k}\hat{H}^G_{k_\pm}e^{+\chi_k}$. Using the identity
\begin{equation}\label{eq:trans-cn}
e^{-\chi}\hat{k}\,e^{+\chi} = \hat{k} - i\chi'(x),
\end{equation}
we obtain:
\begin{equation}\label{eq:transH-cn}
\tilde{H} = v_{k_\pm}\big(\hat{k} + i[A_k(x) - \chi'(x)]\big) + i\Phi(x).
\end{equation}
The transformation therefore acts as:
\begin{equation}\label{eq:gaugetrans-cn}
A_k(x) \;\longrightarrow\; \tilde{A}_k(x) = A_k(x) - \chi'(x), \; \Phi(x) \;\longrightarrow\; \tilde{\Phi}(x) = \Phi(x).
\end{equation}
This is structurally analogous to time-independent gauge transformations in electromagnetism: the vector potential is gauge-dependent, while the scalar potential is gauge-invariant. We choose $\chi'_k(x) = A_k(x)$ to completely eliminate $A_k(x)$, yielding $\chi_k(x) = \int_0^x A_k(x'),dx'$. We now discuss the consequences of this elimination under different boundary conditions.

We consider the case where the system $H_s$ is the original single chain without mirror conjugation, exhibiting only the magnetic skin effect. In this case, $v_{k_\pm}A_k(x)=\mathrm{Im}(E(k))$, where $E(k)$ is the spectrum of $H_s$ under PBC, so $A_k(x)$ is independent of $x$, and we have $\bar{A}_k=A_k(x)$.

\noindent{\textit{Under PBC:}}

The condition $\psi_k(0) = \psi_k(L)$ transforms into $\tilde{\psi}_k(L) = e^{-L\bar{A}_k}\tilde{\psi}_k(0)$, where $\bar{A}_k = L^{-1}\int_0^L A_k(x)dx$ is the spatial average of the imaginary vector potential. The transformed free-particle equation $v_{k_\pm}\hat{k}\tilde{\psi} = E\tilde{\psi}$ has the solution $\tilde{\psi} = Ce^{i\tilde{k}x}$, where $C$ is a normalization coefficient. Substituting $\tilde{\psi}$ into the PBC boundary condition yields:
\begin{equation}
e^{i\tilde{k}L} = e^{-\bar{A}_k L} \implies \tilde{k}_n = \frac{2\pi n}{L} + i\bar{A}_k,
\end{equation}
The wavefunction and energy spectrum are:
\begin{equation}
 \begin{aligned}
\psi_k(x) &= e^{\chi_k(x)}\tilde{\psi}_k(x)\\
          &=Ce^{\int_0^x A_k(x')-\bar{A}_k dx'}e^{i k_n x}=Ce^{i k_n x},\\
E_n^{\text{PBC}} &= v_{k_\pm}\tilde{k}_n = v_{k_\pm}\frac{2\pi n}{L} + iv_{k_\pm}\bar{A}_k,
  \end{aligned}
\end{equation}
where $k_n=\frac{2\pi n}{L}$. Thus, to satisfy the PBC boundary condition after the imaginary gauge transformation, the energy spectrum must be complex, while the wavefunction is a plane wave.

\noindent{\textit{Under OBC:}}

The Dirichlet conditions $\psi_k(0) = \psi_k(L) = 0$ are automatically preserved under the imaginary gauge transformation, since $0 \times e^{-\chi_k} = 0$ holds for any $\chi_k$. After the imaginary vector potential is completely eliminated, one again obtains the free-particle equation $v_{k_\pm}\hat{k}\tilde{\psi} = E\tilde{\psi}$. However, unlike the PBC case, because the solution of the first-order derivative operator $\mathrm{i}\dv{}{x}$ takes the form of a single-parameter exponential function, it cannot simultaneously satisfy both boundary conditions. We therefore adopt the same approach as in Section.~\ref{eq2-obc}.A, namely the coupling of the $k_+$ and $k_-$ modes. The two equations before the gauge transformation are:
\begin{equation}\label{eq:pre1}
 \begin{aligned}
-iv_{k_+}\frac{d\psi_{k_+}}{dx} + iv_{k_+}A_{k_+}\psi_{k_+} &= E\psi_{k_+},\\
-iv_{k_-}\frac{d\psi_{k_-}}{dx} + iv_{k_-}A_{k_-}\psi_{k_-} &= E\psi_{k_-}. 
  \end{aligned}
\end{equation}
The eigenvalue $E$ is the same in both equations (since $k_+$ and $k_-$ are two modes corresponding to the same energy). Performing the gauge transformation on each separately:
\begin{equation}
\psi_{k_+} = e^{A_{k_+}x}\tilde{\psi}_{k_+}, \quad \psi_{k_-} = e^{A_{k_-}x}\tilde{\psi}_{k_-}.
\end{equation}
Since $A_k(x)$ is independent of $x$ for the single-chain case, the exponent becomes $\chi_{k_\pm}(x) = \int_0^x A_{k_\pm}(x')dx'=A_{k_\pm} x$. After the transformation:
\begin{equation}\label{eq:trans1}
 \begin{aligned}
-iv_{k_+}\frac{d\tilde{\psi}_{k_+}}{dx} &= E\tilde{\psi}_{k_+} \quad\Longrightarrow\quad \tilde{\psi}_{k_+} = C_+\,e^{iEx/v_{k_+}},\\
-iv_{k_-}\frac{d\tilde{\psi}_{k_-}}{dx} &= E\tilde{\psi}_{k_-} \quad\Longrightarrow\quad \tilde{\psi}_{k_-} = C_-\,e^{iEx/v_{k_-}}. 
  \end{aligned}
\end{equation}
The OBC solution in physical space is a superposition of the two modes:
\begin{equation}\label{Gauge-Obc-psi}
 \begin{aligned}
\psi(x) &= e^{A_{k_+}x}\tilde{\psi}_{k_+}(x) + e^{A_{k_-}x}\tilde{\psi}_{k_-}(x) \\
    &= C_+\,e^{(A_{k_+} + iE/v_{k_+})x} + C_-\,e^{(A_{k_-} + iE/v_{k_-})x}.
 \end{aligned}
\end{equation}
Substituting the Dirichlet conditions:
\begin{equation}
 \begin{aligned}
\psi(0) &= C_+ + C_- = 0 \quad\Longrightarrow\quad C_- = -C_+, \\
\psi(L) &= C_+\left[e^{(A_{k_+} + iE/v_{k_+})L} - e^{(A_{k_-} + iE/v_{k_-})L}\right] = 0. 
 \end{aligned}
\end{equation}
Nontrivial solutions require:
\begin{equation}
e^{(A_{k_+} + iE/v_{k_+})L} = e^{(A_{k_-} + iE/v_{k_-})L},
\end{equation}
that is:
\begin{equation}
(A_{k_+} - A_{k_-})L + iEL\left(\frac{1}{v_{k_+}} - \frac{1}{v_{k_-}}\right) = 2\pi i n,\; n\in\mathbb{Z},
\end{equation}
Separating real and imaginary parts, with $E = E_r + iE_i$ and $\Delta_v = 1/v_{k_+} - 1/v_{k_-}$:
\begin{equation}
 \begin{aligned}
(A_{k_+} - A_{k_-})L - E_i L\Delta_v &= 0 \\
E_r L\Delta_v &= 2\pi n,
 \end{aligned}
\end{equation}
we finally obtain:
\begin{equation}
E_i = \frac{A_{k_+} - A_{k_-}}{\Delta_v},\; E_r = v_a\frac{2\pi n}{L},
\end{equation}
where $v_a = 1/\Delta_v = v_{k_+}v_{k_-}/(v_{k_-} - v_{k_+})$. As established in the main text, one of the hallmarks of NHMSE is $\operatorname{sgn}(\xi^ {\text{M}}_{k_+}\xi^{\text{M}}_{k_-})>0$, meaning the $k_+$ and $k_-$ modes localize in the same direction. From Eq.~\eqref{Gauge-Obc-psi}, the localization length is determined by the exponential factor $A_{k_\pm}$, which yields $A_{k_+} = A_{k_-}=A_{k}$ (this can be verified for the magnetic skin effect models considered in this work, namely the Hatano--Nelson model and the Bipolar NHSE; even if certain special NHSE models do not satisfy this condition, one finds that the eigenstates and the imaginary part of the spectrum $E_i$ under OBC differ fundamentally from those under PBC, so the conclusion remains unchanged). In this case, $E_i = 0$, the OBC spectrum is purely real, and the wavefunction is:
\begin{equation}
 \begin{aligned}
\psi(x)&=Ce^{A_k x}\left(\exp(i\frac{v_a}{v_{k_+}}kx) + \exp(i\frac{v_a}{v_{k_-}}kx)\right)\\
      &=e^{A_k x}\tilde{\psi}(x),
 \end{aligned}
\end{equation}
where $k=2\pi n/L$. Thus, under OBC, the magnetic skin effect is entirely encoded in the envelope factor $e^{A_k x}$, and the spectrum is purely real.

For a purely electric system ($A_k = 0$), the imaginary gauge transformation reduces to the identity. The Hamiltonian $\hat{H}^G_k$ under both PBC and OBC is $v_k\hat{k} + i\Phi(x)$, and no time-independent imaginary gauge transformation can eliminate or modify $\Phi(x)$. Consequently, the electric skin effect driven by the imaginary scalar potential persists under any boundary condition, which constitutes the gauge-theoretic origin of the distinction between NHESE and NHMSE.

We further note that the imaginary gauge transformation introduced here does not require the GBZ to be circular or a global similarity matrix $S$ to exist \cite{yao_EdgeStatesTopological_2018}, since $\chi_k(x)$ differs for each mode --- a feature that is natural within the mode-dependent framework of GNHSE. This stands in contrast to the well-known global similarity transformation in NHSE.

In summary, by combining the mode-dependent imaginary gauge transformation, the imaginary vector potential $A_k(x)$ in the effective Hamiltonian $\hat{H}^G_k$ can be completely eliminated. Under PBC, this transformation renders the spectrum complex; under OBC, it introduces an additional envelope factor $e^{A_k x}$ in the wavefunction. The mode-dependent imaginary gauge transformation thus reveals the origin of NHMSE boundary sensitivity: PBC and OBC respond to the imaginary gauge transformation in a fundamentally asymmetric manner. In contrast, the scalar potential $\Phi(x)$ is unaffected by the imaginary gauge transformation, so the spectrum and wavefunctions exhibit no essential differences under different boundary conditions, providing the theoretical foundation for the boundary insensitivity of NHESE.

However, imaginary gauge invariance is only a necessary condition for boundary insensitivity, not a sufficient one. The sufficient condition further requires that the spatial average of the potential $\Phi_a = L^{-1}\int_0^L \Phi(x),dx$ be an intrinsic property of $\Phi(x)$, independent of the system size or subsystem size.

When $\Phi(x)$ is spatially extensive (possessing structure on the $O(L)$ scale), $\Phi_a$ is intrinsically determined by the functional form of $\Phi(x)$. For instance, in the RSE with $\Phi(x) = \lambda,\text{sgn}((x-L/4)(x-3L/4))$, the symmetry guarantees $\Phi_a = 0$ for any $L$, and the localization length $\xi = t\sin k/\lambda$ contains no system-size parameter. An observer need not know the system size or boundary positions to determine the local decay rate. This is the microscopic mechanism underlying the boundary insensitivity of the RSE.

When $\Phi(x)$ is spatially localized (as in the SFL of the non-Hermitian electric skin effect, where $\Phi(x) = g\delta(x-d)$), $\Phi_a = g/d$ depends on the subsystem size $d$. The inverse decay length at bulk sites, $\kappa \propto - g/(d v_{k_\pm})$, explicitly contains $d$, and changing the boundary conditions causes $\xi$ to vary with the boundary distance $d$. This is the origin of boundary sensitivity in SFL. The non-Hermitian electric skin effect can therefore be further subdivided as shown in Table~\ref{tab:NHESE}; the two subclasses share the electric-type signature $\text{sgn}(\xi^E_{k_+}\xi^E_{k_-}) < 0$ but respond differently to boundary conditions.
\begin{table}[ptb]
\centering
\renewcommand{\arraystretch}{1.3}
\setlength{\tabcolsep}{12pt}
\caption{Detailed classification of NHESE}
\label{tab:NHESE}
\begin{tabular}{l c c c c}
\toprule
Subclass & Spatial structure of $\Phi(x)$ & Property of $\Phi_a$ & Boundary sensitivity & Representative \\
\midrule
Extensive type & $O(L)$ scale structure & Independent of $L$ & Insensitive & RSE \\
Localized type & $O(1)$ sites & $\sim 1/L$ or $1/d$ & Sensitive & SFL \\
\bottomrule
\end{tabular}
\end{table}

\noindent{\textbf{Remark on the \emph{imaginary gauge transformation}:}}

It should be emphasized that our imaginary gauge transformation does not constitute a gauge theory in the standard sense. In standard $U(1)$ gauge theory, no physically preferred gauge choice exists, and all observables must be gauge-invariant; in our framework, however, the basis defined by the original lattice Hamiltonian $H_s$ constitutes a physically preferred \emph{gauge}, in which observables such as the spatial distribution of the wavefunction are uniquely determined. Unlike standard gauge transformations, the imaginary gauge transformation alters the spatial distribution of $|\psi(x)|^2$, which is precisely the fundamental reason it does not constitute a true gauge symmetry. The role of the imaginary gauge transformation is that of an analytical tool rather than a fundamental symmetry principle: it separates the removable vector potential $A_k(x)$ (corresponding to the boundary-sensitive magnetic skin effect) from the irremovable scalar potential $\Phi(x)$ (corresponding to the boundary-insensitive electric skin effect) in the effective Hamiltonian, thereby providing a more intuitive understanding of the physical origin of boundary sensitivity within the GNHSE framework.

\noindent{\textbf{Conditions for real OBC spectra:}}

The conclusion that ``the spectrum becomes real after the imaginary vector potential is eliminated under OBC'' requires explicit conditions. For a one-dimensional tight-binding system exhibiting the non-Hermitian skin effect (NHSE), when the non-Hermitian strength responsible for the NHSE is perturbative, the OBC spectrum remains entirely real provided the following conditions are satisfied:
\begin{enumerate}
\item[(i)] The system can be obtained from the Hermitian limit by continuous adiabatic deformation, i.e., the non-Hermitian parameters responsible for the NHSE (nonreciprocal hopping asymmetry, gain/loss strength, etc.) are continuously turned on from zero;
\item[(ii)] For multi-band models, the system must possess some form of generalized PT symmetry \cite{hu_GeometricOriginNonBloch_2024}.
\end{enumerate}

For the simple one-dimensional Hatano--Nelson model, the global real-space similarity transformation $S = \text{diag}(1, r, r^2, \ldots, r^{L-1})$ with $r = \sqrt{t_R/t_L}$ maps the non-Hermitian Hamiltonian to a hopping-symmetric Hermitian model \cite{yao_EdgeStatesTopological_2018}. The sole requirement for this mapping is $t_R t_L > 0$ \cite{longhi_NonHermitianSkinEffect_2021}, which is automatically guaranteed by condition (i) when continuously deforming from the Hermitian limit ($t_R = t_L =t > 0$). For the multi-band case, the conclusion follows from two well-established theoretical results. First, the NHSE has been shown to be a general mechanism for realizing PT symmetry: even when the system does not possess PT symmetry in the Bloch (periodic boundary) sense, the GBZ deformation induced by the NHSE can restore an effective symmetry under open boundaries, known as non-Bloch PT symmetry \cite{xiao_ObservationNonBlochParityTime_2021,song_NonBlochPTSymmetry_2022}. Real spectra and PT symmetry are therefore possible only under open boundary conditions, making non-Bloch PT symmetry the sole general mechanism for achieving real spectra in NHSE systems \cite{xiao_ObservationNonBlochParityTime_2021,song_NonBlochPTSymmetry_2022}. Second, in one dimension, the spontaneous breaking threshold $\gamma_c$ of non-Bloch PT symmetry is generically finite and nonzero, in contrast to two and higher dimensions where $\gamma_c$ tends to zero with increasing system size \cite{song_NonBlochPTSymmetry_2022}. The geometric origin of non-Bloch PT breaking has been identified as the formation of cusps in the GBZ, and exact criteria exist for determining $\gamma_c$ without computing the spectrum or the GBZ \cite{hu_GeometricOriginNonBloch_2024}. Therefore, when the non-Hermitian strength $\gamma < \gamma_c$, the system resides in the exact non-Bloch PT phase, and the OBC spectrum is entirely real. Moreover, Ref.~\cite{hu_GeometricOriginNonBloch_2024} has rigorously demonstrated that possessing (generalized) PT symmetry is a necessary condition for a robust exact non-Bloch PT phase, i.e., a nonzero-measure region in parameter space where the OBC spectrum is purely real. This is precisely why condition (ii) is indispensable for multi-band models, and it is naturally satisfied by the vast majority of physically relevant NHSE models.

In summary, the conclusion within the GNHSE framework that ``the magnetic skin effect renders the OBC spectrum real'' can be distilled into a unified physical picture: the NHSE is not merely a consequence of non-Hermiticity but simultaneously feeds back to protect the reality of the spectrum. The exponential envelope of the skin states (equivalent to the imaginary gauge factor $e^{\chi_k(x)}$) re-establishes non-Bloch PT symmetry under open boundaries, thereby maintaining a fully real OBC spectrum below the finite threshold $\gamma_c$. This condition is automatically satisfied under the perturbative non-Hermiticity assumption ($\gamma \ll t$) of the GNHSE theory.

\section{Electric–magnetic localization transition}
In this section, using the GNHSE theory, we present a systematic derivation of the electric--magnetic skin effect localization phase transition for three distinct imaginary electric potential configurations, demonstrating that the electric--magnetic localization phase transition is a universal and robust phenomenon, while the properties of the critical state and the transition pathway depend on the specific spatial structure of the imaginary electric potential.

\subsection{Coincident electric and magnetic IDWs}
We proceed through the following steps: (1) extract the imaginary vector and scalar potentials from the microscopic Hamiltonian, determine the positions of the magnetic and electric IDWs, and solve for the wavefunction; (2) analyze the wavefunction structure at all IDWs, from which the order parameter of the phase transition is naturally defined; (3) derive the energy-dependent transition condition and the mobility edge from the condition that the order parameter vanishes.

\subsubsection{Model and imaginary potential fields}
The Hamiltonian for the phase transition model is [main text Eq.~(7)]:
\begin{equation}\label{eq:Ham-EM}
\hat{H}_{\text{E-M}} = \sum_{j=1}^{2N}\left[t_1\hat{c}^\dagger_{j+1}\hat{c}_j + t_2\hat{c}^\dagger_j\hat{c}_{j+1} + i\Phi(j)\hat{c}^\dagger_j\hat{c}_j\right],
\end{equation}
where $\hat{c}_{2N+1} = \hat{c}_1$ (PBC), $t_1 = t + \gamma s_j$, $t_2 = t - \gamma s_j$, $s_j = -\text{sgn}(j - N - 0.5)$, $\Phi(j) = \lambda,\text{sgn}((j-N)(j-2N))$, and $\lambda, \gamma \ll t$. The system consists of $2N$ sites and is naturally divided into two subintervals as shown in Table.~\ref{tab:EM-intervals}. It is straightforward to see that this model is constructed by adding a piecewise-constant imaginary scalar potential to the model corresponding to Eq.~(6) in the main text (two mirror-conjugated Hatano--Nelson chains under open boundary conditions, i.e., $H_{HN}$ joined with $H_{HN}^\dagger$). The detailed analysis is as follows:

From Eq.~(6) in the main text, the imaginary magnetic vector potential for the Hatano--Nelson model can be expressed as $\text{Im}(E(k)) \propto 2\gamma\sin k_\pm=v_{k_\pm}A(x)$. In the perturbative limit $\gamma \ll t$, the nonreciprocal hoppings $t \pm \gamma$ in interval I correspond to a local imaginary magnetic vector potential $A^{(\text{I})} = \gamma/t$, while the reversed hopping direction ($t \mp \gamma$) in interval II corresponds to $A^{(\text{II})} = -\gamma/t$. Since the spatial average over the two intervals vanishes, $A_a = 0$, we define the step discontinuities of the magnetic vector potential $A(j)$ at $j = N$ and $j = 2N$ as \textbf{magnetic IDWs}.

Similarly, $\Phi(j) = +\lambda$ (interval I) and $\Phi(j) = -\lambda$ (interval II), with spatial average $\Phi_a = 0$. The step discontinuities of the electric scalar potential $\Phi(j)$ at $j = N$ and $j = 2N$ are defined as \textbf{electric IDWs}. The magnetic and electric IDWs are both located at $j = N$ and $j = 2N$, coinciding spatially. This is a special configuration of the model; more general cases will be discussed below.

\begin{table}[hptb]
\centering
\renewcommand{\arraystretch}{1.3}
\setlength{\tabcolsep}{12pt}   
\caption{Subinterval parameters}
\label{tab:EM-intervals}
\begin{tabular}{c c c c c}
\toprule
Subinterval & $s_j$ & Right hopping $t_1$ & Left hopping $t_2$ & Scalar potential $\Phi$ \\
\midrule
I: $j \in [1, N]$ & $+1$ & $t + \gamma$ & $t - \gamma$ & $+\lambda$ \\
II: $j \in [N+1, 2N]$ & $-1$ & $t - \gamma$ & $t + \gamma$ & $-\lambda$ \\
\bottomrule
\end{tabular}
\end{table}

For Eq.~\eqref{eq:Ham-EM}, the dispersion is $\text{Re}(E(k)) = 2t\cos k$, the reduced wavevectors $k_\pm$ satisfy $\text{sgn}(v_{k_+} \cdot v_{k_-}) < 0$, and the effective velocities are $v_{k_\pm} = -2t\sin k_\pm$. Taking $k_+\in(\pi,2\pi)$ and $k_-\in(0,\pi)$, and introducing $s \equiv \sin k_- > 0$, we have $v_{k_+}=2ts > 0$ and $v_{k_-} =-2ts< 0$. According to the GNHSE theory, the unified imaginary potential field is $\Phi_{k_\pm}(x) = v_{k_\pm} \cdot A(x) + \Phi(x)$; the vector and scalar potentials for the two subintervals are listed in Table.~\ref{tab:Phi-kpm}. The spatial average is guaranteed to vanish by symmetry:
\begin{equation}
\Phi_a^{\pm} = \frac{1}{2N}\Big[N \cdot \Phi_{k_\pm}^{(\text{I})} + N \cdot \Phi_{k_\pm}^{(\text{II})}\Big] = 0.
\end{equation}
The accumulated potential deviation $\Phi^s_{k_\pm}(x) = \int_0^x [\Phi_{k_\pm}(x') - \Phi_a]dx'$ is a linear function of $x$ within each subinterval (allowing the localization length to be extracted), and the wavefunction probability distribution is given by Eq.~\eqref{eq:unified wavefunction}:
\begin{equation}\label{E-M-psi}
|\psi_{k_\pm}(x)|^2 = \frac{1}{\mathcal{N}}\exp\!\left(\frac{2\Phi^s_{k_\pm}(x)}{v_{k_\pm}}\right).
\end{equation}
For the $k_+$ mode, taking interval II ($N+1 \leq x \leq 2N$) as an example, the exponential factor in the wavefunction decomposes into magnetic and electric contributions:
\begin{equation}
\frac{2\Phi^s_{k_+}(x)}{v_{k_+}} = 2\underbrace{\int_N^x [A(x') - A_a]\,dx'}_{\mathcal{A}^s(x)} + \frac{2}{v_{k_+}}\underbrace{\int_N^x [\Phi(x') - \Phi_a]\,dx'}_{\tilde{\Phi}^s(x)}.
\end{equation}
Since $A_a = 0$ and $\Phi_a = 0$, substituting the values for interval II, $A = -\gamma/t$ and $\Phi = -\lambda$:
\begin{align}
\mathcal{A}^s(x) &=- \frac{\gamma}{t}(x - N), \quad \tilde{\Phi}^s(x) = -\lambda(x - N),\\
\frac{2\Phi^s_{k_+}}{v_{k_+}} &= \frac{-2\gamma}{t}(x - N) + \frac{-2\lambda}{2ts}(x - N) = \left(-\frac{2\gamma}{t} - \frac{\lambda}{ts}\right)(x - N).
\end{align}
The unified inverse localization length is:
\begin{equation}
\frac{1}{\xi^{\text{E-M}}_{k_+}} = -\left(\underbrace{\frac{2\gamma}{t}}_{1/\xi^M} + \underbrace{\frac{\lambda}{ts}}_{1/\xi^E_{k_+}}\right).
\end{equation}
For the $k_-$ mode in interval II:
\begin{equation}
\frac{2\Phi^s_{k_-}}{v_{k_-}} = \left(-\frac{2\gamma}{t} + \frac{\lambda}{ts}\right)(x - N), \; \frac{1}{\xi^{\text{E-M}}_{k_-}} = -\frac{2\gamma}{t} + \frac{\lambda}{ts}.
\end{equation}
Similarly, in interval I, exploiting the symmetry $\Phi^s_{k_\pm}(x) = \Phi^s_{k_\pm}(2N - x)$ (under PBC), the localization analysis proceeds identically, with the result differing only by a sign.

\begin{table}[ptb]
\centering
\renewcommand{\arraystretch}{1.3}
\setlength{\tabcolsep}{12pt}
\caption{Computation of the unified imaginary potential field $\Phi_{k_\pm}$}
\label{tab:Phi-kpm}
\begin{tabular}{l c c c c c}
\toprule
Subinterval & $v_{k_+} A$ & $\Phi$ & $\Phi_{k_+}$ & $v_{k_-} A$ & $\Phi_{k_-}$ \\
\midrule
I: $[1, N]$ & $(2ts)(\gamma/t) = 2\gamma s$ & $+\lambda$ & $2\gamma s + \lambda$ & $(-2ts)(\gamma/t) = -2\gamma s$ & $-2\gamma s + \lambda$ \\
II: $[N+1, 2N]$ & $(2ts)(-\gamma/t) = -2\gamma s$ & $-\lambda$ & $-2\gamma s - \lambda$ & $(-2ts)(-\gamma/t) = 2\gamma s$ & $2\gamma s - \lambda$ \\
\bottomrule
\end{tabular}
\end{table}

\subsubsection{Eigenstate localization analysis and phase transition order parameter}
We now analyze how the localization behavior of the eigenstates varies with parameters. From Eq.~\eqref{eq:unified wavefunction}, since the exponential factor involves an integral over $x$, the wavefunction is continuous at any point $x_0$; let $\psi(x_0) = \psi_0 \neq 0$. In the neighborhood of $x_0$, since $\psi$ is continuous and nonzero, there exists an interval in which $\psi(x)$ does not change sign. Denote this sign as:
\[
\sigma =  \operatorname{sgn}(\psi(x)) ,\quad \sigma = \pm 1.
\]
Then $|\psi(x)| = \sigma \psi(x)$ and $|\psi|'(x) = \sigma \psi'(x)$ (at differentiable points). We define the logarithmic slope of the wavefunction (at nonzero points):
\begin{equation}
\kappa_{k_\pm}(x) \equiv \frac{d}{dx}\ln|\psi(x)| =\frac{|\psi|'(x)}{|\psi(x)|} =\frac{\psi'(x)}{\psi(x)},
\end{equation}
so $\kappa(x) = \psi'(x)/\psi(x)$ holds at nonzero points and is independent of $\sigma$. At $x_0$, the left and right limits exist (derivative jump):
\[
\kappa^-= \lim_{x\to x_0^-} \kappa(x) = \frac{\psi'(x_0^-)}{\psi(x_0)}, ;
\kappa^+= \lim_{x\to x_0^+} \kappa(x) = \frac{\psi'(x_0^+)}{\psi(x_0)}.
\]
Define the jump:
\[
\Delta \kappa(x_0) = \kappa^+- \kappa^- = \frac{\psi'(x_0^+) - \psi'(x_0^-)}{\psi(x_0)}.
\]
We are interested in the extrema of $|\psi(x)|$ at $x_0$. Since $|\psi|$ is continuous, its extrema can be determined from the signs of the first derivative on either side. For a peak (maximum): $|\psi|'(x_0^-) > 0$ and $|\psi|'(x_0^+) < 0$. Using $|\psi|' = \sigma \psi'$, we obtain $\sigma \psi'(x_0^-) > 0$ and $\sigma \psi'(x_0^+) < 0$, which means $\psi'(x_0^-)$ has the same sign as $\sigma$ (i.e., $\psi'(x_0^-) = \sigma a$ with $a > 0$), and $\psi'(x_0^+)$ has the opposite sign to $\sigma$ (i.e., $\psi'(x_0^+) = -\sigma b$ with $b > 0$). We obtain:
\[
\psi'(x_0^+) - \psi'(x_0^-) = -\sigma b - \sigma a = -\sigma (a + b).
\]
Substituting into $\Delta \kappa$:
\[
\Delta \kappa(x_0) = \frac{-\sigma (a + b)}{\psi_0}.
\]
Since $\psi_0 = \sigma |\psi_0|$ (because $\sigma = \operatorname{sgn}(\psi_0)$):
\[
\Delta \kappa(x_0) = \frac{-\sigma (a + b)}{\sigma |\psi_0|} = -\frac{a + b}{|\psi_0|} < 0.
\]
Therefore, at a peak: $\Delta \kappa(x_0) < 0$. Similarly, at a trough (minimum): $\Delta \kappa(x_0) > 0$. Since the derivation explicitly accounts for the relationship between $\psi_0$ and $\sigma$, the criterion is independent of the sign of $\psi_0$.

For Eq.~\eqref{E-M-psi}, we obtain:
\begin{equation}
\kappa_{k_\pm}(x) = \frac{\Phi_{k_\pm}(x) - \Phi_a}{v_{k_\pm}},
\end{equation}
Since $\Phi_a = 0$, we have $\kappa(x) = \Phi_{k_\pm}(x)/v_{k_\pm}$, so $\kappa(x)$ undergoes a jump at the step discontinuities of $\Phi_{k_\pm}$ (the IDWs). Defining $\mu \equiv 2\gamma s$, the values of $\Delta \kappa_{k_\pm}$ at the two IDWs for the $k_+$ mode ($v_{k_+} = -2ts < 0$) and the $k_-$ mode ($v_{k_-} = +2ts > 0$) are listed in Table.\ref{tab:kink-comparison}. The sign of $\Delta\kappa$ determines whether the IDW corresponds to a wavefunction \textbf{peak} ($\Delta\kappa < 0$: slope changes from positive to negative) or \textbf{trough} ($\Delta\kappa > 0$: slope changes from negative to positive).

\begin{table}[htbp]
\centering
\renewcommand{\arraystretch}{1.3}
\setlength{\tabcolsep}{12pt}   
\caption{$\Delta \kappa_{k_\pm}$ at the IDWs for the $k_\pm$ modes}
\label{tab:kink-comparison}
\begin{tabular}{l c c c c}
\toprule
Mode & IDW & $\kappa(x^-)$ & $\kappa(x^+)$ & $\Delta\kappa$ \\
\midrule
\multirow{2}{*}{$k_+$} & $x = N$($I \rightarrow II$) & $\dfrac{\mu+\lambda}{2ts}$ & $-\dfrac{\mu+\lambda}{2ts}$ & $-\dfrac{\mu + \lambda}{ts}$ \\
\addlinespace[0.25em]
& $x = 2N$($II \rightarrow I$) & $-\dfrac{\mu+\lambda}{2ts}$ & $\dfrac{\mu+\lambda}{2ts}$ & $\dfrac{\mu + \lambda}{ts}$ \\
\addlinespace[0.5em]  
\multirow{2}{*}{$k_-$} & $x = N$($I \rightarrow II$) & $\dfrac{\mu - \lambda}{2ts}$ & $\dfrac{\lambda-\mu }{2ts}$ & $\dfrac{\lambda-\mu}{ts}$ \\
\addlinespace[0.25em]
& $x = 2N$($II \rightarrow I$) & $\dfrac{\lambda-\mu }{2ts}$ & $\dfrac{\mu - \lambda}{2ts}$ & $\dfrac{\mu -\lambda }{ts}$ \\
\bottomrule
\end{tabular}
\end{table}

For the $k_+$ mode, since $\mu + \lambda > 0$ always holds, $\Delta\kappa > 0$ at $x = 2N$ (trough) and $\Delta\kappa < 0$ at $x = N$ (\textbf{peak}, localization center). Therefore, the $k_+$ mode always attains its wavefunction maximum at $x = N$, with the peak height and localization length being the superposition of the magnetic and electric skin effect strengths ($\propto\mu+\lambda$). For the $k_-$ mode, the situation differs. At $x = 2N$:
\begin{itemize}
\item When $\lambda > \mu$ (RSE dominant): $\Delta\kappa_{k_-}(2N) = (\mu - \lambda)/(ts) < 0$ (peak)
\item When $\lambda < \mu$ (NSE dominant): $\Delta\kappa_{k_-}(2N) = (\mu - \lambda)/(ts) > 0$ (trough)
\end{itemize}
while $\Delta \kappa_{k_-}(N)$ has the opposite sign to $\Delta\kappa_{k_-}(2N)$. The complete localization picture is summarized in Table.~\ref{tab:localization-summary}.

\begin{table}[htbp]
\centering
\renewcommand{\arraystretch}{1.3}
\setlength{\tabcolsep}{12pt} 
\caption{Summary of the localization picture}
\label{tab:localization-summary}
\begin{tabular}{ccc}
\toprule
Parameter regime & Peak position of $k_+$ mode & Peak position of $k_-$ mode \\
\midrule
$\lambda > \mu$ (RSE dominated) & $N$ & $2N$ \\
$\lambda = \mu$ (Transition point) & $N$ & Plane wave \\
$\lambda < \mu$ (NSE dominated) & $N$ & $N$ \\
\bottomrule
\end{tabular}
\end{table}

The peak position of the $k_-$ mode jumps from $x = 2N$ to $x = N$ as it crosses the transition point, which naturally defines the order parameter: the jump amplitude $\Delta\kappa_{k_-}(2N)$ of the $k_-$ mode at $x = 2N$ encodes the phase transition information:
\begin{equation}
\eta \equiv \Delta\kappa_{k_-}(2N) =\frac{2\gamma s - \lambda}{ts}=\frac{1}{\xi^{\text{E-M}}_{k_-}} = \frac{1}{\xi^M} - \frac{1}{\xi^E_{k_-}},
\end{equation}
where $1/\xi^M = 2\gamma/t$ is the magnetic inverse localization length and $1/\xi^E_{k_-} = \lambda/(ts)$ is the electric inverse localization length of the $k_-$ mode. The physical meaning of the order parameter $\eta$ is:
\begin{itemize}
\item $\eta > 0$ ($2\gamma s > \lambda$, NSE dominant): $x = 2N$ is a trough, $k_-$ localizes at $x = N$, on the same side as $k_+$ (magnetic-type behavior)
\item $\eta = 0$ ($\lambda = 2\gamma s$): $\kappa_{k_-}(x)=0$ holds identically, $\Delta\kappa(2N)=\Delta\kappa(N)=0$, and $k_-$ is a \textbf{plane wave} within interval I
\item $\eta < 0$ ($2\gamma s < \lambda$, RSE dominant): $x = 2N$ is a peak, $k_-$ localizes at $x = 2N$, on the opposite side from $k_+$ (electric-type behavior)
\end{itemize}

\subsubsection{Mode-selective delocalization and mobility edge}
From the analysis of the order parameter $\eta$, the phase transition occurs when the order parameter of the $k_-$ mode vanishes:
\begin{equation}\label{eq:transition-condition}
\eta= \frac{2\gamma}{t}-\frac{\lambda}{ts} = 0 \quad\Longrightarrow\quad \lambda = 2\gamma s = 2\gamma\sin k_-.
\end{equation}
Since different eigenmodes correspond to different $k_-$ (and hence different $s = \sin k_-$), the transition condition is \textbf{mode-dependent}: for the specific mode satisfying $\sin k_- = \lambda/(2\gamma)$, the corresponding $k_-$ mode is precisely at the transition point, while other modes remain in either the RSE or NSE phase. Using $\text{Re}(E) = 2t\cos k_-$ and $\sin k_- = \sqrt{1 - (\text{Re}(E)/(2t))^2}$, the transition condition $s = \lambda/(2\gamma)$ yields the critical energy:
\begin{equation}\label{eq:Ec}
E_c = \pm\,2t\sqrt{1 - \left(\frac{\lambda}{2\gamma}\right)^2}.
\end{equation}
This expression defines a \textbf{mobility edge} in the energy spectrum, separating the $k_-$ modes into two phases. The three parameter regimes are as follows:
\begin{enumerate}
    \item \textbf{$\lambda > 2\gamma$ (RSE fully dominant).} In this case $\lambda/(2\gamma) > 1$, so $E_c^2 < 0$ and no real solution exists. For all modes, $\eta < 0$ (since $s\leq 1 < \lambda/(2\gamma)$), and all $k_-$ modes reside in the RSE phase, localizing at $x = 2N$ (opposite side from $k_+$). No mobility edge exists.
\item \textbf{$\lambda = 2\gamma$ (critical point).} $E_c = 0$, and the mobility edge passes through the band center. Only the $s = 1$ ($\text{Re}(E) = 0$) mode is precisely at the transition point (plane wave); all other modes are in the electric skin phase.
\item \textbf{$\lambda < 2\gamma$ (mobility edge exists).} $E_c$ is real and divides the $k_-$ modes in the spectrum into two groups as shown in Table.~\ref{tab:kminus-phases}: an electric-dominant region and a magnetic-dominant region, serving as a mobility edge.
    \begin{table}[htbp]
        \centering
        \renewcommand{\arraystretch}{1.3}
        \setlength{\tabcolsep}{12pt} 
        \caption{Localization behavior of $k_-$ modes in different energy ranges}
        \label{tab:kminus-phases}
        \begin{tabular}{c c c c}
\toprule
            Energy range & Parameter relation & Order parameter & Localization behavior \\
\midrule
            $|\text{Re}(E)| > |E_c|$ & $s < \lambda/(2\gamma)$ & $\eta < 0$ & Localized at $x = 2N$ (Electric-type) \\
            $|\text{Re}(E)| = |E_c|$ & $s = \lambda/(2\gamma)$ & $\eta= 0$ & \textbf{Plane wave} \\
            $|\text{Re}(E)| < |E_c|$ & $s > \lambda/(2\gamma)$ & $\eta > 0$ & Localized at $x = N$ (Magnetic-type) \\
\bottomrule
        \end{tabular}
    \end{table}
\end{enumerate}
Unlike $k_-$, taking $j\in[N+1,2N]$ as an example, the $k_+$ mode satisfies $1/\xi^{\text{E-M}}_{k_+} = -(2\gamma/t + \lambda/(ts)) < 0$ throughout the entire parameter space (magnetic and electric contributions add constructively), remaining localized at $x = N$ with no phase transition and no mobility edge. The mobility edge is therefore \textbf{mode-selective}, affecting only the $k_-$ mode, as shown in Fig.~5(e) of the main text.

The above mobility edge and the Anderson localization mobility edge both represent critical energies separating different localization behaviors in the spectrum. Despite their formal similarity, the two have fundamentally different physical origins and properties, as summarized in Table.~\ref{tab:Ec-comparison}. The last point is particularly distinctive: for coincident electric--magnetic IDWs, the mobility edge exhibits \textbf{chiral selectivity} --- for the two degenerate modes at the same energy $E$, one undergoes the phase transition while the other remains unaffected. This behavior is a unique product of the electric--magnetic competition mechanism. Furthermore, the critical state at this mobility edge is a plane wave rather than a fractal wavefunction, implying the absence of critical exponents or scaling behavior. From the perspective of the order parameter $\eta$, the transition is \textbf{continuous} ($\eta$ passes through zero), yet the localization behavior changes qualitatively on either side of the transition point (the peak of $k_-$ jumps from one IDW to the other).
\begin{table}[htbp]
\centering
\renewcommand{\arraystretch}{1.3}
\caption{Comparison between the Anderson mobility edge and the GNHSE mobility edge}
\label{tab:Ec-comparison}
\begin{tabularx}{\textwidth}{X X X} 
\toprule
Feature & Anderson mobility edge & GNHSE mobility edge \\
\midrule
Physical origin & Disorder potential & Magnetic $A(x)$ vs. Electric $\Phi(x)$\\ 
Separated Phases & Localized vs. Extended states & Electric-type ($k_\pm$ opp.) vs. Magnetic-type ($k_\pm$ same)\\
Critical state & Fractal (Multifractal scaling) & \textbf{Plane waves}\\
Dimensionality & $d \geq 3$ (Absent in 1D Hermitian) & \textbf{Possible in 1D} \\
Mode selectivity & All states participate & \textbf{$k_-$ only} ($k_+$ unaffected) \\
\bottomrule
\end{tabularx}
\end{table}

\subsection{Non-coincident electric and magnetic IDWs}
In this section, we keep the magnetic structure (nonreciprocal hoppings) and the spatial envelope of the imaginary electric potential $\Phi(j)$ unchanged, but the IDWs of the two are no longer coincident; instead, they are relatively rotated by an angle (under PBC), i.e., the imaginary scalar potential is rigidly shifted by $\delta$ lattice sites ($0 < \delta < N$):
\begin{equation}\label{eq:Phi}
\Phi(j) = \begin{cases} 
-\lambda, & j \in [1,\,\delta] \\ 
+\lambda, & j \in [\delta+1,\,N+\delta] \\ 
-\lambda, & j \in [N+\delta+1,\,2N] 
\end{cases}
\end{equation}
Here, the regions $[1,\delta]$ and $[N+\delta+1,2N]$ combined form the $-\lambda$ domain (totaling $N$ sites), whereas $[\delta+1, N+\delta]$ defines the $+\lambda$ domain (consisting of $N$ sites).

\subsubsection{Model and imaginary potential fields}
We adopt the notation from the previous section: $s \equiv \sin k_- > 0$, $\mu \equiv 2\gamma s$, $v_{k_+} = 2ts > 0$, $v_{k_-} = -2ts < 0$. The system now possesses four distinct subintervals as shown in Table.~\ref{tab:merged_intervals}, yielding:
\begin{equation}\label{eq:Phi_avg}
\Phi_a^{\pm} = \frac{1}{2N}\Big[\delta \cdot \Phi_{k_\pm}^{(1)} + (N-\delta) \cdot \Phi_{k_\pm}^{(2)} + \delta \cdot \Phi_{k_\pm}^{(3)} + (N-\delta) \cdot \Phi_{k_\pm}^{(4)}\Big] = 0
\end{equation}
It is straightforward to see that the system has four IDWs: the magnetic IDWs ($x = N$ and $x = 2N$) remain at their original positions, while the electric IDWs ($x = N+\delta$ and $x = \delta$) are shifted by $\delta$ relative to the magnetic ones. When $\delta = 0$, subintervals 1 and 3 become empty sets, recovering the case of fully coincident electric and magnetic IDWs.
\begin{table}[htbp]
\centering
\renewcommand{\arraystretch}{1.3}
\setlength{\tabcolsep}{8pt}
\caption{Subinterval partition and imaginary potential field values for each mode}
\label{tab:merged_intervals}
\begin{tabular}{cccccccc}
\toprule
Subinterval & Site range &  Magnetic potential $A$ & Electric potential $\Phi$ & $v_{k_+}A$ & $\Phi_{k_+}$ & $v_{k_-}A$ & $\Phi_{k_-}$ \\ 
\midrule
1 & $[1,\delta]$ & $+\gamma/t$ & $-\lambda$ & $+\mu$ & $\mu - \lambda$ & $-\mu$ & $-\mu - \lambda$ \\
2 & $[\delta+1,N]$ & $+\gamma/t$ & $+\lambda$ & $+\mu$ & $\mu + \lambda$ & $-\mu$ & $-\mu + \lambda$ \\
3 & $[N+1,N+\delta]$ & $-\gamma/t$ & $+\lambda$ & $-\mu$ & $-\mu + \lambda$ & $+\mu$ & $\mu + \lambda$ \\
4 & $[N+\delta+1,2N]$ & $-\gamma/t$ & $-\lambda$ & $-\mu$ & $-\mu - \lambda$ & $+\mu$ & $\mu - \lambda$ \\
\bottomrule
\end{tabular}
\end{table}

\subsubsection{Eigenstate localization analysis and phase transition order parameter}
Since $\Phi_{k_\pm}(x)$ is constant on each subinterval, the accumulated potential deviation $\Phi^s_{k_\pm}(x) = \int_0^x \Phi_{k_\pm}(x'),dx'$ is a piecewise linear function, with the slope of each segment equal to the corresponding subinterval value of $\Phi_{k_\pm}$. For the $k_+$ mode (where $v_{k_+} = 2ts > 0$), we obtain:
\begin{equation}\label{eq:Phi_s_k+}
\Phi^s_{k_+}(x) = \begin{cases}
(\mu - \lambda)\,x, & 0 \leq x \leq \delta, \\[3pt]
(\mu + \lambda)\,x - 2\lambda\delta, & \delta \leq x \leq N, \\[3pt]
(\mu + \lambda)N - 2\lambda\delta + (\lambda - \mu)(x - N), & N \leq x \leq N+\delta, \\[3pt]
(\mu + \lambda)(2N - x), & N+\delta \leq x \leq 2N.
\end{cases}
\end{equation}
The values at the nodes are: $\Phi^s(0) = 0$, $\Phi^s(\delta) = (\mu - \lambda)\delta$, $\Phi^s(N) = (\mu + \lambda)N - 2\lambda\delta$, $\Phi^s(N + \delta) = (\mu + \lambda)(N - \delta)$, $\Phi^s(2N) = 0$. For the $k_-$ mode (where $v_{k_-} = -2ts < 0$), we have:
\begin{equation}\label{eq:Phi_s_k-}
\Phi^s_{k_-}(x) = \begin{cases}
-(\mu + \lambda)\,x, & 0 \leq x \leq \delta, \\[3pt]
(\lambda - \mu)\,x - 2\lambda\delta, & \delta \leq x \leq N, \\[3pt]
(\lambda - \mu)N - 2\lambda\delta + (\mu + \lambda)(x - N), & N \leq x \leq N+\delta, \\[3pt]
(\lambda - \mu)(2N - x), & N+\delta \leq x \leq 2N.
\end{cases}
\end{equation}
The values at the nodes are: $\Phi^s(0) = 0$, $\Phi^s(\delta) = -(\mu + \lambda)\delta$, $\Phi^s(N) = (\lambda - \mu)N - 2\lambda\delta$, $\Phi^s(N + \delta) = (\lambda - \mu)(N - \delta)$, $\Phi^s(2N) = 0$.

Since $\Phi_a^\pm = 0$, the logarithmic slope $\kappa_{k_\pm}(x) = \Phi_{k_\pm}(x)/v_{k_\pm}$ is constant within each subinterval, so the values of $\Delta\kappa$ and the peak/trough assignments at the four IDWs are listed in Table~\ref{tab:kink-shifted}. The magnetic IDWs contribute $\Delta\kappa$ of the same sign for both $k_+$ and $k_-$ (both peaks or both troughs), whereas the electric IDWs contribute $\Delta\kappa$ of opposite signs for the two modes (one mode's peak is the other's trough). This is a direct real-space manifestation of the fundamental distinction between NHMSE and NHESE.
\begin{table}[htbp]
\centering
\renewcommand{\arraystretch}{1.3}
\setlength{\tabcolsep}{12pt}
\caption{$\Delta\kappa_{k_\pm}$ at each IDW for non-coincident electric--magnetic IDWs ($\delta \neq 0$)}
\label{tab:kink-shifted}
\begin{tabular}{cccccccc}
\toprule
Mode & IDW & Type & $\kappa(x^-)$ & $\kappa(x^+)$ & $\Delta\kappa$ & Peak/Trough\\
\midrule
\multirow{4}{*}{$k_+$}
 & $x = \delta$\;($1\!\to\!2$) & Electric & $\dfrac{\mu - \lambda}{2ts}$ & $\dfrac{\mu + \lambda}{2ts}$ & $+\dfrac{\lambda}{ts}$ & Trough \\[4pt]
 & $x = N$\;($2\!\to\!3$) & Magnetic & $\dfrac{\mu + \lambda}{2ts}$ & $\dfrac{\lambda - \mu}{2ts}$ & $-\dfrac{\mu}{ts}$ & \textbf{Peak} \\[4pt]
 & $x = N\!+\!\delta$\;($3\!\to\!4$) & Electric & $\dfrac{\lambda - \mu}{2ts}$ & $-\dfrac{\mu + \lambda}{2ts}$ & $-\dfrac{\lambda}{ts}$ & \textbf{Peak} \\[4pt]
 & $x = 2N$\;($4\!\to\!1$) & Magnetic & $-\dfrac{\mu + \lambda}{2ts}$ & $\dfrac{\mu - \lambda}{2ts}$ & $+\dfrac{\mu}{ts}$ & Trough \\[4pt]
\midrule
\multirow{4}{*}{$k_-$}
 & $x = \delta$\;($1\!\to\!2$) & Electric & $\dfrac{\mu + \lambda}{2ts}$ & $\dfrac{\mu - \lambda}{2ts}$ & $-\dfrac{\lambda}{ts}$ & \textbf{Peak} \\[4pt]
 & $x = N$\;($2\!\to\!3$) & Magnetic & $\dfrac{\mu - \lambda}{2ts}$ & $-\dfrac{\mu + \lambda}{2ts}$ & $-\dfrac{\mu}{ts}$ & \textbf{Peak} \\[4pt]
 & $x = N\!+\!\delta$\;($3\!\to\!4$) & Electric & $-\dfrac{\mu + \lambda}{2ts}$ & $\dfrac{\lambda - \mu}{2ts}$ & $+\dfrac{\lambda}{ts}$ & Trough \\[4pt]
 & $x = 2N$\;($4\!\to\!1$) & Magnetic & $\dfrac{\lambda - \mu}{2ts}$ & $\dfrac{\mu + \lambda}{2ts}$ & $+\dfrac{\mu}{ts}$ & Trough \\[4pt]
\bottomrule
\end{tabular}
\end{table}

Compared with the coincident case ($\delta = 0$, where the electric and magnetic IDWs merge into two), the non-coincident configuration produces four independent IDWs, giving \textbf{each mode two competing peaks}. The two peaks of $k_+$ are located at the magnetic IDW ($x = N$) and the electric IDW ($x = N + \delta$), while the two peaks of $k_-$ are at the electric IDW ($x = \delta$) and the magnetic IDW ($x = N$). Which peak is the dominant one (global maximum) is determined by the magnitude of the exponential factor $f_{k_\pm}(x) = 2\Phi^s_{k_\pm}(x)/v_{k_\pm}$ at the two peak positions. For the $k_+$ mode, comparing the exponential factors at $x = N$ and $x = N + \delta$:
\begin{equation}\label{eq:f_k+_diff}
f_{k_+}(N) - f_{k_+}(N + \delta) = \frac{(\mu + \lambda)N - 2\lambda\delta - (\mu + \lambda)(N - \delta)}{ts} = \frac{(\mu - \lambda)\,\delta}{ts}.
\end{equation}
For the $k_-$ mode, comparing the exponential factors at $x = \delta$ and $x = N$:
\begin{equation}\label{eq:f_k-_diff}
f_{k_-}(N) - f_{k_-}(\delta) = \frac{(\mu - \lambda)N + 2\lambda\delta - (\mu + \lambda)\delta}{ts} = \frac{(\mu - \lambda)(N - \delta)}{ts}.
\end{equation}
The dominant peak position for both modes is determined by the relative magnitude of $\mu$ and $\lambda$ (i.e., $2\gamma s$ versus $\lambda$), which is identical to the transition condition in the coincident case. The complete localization picture is summarized in Table~\ref{tab:loc-shifted}, and the corresponding evolution of the wavefunction distribution is shown in Fig.~\ref{fig:D2phase}.

\begin{table}[htbp]
\centering
\renewcommand{\arraystretch}{1.3}
\setlength{\tabcolsep}{12pt}
\caption{Summary of the localization picture for non-coincident electric--magnetic IDWs}
\label{tab:loc-shifted}
\begin{tabular}{cccc}
\toprule
Parameter regime & $k_+$ peak location & $k_-$ peak location & $k_\pm$ on same side? \\
\midrule
$\mu > \lambda$ (NSE dominated) & $N$ (Magnetic-type) & $N$ (Magnetic-type) & Same side \\
$\mu = \lambda$ (Critical) & Plateau $[N,\,N\!+\!\delta]$ & Plateau $[\delta,\,N]$ & Partial overlap \\
$\mu < \lambda$ (RSE dominated) & $N + \delta$ (Electric-type) & $\delta$ (Electric-type) & Opposite sides \\
\bottomrule
\end{tabular}
\end{table}

\begin{figure}[ptb]
 \centering
{\includegraphics[width=1\linewidth]{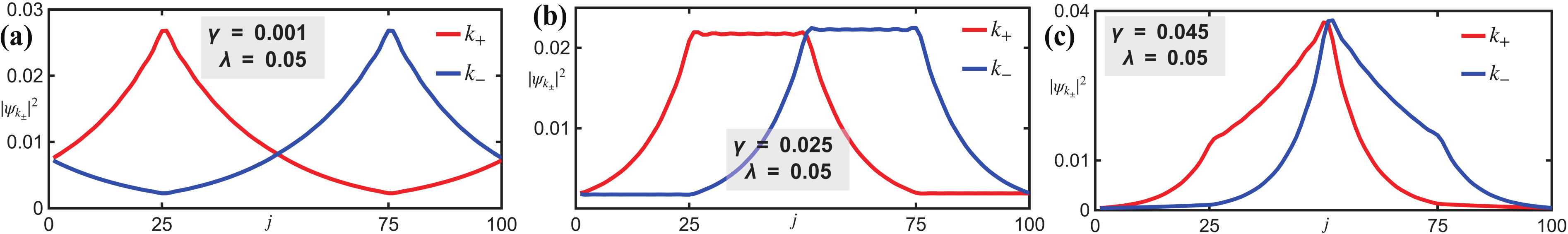}}  %
\caption{Wavefunction probability distributions during the localization phase transition between the relative skin effect (RSE) and the nonreciprocal skin effect (NSE) for $\delta=N/2$ (eigenstates with $s=1$ are shown). (a) Before the transition: since the magnetic skin effect strength is very small, the localization behavior of both $k_\pm$ modes is entirely dominated by the electric skin effect, with localization at the electric IDWs at $N/2$ and $3N/2$. (b) Critical state ($\lambda=2\gamma$): unlike the main text, the critical state here is only partially delocalized --- the $k_+$ and $k_-$ modes are delocalized within the intervals $j\in[N/2, N]$ and $[N, 3N/2]$, respectively, where the electric and magnetic skin effects cancel, forming plateau states. (c) After the transition: the magnetic skin effect dominates, and both $k_\pm$ modes localize on the same side. Notably, unlike the $\delta=0$ case, since the localization positions of both $k_\pm$ modes change, the phase transition does not exhibit chiral selectivity. Parameters: $t=1$, $2N=100$.
}
\label{fig:D2phase}
\end{figure}

Since the dominant peak positions of both the $k_+$ and $k_-$ modes shift simultaneously, for convenience we can still define the \textbf{order parameter} as the difference between the magnetic and electric inverse localization lengths of the $k_-$ mode:
\begin{equation}\label{eq:order}
\eta = \frac{\mu - \lambda}{ts} = \frac{2\gamma s - \lambda}{ts} = \frac{1}{\xi^M} - \frac{1}{\xi^E_{k_-}},
\end{equation}
The transition condition $\eta = 0$ (i.e., $\lambda = 2\gamma s$) and the resulting mobility edge $E_c = \pm,2t\sqrt{1 - (\lambda/2\gamma)^2}$ have expressions identical to those in the coincident case, except that chiral selectivity is lost. The fundamental difference lies in the nature of the critical state: for $\delta = 0$, the exponential factor of $k_-$ vanishes everywhere at the transition point, and the eigenstate is an \textbf{exact plane wave}; for $\delta \neq 0$, the exponential factor equals a nonzero constant $2\lambda\delta/(ts)$ over the interval $[\delta,N]$, and the eigenstate becomes an \textbf{elevated plateau state} (partially delocalized), uniformly raised over $[\delta,N]$ and flat over $[N+\delta,2N]$. The plateau height is continuously tunable as $\propto \delta$: as $\delta \to 0$ the plane wave is recovered, while increasing $\delta$ enhances the localization effect that cannot be completely canceled by the electric--magnetic competition. We now analyze the conditions for \textbf{complete delocalization}.

\subsubsection{Conditions for complete delocalization}
From Eq.~\eqref{eq:unified wavefunction}, the wavefunction probability distribution is
\begin{equation}\label{eq:prob}
|\psi_{k_\pm}(x)|^2 = \frac{1}{\mathcal{N}} \exp\!\left(\frac{2}{v_{k_\pm}} \int_0^x [\Phi_{k_\pm}(x') - \Phi_a^\pm]\,dx'\right).
\end{equation}
If mode $k_\pm$ is \textit{completely delocalized}, then $|\psi_{k_\pm}(x)|^2 = \text{const}$ (i.e., a plane-wave eigenstate). This requires the exponential factor to vanish everywhere, which holds if and only if the unified imaginary potential field equals its spatial average at every point:
\begin{equation}\label{eq:cond}
 \begin{aligned} 
\int_0^x [\Phi_{k_\pm}(x') - \Phi_a^\pm]\,dx' &= 0 \quad \forall\, x \in [0,L],\\
\Phi_{k_\pm}(x) &= \Phi_a^\pm,
  \end{aligned}
\end{equation}
The above condition is automatically satisfied at $x = 0$. Substituting $\Phi_{k_\pm}(x) = v_{k_\pm}\cdot A(x) + \Phi(x)$ and $\Phi_a^\pm = v_{k_\pm} A_a + \Phi_a$ into Eq.~\eqref{eq:cond} yields:
\begin{equation}\label{eq:proportional}
 \begin{aligned} 
v_{k_\pm}[A(x) - A_a] &+ [\Phi(x) - \Phi_a] = 0\\
\tilde{A}(x) &= -\frac{1}{v_{k_\pm}}\,\tilde{\Phi}(x),\quad \forall\, x\in [0,L],
  \end{aligned}
\end{equation}
where $\tilde{A}(x) \equiv A(x) - A_a$ and $\tilde{\Phi}(x) \equiv \Phi(x) - \Phi_a$. That is, the spatial profiles of the magnetic and electric potentials (after subtracting their respective averages) must be \textbf{pointwise proportional}, with proportionality constant $\alpha = -1/v_{k_\pm}$. This leads to three corollaries:

\textbf{(i) Mode selectivity.} Since $\text{sgn}(v_{k_+}\cdot v_{k_-}) < 0$, Eq.~\eqref{eq:proportional} can be satisfied for at most \textbf{one} of $k_+$ and $k_-$. If it holds for $k_-$ ($\alpha = -1/v_{k_-} > 0$), then $\tilde{A}$ and $\tilde{\Phi}$ have the same sign; if it holds for $k_+$ ($\alpha = -1/v_{k_+} < 0$), they have opposite signs. Both cannot be satisfied simultaneously, unless $\tilde{A}(x) \equiv \tilde{\Phi}(x) \equiv 0$ (the trivial case with no skin effect).

\textbf{(ii) Specialization to piecewise-constant potentials.} When both $A(x)$ and $\Phi(x)$ are piecewise constant, the ``pointwise proportionality'' requirement demands that the domain wall positions coincide: all step discontinuities of $A(x)$ and $\Phi(x)$ must be in one-to-one spatial correspondence, and the jump amplitudes at each domain wall must be proportional ($\Delta\tilde{A} = \alpha,\Delta\tilde{\Phi}$), with the same proportionality constant $\alpha = -1/v_{k_\pm}$ at every domain wall. If the domain wall positions do not coincide or the jump amplitudes are not proportional, there necessarily exists some $x$ at which $\tilde{A}(x)$ undergoes a step while $\tilde{\Phi}(x)$ does not (or vice versa), making Eq.~\eqref{eq:proportional} impossible to satisfy and complete delocalization unattainable.

\textbf{(iii) Relation to the transition condition.} Eq.~\eqref{eq:proportional} is a stronger constraint than the transition condition $\eta = 0$. The transition condition only requires the \textbf{slope} of the exponential factor to vanish over a certain interval (determined by the relative magnitude of $\mu$ and $\lambda$), whereas complete delocalization requires the exponential factor to be identically zero over the \textbf{entire space}. Therefore, the phase transition as a codimension-1 phase boundary is universal and robust --- there always exists a reversal of the localization direction of the $k_+$ or $k_-$ mode near the phase boundary --- but this reversal does not necessarily render the wavefunction a plane wave. Complete delocalization is a special property that emerges at the transition point only when Eq.~\eqref{eq:proportional} is satisfied.

\subsubsection{Comparison with the coincident case}
A comparison between the coincident and non-coincident electric--magnetic IDW configurations is presented in Table.~\ref{tab:full-comparison}. For $\delta = 0$ and $\delta = N$, the electric and magnetic IDWs both coincide; the difference lies solely in the \textbf{relative sign} of the electric and magnetic potentials: at $\delta = 0$ they have the same sign, the proportionality constant in Eq.~\eqref{eq:proportional} is $\alpha > 0$, corresponding to $v_{k_-} < 0$ (the $k_-$ mode); at $\delta = N$ they have opposite signs, $\alpha < 0$, corresponding to $v_{k_+} > 0$ (the $k_+$ mode). Therefore, reversing the sign of the electric potential is equivalent to exchanging which chiral mode undergoes the phase transition, while the transition condition itself, the mobility edge, and the form of the order parameter all remain unchanged. For $0 < \delta < N$, the non-coincidence of the domain walls prevents Eq.~\eqref{eq:proportional} from being satisfied: the electric and magnetic steps occur at different positions, the spatial profiles no longer match, and exact cancellation is impossible. The critical state degrades to a plateau state of finite height, but the phase transition as a codimension-1 phase boundary remains robust.

\begin{table}[htbp]
\centering
\renewcommand{\arraystretch}{1.3}
\setlength{\tabcolsep}{10pt}
\caption{Comparison between coincident ($\delta = 0$ and $\delta = N$) and non-coincident ($0 < \delta < N$) electric--magnetic IDWs}
\label{tab:full-comparison}
\begin{tabular}{lccc}
\toprule
Feature & $\delta = 0$ & $\delta = N$ & $0 < \delta < N$ \\
\midrule
Electric--magnetic IDWs coincident? & Yes (same-sign jump) & Yes (opposite-sign jump) & No\\
Transition condition & $\mu = \lambda$ & $\mu = \lambda$ & $\mu = \lambda$\\
Modes undergoing transition & $k_-$ & $k_+$ & Both $k_\pm$ shift peaks \\
Modes not undergoing transition & $k_+$ (always at $N$) & $k_-$ (always at $N$) & — \\
Satisfies Eq.~\eqref{eq:proportional}? & Yes ($\alpha > 0$) & Yes ($\alpha < 0$) & No \\
Critical state nature & Plane wave & Plane wave & Plateau state \\
Complete delocalization & Achievable & Achievable & Not achievable \\
\bottomrule
\end{tabular}
\end{table}

\subsection{Sinusoidal imaginary electric potential}
In this section, we keep the magnetic structure (nonreciprocal hoppings) unchanged but replace the imaginary electric potential with a slowly varying smooth sinusoidal function, while ensuring that the IDWs of the two coincide. We analyze the characteristics of the electric--magnetic localization phase transition in this configuration through both theoretical and numerical methods.

\subsubsection{Model and imaginary potential fields}
Consider the same lattice structure as for the piecewise-constant potential, with the Hamiltonian on $2N$ sites:
\begin{equation}\label{eq:H}
\hat{H}_{\text{E-M}} = \sum_{j=1}^{2N}\left[t_1\hat{c}^\dagger_{j+1}\hat{c}_j + t_2\hat{c}^\dagger_j\hat{c}_{j+1} + i\Phi(j)\hat{c}^\dagger_j\hat{c}_j\right],
\end{equation}
where $\hat{c}_{2N+1} = \hat{c}_1$ (periodic boundary conditions), $t_1 = t + \gamma s_j$, $t_2 = t - \gamma s_j$, $s_j = -\text{sgn}(j - N - 0.5)$. The only difference from the piecewise-constant potential is the imaginary scalar potential:
\begin{equation}\label{eq:Phi}
\Phi(j) = \lambda\sin\!\left(\frac{\pi j}{N}\right).
\end{equation}
where the parameters $\lambda, \gamma \ll t$ ensure that the non-Hermitian terms are perturbative. The system is divided into two subintervals: interval I ($j \in [1, N]$) with nonreciprocal hoppings $t \pm \gamma$ and $\Phi > 0$, and interval II ($j \in [N+1, 2N]$) with reversed nonreciprocal hoppings $t \mp \gamma$ and $\Phi < 0$. The zeros of the scalar potential satisfy $\sin(\pi x/N) = 0$, i.e., $x = N$ and $x = 2N$ ($= 0$). These two points are also the step discontinuities of the vector potential $A(x)$ (jumping from $+\gamma/t$ to $-\gamma/t$). Therefore, \textbf{the electric and magnetic IDWs coincide spatially}, as in the piecewise-constant potential case.

We again adopt the notation from the previous two sections: $\mu \equiv 2\gamma s$ ($s = \sin k_- > 0$), $v_{k_+} = 2ts > 0$, $v_{k_-} = -2ts < 0$. The unified imaginary potential field $\Phi_{k_\pm}(x) = v_{k_\pm} A(x) + \Phi(x)$ in the two intervals is given in Table.~\ref{tab:Phi_kpm}. From the antisymmetry of the sine function $\sin(\pi(x+N)/N) = -\sin(\pi x/N)$ and the sign reversal of the vector potential, one can verify:
\begin{equation}\label{eq:antisym}
\Phi_{k_-}(x + N) = -\Phi_{k_-}(x), \quad \Phi_{k_+}(x + N) = -\Phi_{k_+}(x).
\end{equation}
Integrating over $[0, 2N]$ immediately yields the spatial average $\Phi_a^\pm = 0$, so $\Phi^s_{k_\pm}(x) = \int_0^x \Phi_{k_\pm}(x'),dx'$. Using $\int_0^x \sin(\pi x'/N),dx' = \frac{N}{\pi}[1-\cos(\pi x/N)]$, we define two envelope functions:
\begin{equation}
g_V(x) = \begin{cases} x, & 0 \leq x \leq N, \\ 2N - x, & N \leq x \leq 2N, \end{cases} \qquad g_E(x) = 1 - \cos\frac{\pi x}{N}.
\end{equation}
where $g_V$ is a tent function ($g_V(0)=g_V(2N)=0$, $g_V(N)=N$) and $g_E$ is a smooth cosine envelope ($g_E(0)=g_E(2N)=0$, $g_E(N)=2$). After piecewise integration, the accumulated potential deviations for the two modes can be written in unified form as:
\begin{equation}\label{eq:Phi_s}
\Phi^s_{k_+}(x) = \mu\, g_V(x) + \frac{\lambda N}{\pi}\, g_E(x), \qquad \Phi^s_{k_-}(x) = -\mu\, g_V(x) + \frac{\lambda N}{\pi}\, g_E(x).
\end{equation}
One can directly verify that these expressions yield consistent values at $x = N$ from the two subintervals (continuity), as well as $\Phi^s(0)=\Phi^s(2N)=0$ (periodic boundary self-consistency). Furthermore, from the symmetry of the envelope functions $g_V(2N-x) = g_V(x)$ and $g_E(2N-x) = g_E(x)$, it follows that $\Phi^s_{k_\pm}(2N-x) = \Phi^s_{k_\pm}(x)$.

\begin{table}[htbp]
\centering
\renewcommand{\arraystretch}{1.3}
\setlength{\tabcolsep}{12pt}
\caption{Unified imaginary potential field}
\label{tab:Phi_kpm}
\begin{tabular}{ccccc}
\toprule
Subinterval & $v_{k_\pm}A$ & $\Phi(x)$ & $\Phi_{k_+}$ & $\Phi_{k_-}$ \\
\midrule
I: $[1, N]$ & $\pm\mu$ & $\lambda\sin(\pi x/N)\;(\geq 0)$ & $\mu + \lambda\sin\frac{\pi x}{N}$ & $-\mu + \lambda\sin\frac{\pi x}{N}$ \\
II: $[N\!+\!1, 2N]$ & $\mp\mu$ & $\lambda\sin(\pi x/N)\;(\leq 0)$ & $-\mu + \lambda\sin\frac{\pi x}{N}$ & $\mu + \lambda\sin\frac{\pi x}{N}$ \\
\bottomrule
\end{tabular}
\end{table}

\subsubsection{Eigenstate localization analysis}
The wavefunction probability distribution is given by $|\psi_{k_\pm}(x)|^2 \propto e^{f_{k_\pm}(x)}$, where the exponential factor is $f_{k_\pm}(x) = 2\Phi^s_{k_\pm}(x)/v_{k_\pm}$. For the $k_+$ mode, substituting $v_{k_+} = 2ts$:
\begin{equation}\label{eq:f_k+}
f_{k_+}(x) = \frac{\mu}{ts}\,g_V(x) + \frac{\lambda N}{\pi ts}\,g_E(x).
\end{equation}
Both contributions are non-negative and attain their maxima at $x = N$. Therefore, $f_{k_+}$ has a unique global maximum at $x = N$ over the entire interval $[0, 2N]$, with the magnetic and electric skin effects reinforcing each other. The $k_+$ mode is always localized at $x = N$, and no phase transition occurs.

Similarly, for the $k_-$ mode, substituting $v_{k_-} = -2ts$:
\begin{equation}\label{eq:f_k-}
f_{k_-}(x) = \frac{\mu}{ts}\,g_V(x) - \frac{\lambda N}{\pi ts}\,g_E(x)
\end{equation}
This expression is the difference of two envelope functions, i.e., the magnetic contribution (tent function) minus the electric contribution (cosine envelope). Both attain their maxima at $x = N$, but with different shapes. From the symmetry of the envelope functions $g_V(2N-x) = g_V(x)$ and $g_E(2N-x) = g_E(x)$, the exponential factor is symmetric about $x = N$:
\begin{equation}
f_{k_-}(2N - x) = f_{k_-}(x),
\end{equation}
Therefore, \textbf{it suffices to analyze the behavior on $[0, N]$}; interval II is completely determined by symmetry. Introducing the dimensionless parameter $\alpha \equiv \mu/\lambda = 2\gamma s/\lambda$ (the ratio of magnetic to electric strengths), the derivative of $f_{k_-}(x)$ on interval I is:
\begin{equation}
f'_{k_-}(x)\big|_{\rm I} = \frac{\mu}{ts} - \frac{\lambda}{ts}\sin\frac{\pi x}{N} = \frac{\lambda}{ts}\!\left(\alpha - \sin\frac{\pi x}{N}\right).
\end{equation}
Setting $f' = 0$ gives $\sin(\pi x/N) = \alpha$. Since $\sin(\pi x/N)$ has range $[0, 1]$ on $[0, N]$, three cases arise:

\textbf{Case A:} $\alpha > 1$ ($\lambda < 2\gamma s$, magnetic dominant). The equation has no solution, $f' > 0$ holds throughout, and $f$ is strictly monotonically increasing on $[0, N]$. Combined with symmetry, $f_{k_-}$ is unimodal on $[0, 2N]$ with a global maximum at $x = N$.

\textbf{Case B:} $\alpha = 1$ ($\lambda = 2\gamma s$). The unique solution is $x = N/2$. The second derivative $f'' = -\frac{\lambda\pi}{Nts}\cos(\pi/2) = 0$, and the third derivative $f''' = \frac{\lambda\pi^2}{N^2 ts}\sin(\pi/2) > 0$, so $x = N/2$ is an \textbf{inflection point} (not an extremum). $f' \geq 0$ still holds throughout $[0, N]$ (vanishing only at a single point), and $f$ remains unimodal.

\textbf{Case C:} $0 < \alpha < 1$ ($\lambda > 2\gamma s$, competition emerges). The equation has two solutions in $[0, N]$. Let $\theta_0 = \arcsin\alpha \in (0, \pi/2)$; then:
\begin{equation}
x_1 = \frac{N\theta_0}{\pi}, \qquad x_2 = N - x_1 = \frac{N(\pi - \theta_0)}{\pi},
\end{equation}
satisfying $0 < x_1 < N/2 < x_2 < N$. Using $f''(x) = -\frac{\lambda\pi}{Nts}\cos(\pi x/N)$, the nature of the extrema can be determined: at $x_1$, $\cos(\pi x_1/N) = \cos\theta_0 = \sqrt{1-\alpha^2} > 0$, so $f''(x_1) < 0$ and $x_1$ is a \textbf{local maximum}; at $x_2$, $\cos(\pi x_2/N) = \cos(\pi - \theta_0) = -\sqrt{1-\alpha^2} < 0$, so $f''(x_2) > 0$ and $x_2$ is a \textbf{local minimum}. The sign distribution of $f'$ is: $f' > 0$ on $(0, x_1)$ (increasing), $f' < 0$ on $(x_1, x_2)$ (decreasing), $f' > 0$ on $(x_2, N)$ (increasing).

From the antisymmetry relation $\Phi_{k_-}(x+N) = -\Phi_{k_-}(x)$, we obtain $f'_{k-}(x+N) = -f'_{k-}(x)$. Therefore, the solutions of $f' = 0$ in interval II are $x_3 = N + x_1$ and $x_4 = 2N - x_1$, with the sign distribution strictly reversed from interval I. From the symmetry $f(2N-x) = f(x)$: $f(N+x_1) = f(N-x_1) = f(x_2)$ (local minimum), and $f(2N-x_1) = f(x_1)$ (local maximum).

When $\alpha < 1$, the extrema of $f_{k_-}$ on $[0, 2N]$ are summarized in Table.~\ref{tab:extrema}. The complete behavior of the wavefunction on $[0, 2N]$ can be schematically represented as:
\[
0 \xrightarrow{\nearrow} f(x_1) \xrightarrow{\searrow} f(x_2) \xrightarrow{\nearrow} f(N) \xrightarrow{\searrow} f(x_3) \xrightarrow{\nearrow} f(x_4) \xrightarrow{\searrow} 0
\]
Therefore, only three independent extremal values need to be computed: $f(x_1)$, $f(x_2)$, and $f(N)$, and the assignment of the global maximum depends on the relative magnitudes of $f(N)$ and $f(x_1)$.

\begin{table}[htbp]
\centering
\renewcommand{\arraystretch}{1.3}
\setlength{\tabcolsep}{12pt}
\caption{Extrema of $f_{k_-}$}
\label{tab:extrema}
\begin{tabular}{cccc}
\toprule
 Position & Type & $f$ value & Origin\\
\midrule
$0,\, 2N$ & Boundary minimum & $0$ & $f(0) = f(2N) = 0$ \\
$x_1$ & Smooth local maximum & $f(x_1)$ & $f' = 0$, $f'' < 0$ \\
$x_2=N-x_1 $ & Smooth local minimum & $f(x_2)$ & $f' = 0$, $f'' > 0$ \\
$N$ & Kink local maximum & $f(N)$ & $f'$ jumps from positive to negative \\
$x_3=N+x_1$ & Smooth local minimum & $f(x_2)$ & $f' = 0$, $f'' > 0$ \\
$x_4=2N-x_1$ & Smooth local maximum & $f(x_1)$ & $f' = 0$, $f'' < 0$ \\
\bottomrule
\end{tabular}
\end{table}

\noindent{\textbf{Explicit computation of key extremal values and the switching condition for the global maximum}}\\
Using $f_{k_-}(x)\big|_{\rm I} = \frac{\mu x}{ts} - \frac{\lambda N}{\pi ts}(1 - \cos\frac{\pi x}{N})$, together with $\theta_0 = \arcsin\alpha$ and $\cos\theta_0 = \sqrt{1-\alpha^2}$, we obtain:

\textbf{$f(N)$}: Evaluating at $\cos\pi = -1$,
\begin{equation}
f(N) = \frac{\mu N}{ts} - \frac{2\lambda N}{\pi ts} = \frac{\lambda N}{\pi ts}(\pi\alpha - 2).
\end{equation}

\textbf{$f(x_1)$}: Setting $x_1 = N\theta_0/\pi$, we have $\cos(\pi x_1/N) = \cos\theta_0 = \sqrt{1-\alpha^2}$,
\begin{equation}
f(x_1) = \frac{\mu N\theta_0}{\pi ts} - \frac{\lambda N}{\pi ts}(1 - \sqrt{1-\alpha^2}) = \frac{\lambda N}{\pi ts}\!\left(\alpha\theta_0 - 1 + \sqrt{1-\alpha^2}\right).
\end{equation}

\textbf{$f(x_2)$}: Setting $x_2 = N(\pi-\theta_0)/\pi$, where $\cos(\pi x_2/N) = -\sqrt{1-\alpha^2}$,
\begin{equation}
f(x_2) = \frac{\mu N(\pi-\theta_0)}{\pi ts} - \frac{\lambda N}{\pi ts}(1 + \sqrt{1-\alpha^2}) = \frac{\lambda N}{\pi ts}\!\left[\alpha(\pi-\theta_0) - 1 - \sqrt{1-\alpha^2}\right].
\end{equation}

Adding $f(x_1)$ and $f(x_2)$:
\begin{equation}
f(x_1) + f(x_2) = \frac{\lambda N}{\pi ts}\big[\alpha\theta_0 - 1 + \sqrt{1-\alpha^2} + \alpha(\pi-\theta_0) - 1 - \sqrt{1-\alpha^2}\big] = \frac{\lambda N}{\pi ts}(\pi\alpha - 2) = f(N).
\end{equation}
This yields the identity:
\begin{equation}
f(x_1) + f(x_2) = f(N),
\end{equation}
meaning the height difference between the kink peak $f(N)$ and the smooth peak $f(x_1)$ is exactly equal to the minimum value $f(x_2)$. Therefore, the condition for the global maximum to switch from $x = N$ (magnetic IDW) to $x = x_1$ (internal smooth peak) is $f(N) = f(x_1)$, which is equivalent to $f(x_2) = 0$.

\subsubsection{Analysis of the order parameter behavior}
For convenience in evaluating the condition $f(x_2) = 0$, we define the dimensionless function ($\theta_0 = \arcsin\alpha$):
\begin{equation}
G(\alpha) \equiv \frac{\pi ts}{\lambda N}\,f(x_2) = \alpha(\pi - \arcsin\alpha) - 1 - \sqrt{1-\alpha^2}.
\end{equation}
The critical condition is then $G(\alpha_c) = 0$. Differentiating $G$ and using $d\theta_0/d\alpha = 1/\sqrt{1-\alpha^2}$:
\begin{equation}
G'(\alpha) = (\pi - \arcsin\alpha) + \alpha\!\left(\frac{-1}{\sqrt{1-\alpha^2}}\right) + \frac{\alpha}{\sqrt{1-\alpha^2}} = \pi - \arcsin\alpha > 0, \quad \forall\,\alpha \in (0,1).
\end{equation}
so $G$ is strictly monotonically increasing on $(0,1)$.

As $\alpha \to 0^+$, $\arcsin\alpha \to \alpha$ and $\sqrt{1-\alpha^2} \to 1$, giving $G \to \alpha(\pi - \alpha) - 1 - 1 \to -2 < 0$. At $\alpha = 1$, $G(1) = 1 \cdot (\pi - \pi/2) - 1 - 0 = \pi/2 - 1 \approx 0.571 > 0$. Combined with the strict monotonicity of $G(\alpha)$, the zero $\alpha_c$ \textbf{exists and is unique}. This means the migration of the global maximum must occur at a specific parameter value, which is precisely the hallmark of the sinusoidal electric--magnetic localization phase transition. We now proceed with a more detailed calculation to determine the zero $\alpha_c$.

\begin{figure}[ptb]
 \centering
{\includegraphics[width=1\linewidth]{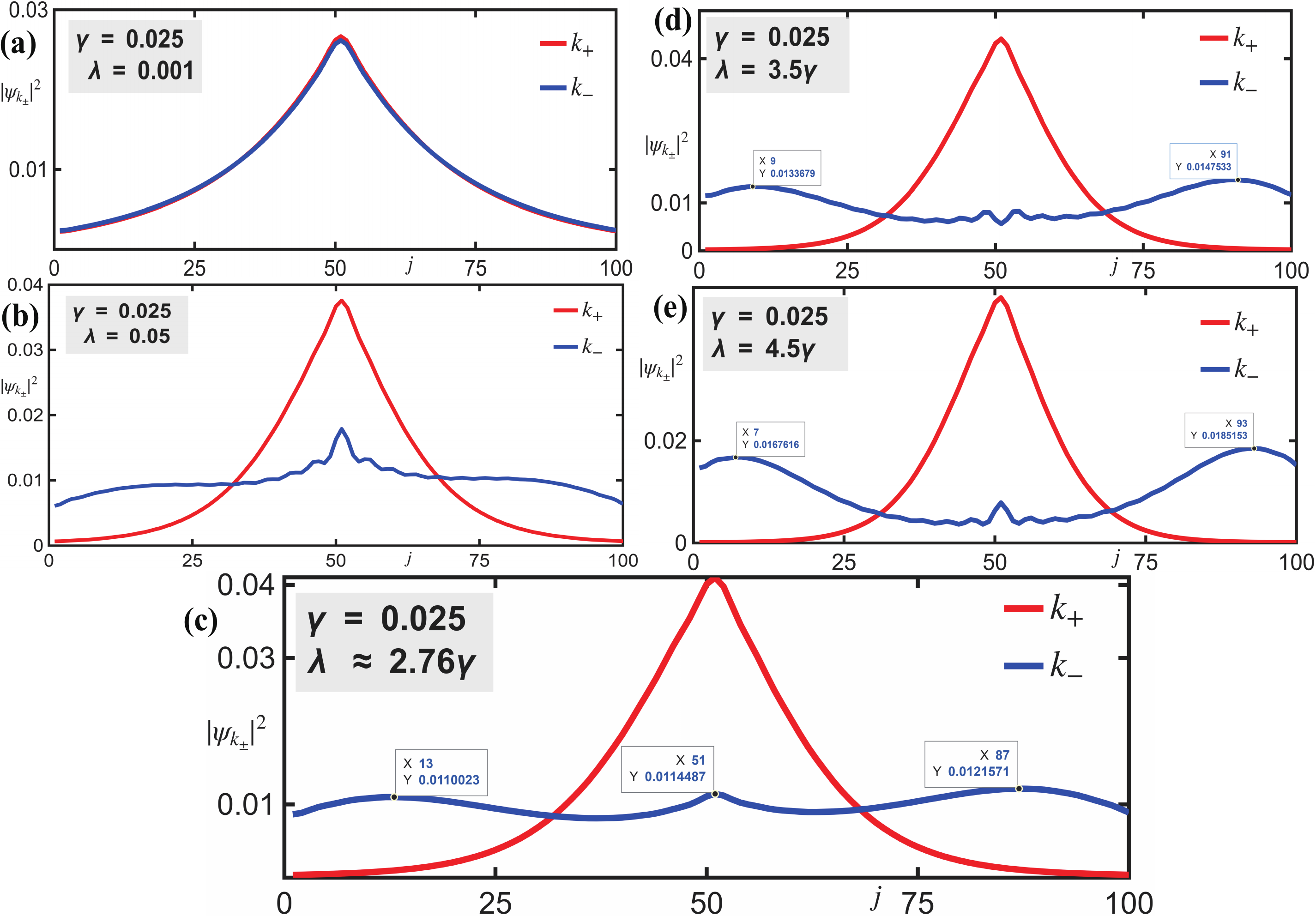}}  
\caption{Wavefunction probability distributions during the localization phase transition between the relative skin effect (RSE) and the nonreciprocal skin effect (NSE) under sinusoidal imaginary electric potential (eigenstates with $s=1$ are shown). (a) Before the transition: since the electric skin effect strength is very small, the localization behavior of both $k_\pm$ modes is entirely dominated by the magnetic skin effect, with localization at the magnetic IDW at $N$. (b) Upper boundary of the coexistence region ($\alpha=\lambda/(2\gamma)=1$): according to the theoretical analysis, $x=N/2$ is an inflection point, and the wavefunction distribution remains monotonically increasing on $[0,N]$, but becomes extremely flat near the inflection points ($x=25,75$). The numerical results for the $k_-$ mode confirm this prediction. (c) Lower boundary of the coexistence region (critical point $\lambda_c \approx 2.76\gamma s$): according to the theoretical analysis, $f(x_1) = f(N) = f(2N-x_1)$ at this point, and the wavefunction distribution exhibits a modulated wave pattern with three peaks of equal height. The numerical results for the $k_-$ mode likewise confirm this within the allowed error margin. (d)-(e) Electric phase ($\alpha < \alpha_c$): the global maximum clearly migrates to $x_1$ and $2N - x_1$, and the peak position $x_1$ drifts smoothly toward $j = 0$ and $2N$ as $\lambda$ continues to increase, i.e., continuous drift of the peak position. This is markedly distinct from the piecewise-constant imaginary electric potential, where the peak position switches abruptly between fixed locations. Parameters: $t=1$, $N=50$.
}
\label{fig:sin-phase}
\end{figure}

\noindent{\textbf{Numerical solution of the critical equation}}\\
Let $\alpha = \sin\theta_0$; the condition $G = 0$ becomes:
\begin{equation}
\sin\theta_0(\pi - \theta_0) = 1 + \cos\theta_0.
\end{equation}
Using the half-angle identities $\sin\theta_0 = 2\sin\frac{\theta_0}{2}\cos\frac{\theta_0}{2}$ and $1 + \cos\theta_0 = 2\cos^2\frac{\theta_0}{2}$, dividing both sides by $2\cos\frac{\theta_0}{2} > 0$:
\begin{equation}
\sin\frac{\theta_0}{2}\cdot(\pi - \theta_0) = \cos\frac{\theta_0}{2}.
\end{equation}
Let $\phi = \theta_0/2$ ($\theta_0 = 2\phi$); rearranging gives:
\begin{equation}
(\pi - 2\phi)\,\tan\phi = 1.
\end{equation}
This transcendental equation can be solved by numerical iteration, yielding to ten significant digits: $\phi_c\approx0.4052351416$, $\theta_c = 2\phi_c \approx 0.8104702832$. The final result is:
\begin{equation}
\alpha_c = \sin\theta_c \approx 0.7246113538, \qquad \lambda_c = \frac{2\gamma s}{\alpha_c} \approx 2.760100279\,\gamma s.
\end{equation}
The position of the internal smooth peak at criticality is $x_1^{(c)} = N\theta_c/\pi \approx 0.2579807036,N$, and the trough position is $x_2^{(c)}\approx 0.7420192964,N$.

In the piecewise-constant potential, the jump $\Delta\kappa(2N) = (\mu - \lambda)/(ts)$ at the IDW simultaneously contains both magnetic and electric contributions, naturally serving as the order parameter that distinguishes the two phases. However, for the sinusoidal potential, since $\sin(\pi N/N) = 0$, the jump at the IDW degenerates to $\pm\mu/ts$ (purely magnetic), independent of $\lambda$, and loses its ability to resolve the electric--magnetic competition. A new order parameter is therefore needed. From the theoretical analysis, the essence of the sinusoidal potential phase transition is the \textbf{migration of the global maximum}: the wavefunction redistributes from the magnetic IDW ($x = N$) to the internal smooth peaks ($x_1$ and $2N - x_1$). The natural order parameter for this process is the height difference between the kink peak and the smooth peak, i.e., the previously defined $G(\alpha)$:
\begin{equation}
\eta \equiv G(\alpha) = \alpha\big(\pi - \arcsin\alpha\big) - 1 - \sqrt{1-\alpha^2}, \qquad \alpha = \frac{2\gamma s}{\lambda}.
\end{equation}
The properties of $G(\alpha)$ are known. Combined with the fact that no internal extrema exist for $\alpha > 1$, the complete picture of the four phases is as follows:

\textbf{Magnetic phase} ($\alpha > 1$, i.e., $\lambda < 2\gamma s$): No internal extrema exist, and $f_{k_-}$ is unimodal on $[0, 2N]$. The global maximum of the $k_-$ mode is at $x = N$, on the same side as $k_+$, exhibiting purely magnetic skin behavior.

\textbf{Coexistence region} ($\alpha_c < \alpha < 1$, i.e., $2\gamma s < \lambda < \lambda_c$): The order parameter $\eta > 0$, i.e., $f(N) > f(x_1)$. Internal smooth peaks appear as satellite peaks at $x_1$ and $2N - x_1$, but the dominant peak remains at $x = N$. The magnetic effect still governs the global localization position. As $\lambda$ increases ($\alpha$ decreases), the satellite peaks gradually rise.

\textbf{Critical point} ($\alpha = \alpha_c$, i.e., $\lambda = \lambda_c$): The order parameter $\eta = 0$, and the three maxima are of equal height: $f(N) = f(x_1) = f(2N-x_1)$. The wavefunction exhibits a modulated wave pattern with three peaks of equal height, which is \textbf{not a plane wave} ($f$ retains residual spatial modulation).

\textbf{Electric phase} ($\alpha < \alpha_c$, i.e., $\lambda > \lambda_c$): The order parameter $\eta < 0$, i.e., $f(x_1) > f(N)$. The global maximum migrates to $x_1$ and $2N - x_1$. The peak position $x_1(\alpha) = \frac{N}{\pi}\arcsin\alpha$ drifts smoothly toward $x = 0$ and $x = 2N$ (the electric IDWs) as $\lambda$ continues to increase. In the limit $\alpha \to 0$ ($\lambda \gg \gamma$), $x_1 \to 0$, and $k_-$ becomes fully localized at the other IDW, on the opposite side from $k_+$, exhibiting purely electric behavior.

Compared with the two piecewise-constant imaginary electric potential configurations discussed previously, the most distinctive feature of the sinusoidal potential phase transition is the existence of a coexistence region, where the localization peak transitions via continuous drift. A detailed comparison is presented in Table.~\ref{tab:comparison}, and the corresponding evolution of the wavefunction distribution is shown in Fig.~\ref{fig:sin-phase}.

From a physical perspective, the origin of this delayed switching is as follows: for the piecewise-constant imaginary electric potential, the magnetic tent $g_V$ and the electric envelope have the same shape, so their pointwise cancellation is achieved simultaneously at $\lambda = 2\gamma s$, and the instant the smooth peak emerges from zero coincides with the peak switching. For the sinusoidal potential, the electric envelope $g_E$ is more concentrated near $x = N$ than the tent $g_V$ (different second derivatives), so even when the smooth peak has already nucleated at $\lambda = 2\gamma s$, the magnetic contribution at the kink peak still dominates. Only when $\lambda$ increases further to $\lambda_c \approx 2.76\gamma s$ does the cumulative advantage of the electric envelope far from $x = N$ become sufficient to overtake the kink peak, completing the switch of the global maximum. This \textbf{delay mechanism} is a direct consequence of the shape mismatch between the magnetic tent envelope and the electric smooth envelope.

\begin{table}[htbp]
\centering
\renewcommand{\arraystretch}{1.3}
\caption{Comparison between the sinusoidal and piecewise-constant potentials}
\label{tab:comparison}
\begin{tabularx}{\textwidth}{X X X} 
\toprule
\textbf{Feature} & 	\textbf{Piecewise-constant} $\Phi = \pm\lambda$ & \textbf{Sinusoidal} $\Phi = \lambda\sin(\pi x/N)$ \\
\midrule
Electric envelope shape & Tent function & $g_E(x) = 1-\cos(\pi x/N)$ (smooth)\\
Existence of exact delocalization & Yes ($f \equiv 0$, strict plane wave)	& No ($g_V \neq c,g_E$, pointwise cancellation impossible)\\
Coexistence region	& Absent (single-step transition) &	$\alpha_c < \alpha < 1$ (three-peak coexistence, satellite peaks gradually rise)\\
Peak switching point &	$\lambda_0 = 2\gamma s$ & $\lambda_c \approx 2.76\gamma s $ (delayed switching)\\
Critical state & Delocalized state & 3-peak equal-height wave \\
Peak migration mechanism & Abrupt jump ($N \leftrightarrow 2N$) & Continuous drift ($x_1(\alpha)$)\\
Order parameter & $\Delta\kappa(2N)$ (IDW jump) & $G(\alpha)$ (peak height difference function)\\
\bottomrule
\end{tabularx}
\end{table}

\begin{figure}[ptb]
 \centering
{\includegraphics[width=1\linewidth]{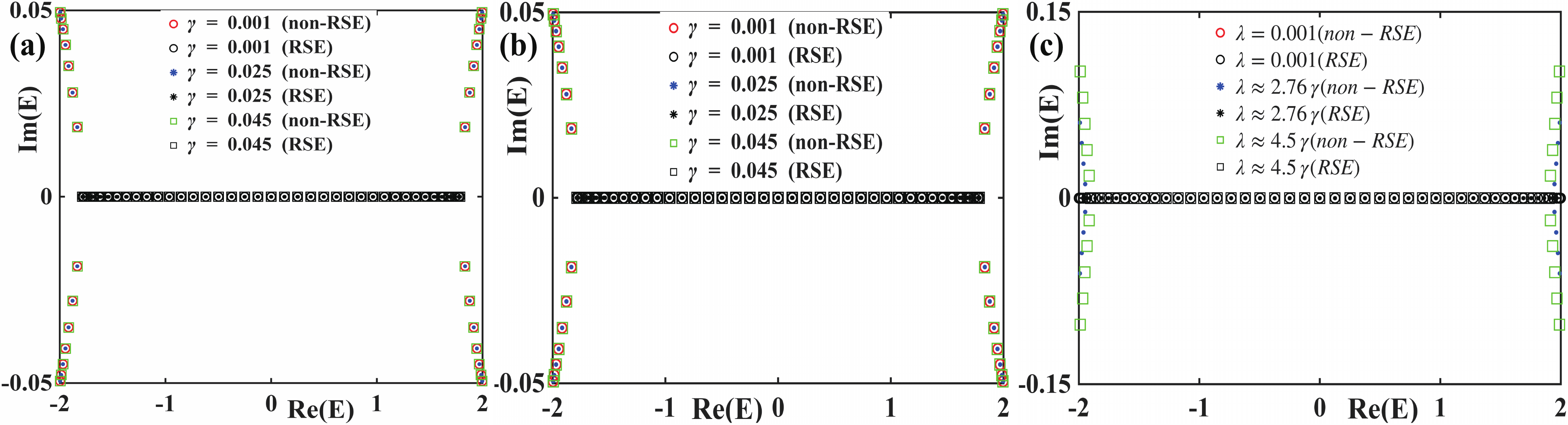}}  %
\caption{Complex energy spectra during the electric--magnetic phase transition. Black markers represent modes subject to $\Phi_a$ and involved in the electric--magnetic phase transition; colored markers represent modes without RSE. (a) Piecewise-constant imaginary electric potential with coincident electric--magnetic IDWs. (b) Piecewise-constant imaginary electric potential with non-coincident electric--magnetic IDWs. Both panels show spectra for $\gamma = 0.001, 0.025, 0.045$ (with imaginary electric potential strength $\lambda=0.05$); the modes undergoing the transition consistently lie on the real axis ($\text{Im}(E) \approx 0$). (c) Sinusoidal imaginary electric potential model, with fixed $\gamma=0.025$, showing three cases: $\lambda=0.001$, $\lambda \approx 2.76\gamma$, and $4.5\gamma$. The imaginary part of the spectrum for the transitioning modes remains pinned near zero. Parameters: $t=1$, $N=100$.
}
\label{fig:E-phase}
\end{figure}

\subsection{Complex energy spectra}
This section presents the complex energy spectra for the three imaginary electric potential configurations discussed in the preceding three sections, providing a direct visualization of the spectral signatures of the electric--magnetic phase transition. In Fig.~1(c) of the main text, we have already noted that the energy spectrum can be divided into two categories: RSE modes dominated by the spatial average $\Phi_a$ of the imaginary potential field, and non-RSE modes not dominated by $\Phi_a$ (corresponding to the red and blue markers in Fig.~1(c) of the main text). Since the electric--magnetic phase transition is fundamentally a competition between RSE (electric type) and NSE (magnetic type), the transition phenomenon occurs only among eigenstates exhibiting RSE modes, i.e., the eigenstates marked in black in Fig.~\ref{fig:E-phase}. The modes without RSE (colored markers) are not of concern and are unrelated to the electric--magnetic phase transition.

According to the GNHSE theory, the imaginary part of the eigenenergy is primarily determined by the spatial average $\Phi_a^\pm$ of the unified imaginary potential field $\Phi_{k_\pm}(x)$ (i.e., $E_n = v_{k_\pm} k - i\Phi_a^\pm$). For the piecewise-constant imaginary electric potentials shown in Figs.~\ref{fig:E-phase}(a)-(b), regardless of whether the electric--magnetic IDWs coincide or not, the spatial averages of both the electric scalar potential and the magnetic vector potential vanish ($\Phi_{a,E}=0$, $\Phi_{a,M}=0$), so the spatial average of the unified imaginary potential field is $\Phi_a^\pm = 0$. This causes the imaginary part of the spectrum for the transitioning modes to be extremely small (theoretically zero, strictly on the real axis within the perturbative approximation), manifesting as a purely real spectrum in the complex plane. Even as the nonreciprocal parameter $\gamma$ is varied, the transitioning modes (black markers) remain tightly pinned to the real axis.

This spectral behavior reveals an important distinction between the GNHSE theory and the conventional GBZ theory in characterizing non-Hermitian phase transitions: within the conventional GBZ framework, skin effect phase transitions are typically accompanied by changes in the topology of the complex spectrum (such as winding numbers) and the closing of point gaps \cite{yao_EdgeStatesTopological_2018}, with the complex spectrum itself carrying key information about the transition. In contrast, in the electric--magnetic phase transition within the GNHSE framework, since the value of $\Phi_a^\pm$ does not undergo a fundamental change across the transition, the critical information of the transition is entirely encoded in the envelope behavior of the eigenstates in real space. It should be emphasized that the GNHSE theory studied in this work concerns bulk states, not zero modes or boundary states residing in band gaps. The interplay between GNHSE and topological boundary states is one of the directions for future research.

For the sinusoidal imaginary electric potential transition shown in Fig.~\ref{fig:E-phase}(c), we fix the magnetic potential $\gamma=0.025$ and vary the electric potential strength $\lambda$. One can observe that the imaginary part of the spectrum for the transitioning modes of interest ($\Phi_a^\pm \approx 0$) remains pinned near zero, with little variation as $\lambda$ changes. This once again corroborates the central tenet of GNHSE: as long as the global spatial average of the imaginary potential field remains unchanged, the imaginary part of the spectral structure remains robust. The diversity of transition pathways (such as the abrupt jump for coincident IDWs versus the continuous drift for the sinusoidal potential) is determined solely by the local envelope structure of $\Phi(x)$ in real space, without requiring any change in the topology of the complex spectrum.

\section{Theoretical analysis of the Fourier filtering separation method}
In the study of non-Hermitian systems containing spatially inhomogeneous imaginary potentials $\mathrm{i}\Phi(x)$, the relative skin effect (RSE) exhibits boundary-insensitive characteristics that are fundamentally distinct from the conventional non-Hermitian skin effect. According to the GNHSE theory, the system possesses a pair of nearly degenerate eigenmodes ($k_+$ and $k_-$ states) with opposite group velocities in the real-energy spectral region, which localize at different imaginary domain walls (IDWs). However, when obtaining numerical solutions for finite-size systems via exact diagonalization, the numerical eigenstates often appear as hybridized distributions localized simultaneously at both IDWs, obscuring the independent intrinsic localization characteristics of the $k_\pm$ modes. To recover the physical profiles of these two classes of modes as rigorously and losslessly as possible, we establish the theoretical mechanism for separating degenerate states via momentum-space Fourier filtering, based on the physical picture of frequency-division multiplexing of wavefunctions. Figs.~5(b)-(d) in the main text employ this method.

\noindent{\textbf{Spatial structure of RSE eigenmodes and the momentum--velocity correspondence}}\\
We first construct the analytical forms of the $k_+$ and $k_-$ modes from theory. Two states with opposite group velocities $v_{k_\pm} = \frac{\partial \mathrm{Re}(E(k))}{\partial k}\big|{k\pm}$ are known to exist. We define: the $k_+$ mode (right-moving mode) with group velocity $v_{k_+} > 0$, and the $k_-$ mode (left-moving mode) with group velocity $v_{k_-} < 0$. The physical momenta carried by the two must be opposite. Taking the simple hopping model as an example: the Hermitian Hamiltonian is $H_h(k) = 2t\cos k$, and at real energy $E = 2t\cos k_0$ (with $k_0 \in (0, \pi)$), two states with opposite group velocities $v_k = -2t\sin k$ exist, giving $k_+ = -k_0$ and $k_- = +k_0$.

Under weak non-Hermitian perturbation, the real-space wavefunctions of these two modes can be written as the product of a high-frequency momentum carrier and a slowly varying localization envelope:
\begin{equation}\label{eq:wavefunction_ansatz}
\psi_{k_+}(x) = f_+(x) e^{-i k_0 x}, \quad \psi_{k_-}(x) = f_-(x) e^{+i k_0 x}
\end{equation}
where the real envelope functions $f_\pm(x) \propto \exp\left( \frac{2}{v_{k_\pm}} \int_0^x [\Phi(x') - \Phi_a] dx' \right)$. Since $v_{k_+}$ and $v_{k_-}$ have opposite signs, these two envelope functions inevitably separate in space, each exponentially localizing at different nodes of $\Phi(x)$ (i.e., different IDWs).

In finite-size systems, because the distance between the two IDWs is finite, the exponentially decaying tails of the envelope functions $f_+(x)$ and $f_-(x)$ overlap in the bulk of the system. This spatial overlap of the wavefunctions triggers a finite-size tunneling effect in quantum mechanics, causing the originally strictly degenerate macroscopic energy levels to undergo a small level repulsion, splitting into a nearly degenerate pair $E_0 \pm \delta E$. Due to this double-well-type hybridization, the numerical eigenstates $\Psi^{\text{num}}(x)$ obtained from exact diagonalization become linear superpositions of the $k_+$ and $k_-$ modes:
\begin{equation}\label{eq:numerical_state}
\Psi^{\text{num}}(x) = \alpha \psi_{k_+}(x) + \beta \psi_{k_-}(x) = \alpha f_+(x) e^{-i k_0 x} + \beta f_-(x) e^{+i k_0 x},
\end{equation}
This is physically equivalent to a ``standing-wave-like'' state formed by the interference of two traveling waves propagating in opposite directions. Consequently, the numerically observed wavefunction modulus squared $|\Psi^{\text{num}}(x)|^2$ exhibits peaks at both IDWs simultaneously, as shown in Fig.~1(d) of the main text. To observe the pure RSE localization modes, one must mathematically deconstruct this hybridized basis.

\noindent{\textbf{Spectral deconstruction in momentum space and the frequency shift theorem}}\\
To separate the above hybridized states, we apply a Fourier transform to momentum space ($q$-space). Let $\tilde{f}_\pm(q) = \int f_\pm(x) e^{-iqx} dx$ denote the momentum spectrum of the purely real envelope functions. By the properties of the Fourier transform, $\tilde{f}_\pm(q)$ is necessarily a broadened peak centered near $q=0$. Performing a global Fourier transform on the hybridized wavefunction $\Psi^{\text{num}}(x)$, the \textbf{frequency shift theorem} dictates that the complex exponential carriers $e^{\mp i k_0 x}$ rigidly shift the envelope spectrum:
\begin{equation}\label{eq:FT_decomposition}
\begin{aligned}
\tilde{\Psi}(q) &= \int_{-\infty}^{\infty} \left[ \alpha f_+(x) e^{-i k_0 x} + \beta f_-(x) e^{+i k_0 x} \right] e^{-iqx} dx \\
&= \alpha \tilde{f}_+(q + k_0) + \beta \tilde{f}_-(q - k_0),
\end{aligned}
\end{equation}
This reveals the momentum-space structure of the RSE degenerate states: the spectral profile $\tilde{f}_+(q + k_0)$ of the $k_+$ mode is shifted to $q = -k_0$ (\textbf{negative momentum region}), while the spectral profile $\tilde{f}_-(q - k_0)$ of the $k_-$ mode is shifted to $q = +k_0$ (\textbf{positive momentum region}).

This bipartition of momentum space provides the theoretical basis for separating the wavefunctions using momentum projection operators. We can define half-space momentum projection operators $P_{q<0}$ and $P_{q>0}$, but whether such frequency-domain truncation can losslessly recover the real-space wavefunction (i.e., avoid Gibbs artifacts) depends entirely on the degree of overlap between the two peaks in momentum space. According to the uncertainty principle, the localization length $\xi$ of the envelope function in real space and its spectral width $\Delta q$ in momentum space satisfy $\Delta q \sim 1/\xi$. Perfect lossless separation therefore requires that the center-to-center distance of the two peaks far exceed their individual spectral widths:
\begin{equation}\label{eq:separation_condition}
2k_0 \gg \Delta q \implies k_0 \gg \frac{1}{\xi}.
\end{equation}
In the effective model of the RSE, the localization exponent is $2\Phi^s(x) / v_{k_0} = x / \xi$. Given that the group velocity $v_{k_0} \sim \mathcal{O}(t)$ and the potential integral $\Phi^s(x) \sim \mathcal{O}(\lambda x)$, we can derive:
\begin{equation}\label{eq:localization_length}
\frac{1}{\xi} \sim \mathcal{O}\left(\frac{\lambda}{t}\right).
\end{equation}
This indicates that the spectral broadening width in momentum space is directly determined by the ratio of the non-Hermitian strength to the Hermitian hopping energy. In the \textbf{perturbative non-Hermitian limit ($\lambda \ll t$)} adopted in this work, the localization length is macroscopically large, leading to $\Delta q \to 0$, while the carrier momentum $k_0 \sim \mathcal{O}(1)$ remains finite (except at the band edges). Therefore, the inequality $k_0 \gg 1/\xi$ is naturally satisfied, and the two modes appear in momentum space as two sharp, completely orthogonal, non-interfering wave packets. Based on this theoretical analysis, we can apply filtering operations to the numerical wavefunction:
\begin{equation}\label{eq:filtering}
\begin{aligned}
\psi_{k_+}^{\text{filtered}}(x) &\propto \mathcal{F}^{-1} \left\{ P_{q < 0} \left[ \mathcal{F}(\Psi^{\text{num}}) \right] \right\}, \\
\psi_{k_-}^{\text{filtered}}(x) &\propto \mathcal{F}^{-1} \left\{ P_{q > 0} \left[ \mathcal{F}(\Psi^{\text{num}}) \right] \right\},
\end{aligned}
\end{equation}
successfully deconstructing an entangled ``standing wave'' basis into two eigenstates with opposite momenta and opposite group velocities.

In summary, the Fourier filtering separation of numerically degenerate states is founded on the rigorous group velocity--momentum correspondence and the wavefunction basis deconstruction under the perturbative non-Hermitian limit. This method is not only mathematically rigorous and self-consistent, but also confirms from the momentum-space perspective the core physical prediction of the GNHSE theory that the $k_\pm$ modes, as independent degrees of freedom, localize at different IDWs.\\
\emph{Note: This theory breaks down at the band edges $E \to \pm 2t$, where $v_k \to 0$ and non-Hermitian exceptional point transitions occur. This is fully consistent with the numerical observation that the band-edge wavefunction behavior deviates from the GNHSE theoretical predictions.}

\section{MATLAB Code for Figures}
This section presents the MATLAB code used to generate Figures in the main text.

\subsection{Figs.~1}

\subsubsection{Figs.~1(a)-(b)}
\begin{lstlisting}
clear
clc
mp.Digits(64) %high-precision computation plugin: https://www.advanpix.com
N=200;

r3=0.1;  %nonreciprocal hopping amplitude
t1=1; 

[Hs,Hop]=Hsum(N,t1,r3);
[VR,VE]=eig(mp(Hs));%
me=diag(VE);
[~,ind]=sort(real(me),'descend'); %
me=me(ind);
VR=VR(:,ind);

[VRo,VEo]=eig(mp(Hop));%
meo=diag(VEo);
[~,ind]=sort(real(meo),'descend'); %
meo=meo(ind);
VRo=VRo(:,ind);

figure
plot(real(me),imag(me),'.r','MarkerSize',27)%, 
hold on 
plot(real(meo),imag(meo),'*b','MarkerSize',8)%,
legend({'Eigenstates under PBC', 'Eigenstates under OBC'},...
       'fontname','Times New Roman',...
       'FontSize',18,...
       'fontweight','bold',...
       'Location','best');
xlabel('Re(E)','fontname','Times New Roman','FontSize',22,'fontweight','bold');
ylabel('Im(E)','fontname','Times New Roman','FontSize',22,'fontweight','bold');
set(gca,'fontname','Times New Roman','FontSize',20,'fontweight','bold');%
box on;

figure
for i=1:N
ve=VR(:,i);   
plot(L, ve.*conj(ve), 'r', 'LineWidth', 2);  
hold on

veo=VRo(:,i);   
plot(L, veo.*conj(veo), 'b', 'LineWidth', 2); 
hold on
end

legend('PBC', 'OBC', 'Location', 'best');
xlabel('$j$','Interpreter','latex','fontsize',13)
ylabel('$|\psi|^{2}$','Interpreter','latex','fontsize',13)
set(gca,'FontName','Times New Roman','FontSize',20)

function[Hs,Hop]=Hsum(N,t1,r3)
Hk0=zeros(N);Hp=Hk0;
for i=1:N-1  
    ip=rem(i,N)+1;
      Hk0(ip,i)=t1+r3; %
      Hk0(i,ip)=t1-r3; %
end
Hp(1,N)=t1+r3; %
Hp(N,1)=t1-r3; %

Hs=Hk0+Hp;   
Hop=Hk0;  
end
\end{lstlisting}

\subsubsection{Figs.~1(c)-(d)}
\begin{lstlisting}
clear
clc
mp.Digits(34) %high-precision computation plugin: https://www.advanpix.com

N=200;

r3=0.02; %imaginary scalar potential  field strength
t1=1; 
phia=0;   %  spatial average of the potential \Phi_a

[Hs,Hop]=Hsum(N,t1,r3);
[VR,VE]=eig(mp(Hs));%
me=diag(VE);
[~,ind]=sort(real(me),'descend'); %
me=me(ind);
VR=VR(:,ind);
ind_logic = abs(imag(me) - phia) < 1.5e-3;
mem=me( ind_logic);
VRm=VR(:, ind_logic);

[VRo,VEo]=eig(mp(Hop));%
meo=diag(VEo);
[~,ind]=sort(real(meo),'descend'); %
meo=meo(ind);
VRo=VRo(:,ind);
ind_logic = abs(imag(meo) - phia) < 1.5e-3;  
memo=meo(ind_logic );
VRmo=VRo(:, ind_logic);

figure
plot(real(me),imag(me),'.r','MarkerSize',27)%, 
hold on 
plot(real(meo),imag(meo),'*b','MarkerSize',7)%,
hold on 
plot(real(mem),imag(mem),'.k','MarkerSize',27)
hold on 
plot(real(memo),imag(memo),'*k','MarkerSize',7)
xlabel('Re(E)','fontname','Times New Roman','FontSize',22,'fontweight','bold');
ylabel('Im(E)','fontname','Times New Roman','FontSize',22,'fontweight','bold');
set(gca,'fontname','Times New Roman','FontSize',20,'fontweight','bold');%
box on;
legend({'Eigenstates under PBC', 'Eigenstates under OBC', 'Skin states under PBC', 'Skin states under OBC'},...
       'fontname','Times New Roman',...
       'FontSize',26,...
       'fontweight','bold',...
       'Location','best');

figure
for i=1:2
ve=VRm(:,i);   
plot(1:N, ve.*conj(ve), 'r', 'LineWidth', 2);  
hold on

veo=VRmo(:,i);   
plot(1:N, veo.*conj(veo), 'c', 'LineWidth', 2); 
hold on
end
xlabel('$j$','Interpreter','latex','fontsize',13)
ylabel('$|\psi|^{2}$','Interpreter','latex','fontsize',13)
set(gca,'FontName','Times New Roman','FontSize',20)
legend('PBC', 'OBC', 'Location', 'best');

function[Hs,Hop]=Hsum(N,t1,r3)
Hk0=zeros(N);Hp=Hk0;
for i=1:N-1  
    ip=rem(i,N)+1;
      Hk0(ip,i)=t1; %
      Hk0(i,ip)=t1; %
end
Hp(1,N)=t1; %
Hp(N,1)=t1; %
Hx=diag([-1i*r3*ones(N/4,1);1i*r3*ones(N/2,1);-1i*r3*ones(N/4,1)]);

Hs=Hk0+Hp+Hx;  
Hop=Hk0+Hx;  
end
\end{lstlisting}

\subsection{Figs.~2}
\begin{lstlisting}
clear
clc
mp.Digits(34) %high-precision computation plugin: https://www.advanpix.com

N=200; 

t1=1;
r3=0.02;  % imaginary scalar potential  field strength
phia=0;   %  spatial average of the potential \Phi_a

[Hs,Hop]=Hsum(N,t1,r3);
[VR,VE]=eig(mp(Hs));%
me=diag(VE);
[~,ind]=sort(real(me),'descend'); %
me=me(ind);VR=VR(:,ind);
mem=me( abs( (imag(me)) )< 1e-13);
VRs=VR(:,abs( (imag(me)) )< 1e-13);

[VRo,VEo]=eig(mp(Hop));%
meo=diag(VEo);
[~,ind]=sort(real(meo),'descend'); %
meo=meo(ind);
VRo=VRo(:,ind);
ind_logic = abs(imag(meo) - phia) < 1.5e-3;  
memo=meo(ind_logic );
VRmo=VRo(:, ind_logic);

figure
for i=1:2
ve=VRs(:,i);   
plot(1:N, ve.*conj(ve), 'r', 'LineWidth', 2);  
hold on

veo=VRmo(:,i);   
plot(1:N, veo.*conj(veo), 'c', 'LineWidth', 2); 
hold on
end
xlabel('$j$','Interpreter','latex','fontsize',13)
ylabel('$|\psi|^{2}$','Interpreter','latex','fontsize',13)
set(gca,'FontName','Times New Roman','FontSize',20)
ax = gca; % 
ax.LineWidth = 2; %
legend('PBC', 'OBC', 'Location', 'best');

xiR=zeros(1,size(mem,1)/2);kt=xiR; 
for ir=1:size(mem,1)/2
ve=VRs(:,ir);
xiR(ir)=funcfit(N,ve);   
kt(ir)=acos( mem(ir)/(2*t1) );
end
kt=kt.'; xiR=xiR.';
theo=r3./( t1.*sin(kt));

[~,ind]=sort(real(1./xiR),'descend'); %
xiR=xiR(ind);
[~,ind]=sort(real(1./theo),'descend'); %
theo=theo(ind);

figure
plot(1:size(mem,1)/2,xiR,'.r','MarkerSize', 25)
hold on
plot(1:size(mem,1)/2,theo,'b','LineWidth', 2)
legend('Numerical','Theoretical','fontsize',24)
xlabel('$States$','Interpreter','latex','fontsize',24)
ylabel('$\xi_{RSE}$','Interpreter','latex','fontsize',24)
set(gca,'FontName','Times New Roman','FontSize',19)
ax = gca; 
ax.LineWidth = 1.5; % 
ax.YAxis.Exponent=-2;

%%
function [xiR]= funcfit(N,ve)
 x =(1:N/2).' ; 
 y=double( ve(N/2+1:N).*conj(  ve(N/2+1:N)  ) ); 

fit_model = @(params, x) params(1)*exp( -x*( params(2) )) + params(1)*exp(  (x-N/2)*( params(2) ) )  ;
 
initial_guess = [max(y)/2, 0]; % 
 
lb = [0, 0]; % 
ub = [Inf, Inf]; % 
options = optimoptions('lsqcurvefit', 'Display', 'off', ... % 
    'MaxIterations', 4000, ... % 
    'FunctionTolerance', 1e-15, ... % 
    'StepTolerance', 1e-10, ... % 
   'Algorithm', 'levenberg-marquardt'); %,
[params, ~] = lsqcurvefit(fit_model, initial_guess, x, y,lb,ub,options);
xiR=params(2); 
end

function[Hs,Hop]=Hsum(N,t1,r3)
Hk0=zeros(N); Hp=Hk0;
for i=1:N-1  
    ip=rem(i,N)+1;
      Hk0(ip,i)=t1; %
      Hk0(i,ip)=t1; %
end
Hp(1,N)=t1; %
Hp(N,1)=t1; %

Hx=diag( [-1i*r3*ones(fix(N/2),1);1i*r3*ones(fix(N/2),1)] );

Hs=Hk0+Hp+Hx; % 
Hop=Hk0+Hx; %
end
\end{lstlisting}

\subsection{Fig.~3(b)}
\begin{lstlisting}
clear
clc
mp.Digits(64) %high-precision computation plugin: https://www.advanpix.com

N=200;

r1=0.05;r2=r1;  %nonreciprocal hopping amplitude
t1=1; 

Hs=Hsum(N,t1,r1,r2);
[VR,VE]=eig(mp(Hs));%
me=diag(VE);
[~,ind]=sort(real(me),'descend'); %
me=me(ind);
VR=VR(:,ind);

figure
for ix=100:110  
    ve=VR(:,ix);   
plot(1:N, ve.*conj(  ve  ),'b');  
hold on
end
xlabel('$x$','Interpreter','latex','fontsize',13)
ylabel('$|\psi|^{2}$','Interpreter','latex','fontsize',13)
set(gca,'FontName','Times New Roman','FontSize',20)
ax = gca; 
ax.LineWidth = 2; 

function[Hk0]=Hsum(N,t1,r1,r2)
Hk0=zeros(N);
for i=1:N/2 
      Hk0(i+1,i)=t1+r1; %
      Hk0(i,i+1)=t1-r1; %
end

for i=N/2+1:N  
    ip=rem(i,N)+1;
      Hk0(ip,i)=t1-r2; %
      Hk0(i,ip)=t1+r2; %
end

end
\end{lstlisting}

\subsection{Figs.~5(a)}
The programming logic is as follows:
 
1. First, set $\gamma = 0$ and vary $\lambda$, then fit the eigenstate distribution to obtain the change in the inverse decay length of the eigenstates under pure RSE conditions. This result serves as the baseline, denoted as "xiR".

\begin{lstlisting}
clear
clc
mp.Digits(34) %High-precision computation library provided by https://www.advanpix.com/
%% Parameter settings
%r_1 and r_2 represent the non-reciprocal hopping terms in the left and right subsystems, respectively.
%r_3 denotes the on-site imaginary potential,t_1 represents the nearest-neighbor hopping strength.
%xiR inverse decay lengths of RSE 

N=100; %Number of lattice sites

t1=1; r1=-0.00;r2=r1;

r3=linspace(0.01, 0.075, 300);
xiR=zeros(1,size(r3,2));
nr=size(r3,2);

parfor ir=1:nr
Hs=Hsum(N,t1,r1,r2,r3(ir));
[VR,VE]=eig(mp(Hs));%
me=diag(VE);
[~,ind]=sort(real(me),'descend'); %
me=me(ind);VR=VR(:,ind);
ve=VR(:,N/2);
[xiR(ir)]=funcfit(N,ve);
end

save('basic.mat','xiR','t1','r1','r3')

function [xiR,xiG]= funcfit(N,ve)
%This function is a numerical fitting of the modulus square distribution of the eigenstates, which is used to determine the inverse decay lengths for GSE and RSE.
%Since the eigenstates are symmetric about the lattice point j = 50, for convenience, only half of the modulus square distribution is taken.
x =(1:N/2).' ; 
y=double( ve(N/2+1:N).*conj(  ve(N/2+1:N)  ) ); 

fit_model = @(params, x) params(1)*exp( -x*( params(4) +params(2) )) + params(3)*exp(  (x-N/2)*( params(4)- params(2) ) )  ;
initial_guess = [max(y), 0, max(y),0]; 
  
lb = [0, 0, 0 ,0];  
ub = [Inf, Inf, Inf ,Inf]; 
options = optimoptions('lsqcurvefit', 'Display', 'off', ...
    'MaxIterations', 4000, ...  
    'FunctionTolerance', 1e-15, ... 
    'StepTolerance', 1e-10, ... 
   'Algorithm', 'levenberg-marquardt'); 
[params, ~] = lsqcurvefit(fit_model, initial_guess, x, y,lb,ub,options);
xiR=params(4);  xiG=params(2);
end

function[Hs]=Hsum(N,t1,r1,r2,r3)
H0=zeros(N);
for i=1:N/2  
      H0(i+1,i)=t1+r1; 
      H0(i,i+1)=t1-r1; 
end

for i=N/2+1:N  
    ip=rem(i,N)+1;
      H0(ip,i)=t1-r2; 
      H0(i,ip)=t1+r2; 
end

Hx=diag( [1i*r3*ones(fix(N/2),1);-1i*r3*ones(fix(N/2),1)] );
Hs=H0+Hx;  
end
\end{lstlisting}

2. Next, starting from this baseline ("xiR"), gradually increase $\gamma$, and fit the eigenstate distribution again to obtain the change in the inverse decay length of the eigenstates after introducing NSE. This data is recorded as "xiG".

\begin{lstlisting}
clear
load("basic.mat")
b=xiR;
mp.Digits(34) %High-precision computation library provided by https://www.advanpix.com/
%% Parameter settings
%r_1 and r_2 represent the non-reciprocal hopping terms in the left and right subsystems, respectively.
%r_3 denotes the on-site imaginary potential,t_1 represents the nearest-neighbor hopping strength.
%xiG inverse decay lengths of NSE

t1=1;
r1=linspace(1e-4, 0.05, 300); nrf=size(r1,2);
r3=linspace(0.01, 0.075, 300); nrt=size(r3,2);

xiG=zeros(size(r3,2),size(r1,2));
total_points = nrf * nrt;
parfor k = 1:total_points
    r11=r1; r31=r3; b1=b;
    [irt, irf] = ind2sub([nrt, nrf], k);
    r1_val = r11(irf); r3_val = r31(irt);
    Hs = Hsum(N, t1, r1_val, r1_val, r3_val);
    [VR, VE] = eig(mp(Hs));
    me=diag(VE);
    [~,ind]=sort(real(me),'descend'); %
    me=me(ind);VR=VR(:,ind);
    ve = VR(:, N/2);
    xiG(k) = funcfit(N, ve, b1(irt));
end
save('cal.mat','xiG','t1','r1','r3')

function [xiG]= funcfit(N,ve,b)
x =(1:N/2).' ;
y=double( ve(N/2+1:N).*conj(  ve(N/2+1:N)  ) );

fit_model = @(params, x) params(1)*exp( -x*( b +params(2) ) ) + (params(1)-params(3))*exp(  (x-N/2)*( b- params(2) ) )  ;
initial_guess = [max(y)/2, 0, 0];

lb = [0, 0, 0]; ub = [Inf, Inf, Inf];
options = optimoptions('lsqcurvefit', 'Display', 'off', ...
    'MaxIterations', 4000, ...
    'FunctionTolerance', 1e-10, ...
    'StepTolerance', 1e-7, ...
    'Algorithm', 'levenberg-marquardt');
[params, ~] = lsqcurvefit(fit_model, initial_guess, x, y,lb,ub,options);
xiG=params(2);
end

function[Hs]=Hsum(N,t1,r1,r2,r3)
H0=zeros(N);
for i=1:N/2
    H0(i+1,i)=t1+r1;
    H0(i,i+1)=t1-r1;
end

for i=N/2+1:N
    ip=rem(i,N)+1;
    H0(ip,i)=t1-r2;
    H0(i,ip)=t1+r2;
end

Hx=diag( [1i*r3*ones(fix(N/2),1);-1i*r3*ones(fix(N/2),1)] );
Hs=H0+Hx;
end
\end{lstlisting}

3. Finally, subtract the two results ("xiR - xiG") to compute the order parameter $\eta$ introduced in the main text. By identifying the zero points of $\eta$, the numerical phase boundary can be determined.

\begin{lstlisting}
clear 
load("basic.mat")
n=size(xiR,1);
xiRb = repmat(xiR,1,n);
load("cal.mat")

[X, Y] = meshgrid(r1,r3); % 1:n,1:n 
pic=xiRb-xiG;  %Inverse decay length during the competition between RSE and NSE

figure;
surf(X, Y, pic, 'EdgeColor', 'none');
colormap(jet(256));                         
hCB = colorbar;
set(get(hCB, 'Label'), 'String', '$Order\;Parameter(\eta)$',... 
    'FontSize', 24,...                         
    'FontName', 'Times New Roman',...                   
    'Interpreter', 'latex');                    
set(gca,'FontName','Times New Roman','FontSize',16)
ax = gca;  
ax.LineWidth = 1.5; 
ax.XTick = [1e-4 0.015 0.035 max(r1)]; % 
ax.YTick = [0.015 0.035 0.055 max(r3)]; % 
ax.XAxis.Exponent=-3;
ax.YAxis.Exponent=-3;

hold on;
[hCont, hContGroup] = contour3(X, Y, pic, [0 0],... 
    'LineWidth', 3.5,...
    'LineColor', 'w',...
    'LineStyle', '--',...
    'DisplayName', 'Numerical (\eta=0)'); 
 
r1_line = linspace(min(r1), max(r1), 100);
r3_line = 2 * r1_line;
hTheory = plot3(r1_line, r3_line,...
    ones(size(r3_line))*(max(pic(:)) ),...
    'k-', 'LineWidth', 1.5,...
    'DisplayName', 'Theoretical (\lambda=2\gamma)');  
 
legend([hContGroup, hTheory],...             
    'FontSize', 26,...
    'Color' ,[0.7 0.7 0.7]);             

view(2);
xlabel('$Strength\; of\; NSE(\gamma)$','Interpreter','latex','fontsize',24)
ylabel('$Strength\; of\; RSE(\lambda)$','Interpreter','latex','fontsize',24)
shading interp;

axis([ min(r1) max(r1) min(r3) max(r3)]); 
\end{lstlisting}

\subsection{Fig.~6}
\begin{lstlisting}
clear
clc
mp.Digits(34) %high-precision computation plugin: https://www.advanpix.com

N=100;

r3=0.0001; %imaginary scalar potential  field strength
t1=1; 
phia=-r3*(N^2)/(12); %  spatial average of the potential \Phi_a

[Hs,Hop]=Hsum(N,t1,r3);
[VR,VE]=eig(mp(Hs));%
me=diag(VE);
[~,ind]=sort(real(me),'descend'); %
me=me(ind);
VR=VR(:,ind);
ind_logic = abs(imag(me) - phia) < 2e-3;
mem=me( ind_logic);
VRm=VR(:, ind_logic);

[VRo,VEo]=eig(mp(Hop));%
meo=diag(VEo);
[~,ind]=sort(real(meo),'descend'); %
meo=meo(ind);
VRo=VRo(:,ind);
ind_logic = abs(imag(meo) - phia) < 2e-3;   
memo=meo(ind_logic );
VRmo=VRo(:, ind_logic);

figure
plot(real(me),imag(me),'.r','MarkerSize',27)%, 
hold on 
plot(real(meo),imag(meo),'*b','MarkerSize',7)%,
hold on 
plot(real(mem),imag(mem),'.k','MarkerSize',27)
hold on 
plot(real(memo),imag(memo),'*k','MarkerSize',7)

xlabel('Re(E)','fontname','Times New Roman','FontSize',22,'fontweight','bold');
ylabel('Im(E)','fontname','Times New Roman','FontSize',22,'fontweight','bold');
set(gca,'fontname','Times New Roman','FontSize',20,'fontweight','bold');%
box on;
legend({'Eigenstates under PBC', 'Eigenstates under OBC', 'Skin states under PBC', 'Skin states under OBC'},...
       'fontname','Times New Roman',...
       'FontSize',26,...
       'fontweight','bold',...
       'Location','best');

figure
for i=1:5
ve=VRm(:,i);   
plot(1:N, ve.*conj(ve), 'r', 'LineWidth', 2);  
hold on

veo=VRmo(:,i);   
plot(1:N, veo.*conj(veo), 'b', 'LineWidth', 2); 
hold on
end
xlabel('$j$','Interpreter','latex','fontsize',13)
ylabel('$|\psi|^{2}$','Interpreter','latex','fontsize',13)
set(gca,'FontName','Times New Roman','FontSize',20)
legend('PBC', 'OBC', 'Location', 'best');

function[Hs,Hop]=Hsum(N,t1,r3)
Hk0=zeros(N);Hp=Hk0;
for i=1:N-1  
    ip=rem(i,N)+1;
      Hk0(ip,i)=t1; %
      Hk0(i,ip)=t1; %
end

Hp(1,N)=t1; %
Hp(N,1)=t1; %
x=1:N;Hx=diag( -1i*r3*(x-N/2-0.5).^2);  

Hs=Hk0+Hp+Hx;   
Hop=Hk0+Hx;  
end
\end{lstlisting}